%% file: ms3.tex
\documentclass[iop]{emulateapj}
\usepackage[utf8]{inputenc}
\usepackage[T1]{fontenc}
\usepackage{lmodern}
\usepackage{graphicx}
\usepackage{longtable}
\usepackage{subfigure}

\newcommand{\noc}{111}
\newcommand{\nv}{567}
\newcommand{\saturated}{255}
\newcommand{\vp}{70}
\newcommand{\offchip}{20}
\newcommand{\duplicates}{18}
\newcommand{\vble}{165}
\newcommand{\ecl}{27}
\newcommand{\ob}{11}

\newcommand{\lpv}{18}

\newcommand{\ngood}{843}
\newcommand{\chancesat}{98}
\newcommand{\rscvns}{27}
\newcommand{\WUMa}{11}
\newcommand{\shortper}{19}
\newcommand{\ellip}{5}

\newcommand{\flicker}{76}
\newcommand{\gae}{\lower 2pt \hbox{$\, \buildrel {\scriptstyle >}\over {\scriptstyle
\sim}\,$}}
\newcommand{\lae}{\lower 2pt \hbox{$\, \buildrel {\scriptstyle <}\over {\scriptstyle
\sim}\,$}}
\newcommand{\half}{\frac{1}{2}}

\revised{\today}

\shorttitle{Optical Variability of GBS sources}
\shortauthors{Britt et al.}

\begin{document}
\title{Variability of Optical Counterparts in the Chandra Galactic Bulge Survey}

\author{Britt, C.T.\altaffilmark{1,2,3}}
\author{Hynes, R.I.\altaffilmark{1,3}}
\author{Johnson, C.B.\altaffilmark{1}}
\author{Baldwin, A.\altaffilmark{1}}
\author{Jonker, P.G.\altaffilmark{4,5,6}} 
\author{Nelemans, G.\altaffilmark{5}}
\author{Torres, M.A.P.\altaffilmark{4,5,6}} 
\author{Maccarone, T.\altaffilmark{2}}
\author{Steeghs, D.\altaffilmark{7}} 
\author{Greiss, S.\altaffilmark{7}} 
\author{Heinke, C.\altaffilmark{8}}
\author{Bassa, C.G.\altaffilmark{9}}
\author{Collazzi, A.\altaffilmark{1}}
\author{Villar, A.\altaffilmark{10}}
\author{Gabb, M.\altaffilmark{11}}
\author{Gossen, L.\altaffilmark{1,3}}

\altaffiltext{1}{Louisiana State University, Department of Physics and
  Astronomy, Baton Rouge LA 70803-4001, U.S.A.}
\altaffiltext{2}{Department of Physics, Texas Tech University, Box
  41051, Science Building, Lubbock, TX 79409-1051, USA}
\altaffiltext{3}{Visiting astronomer, Cerro Tololo Inter-American Observatory, National Optical Astronomy Observatory, which are operated by the Association of Universities for Research in Astronomy, under contract with the National Science Foundation.}
\altaffiltext{4}{SRON, Netherlands Institute for Space Research, Sorbonnelaan 2, 3584~CA, Utrecht, The Netherlands}
\altaffiltext{5}{Dept. of Astrophysics, IMAPP, Radboud UniversityNijmegen, Heyendaalseweg 135, 6525 AJ, Nijmegen, The Netherlands}
\altaffiltext{6}{Harvard-Smithsonian Center for Astrophysics, 60 Garden Street, Cambridge, MA~02138, U.S.A.}
\altaffiltext{7}{Astronomy and Astrophysics, Dept. of Physics, University of Warwick, Coventry, CV4~7AL, United Kingdom}
\altaffiltext{8}{Department of Physics, University of Alberta, CCIS 4-183, Edmonton, AB T6G 2E1}
\altaffiltext{9}{Jodrell Bank Centre for Astrophysics, School of Physics and Astronomy, University of Manchester, Manchester M13 9PL, United Kingdom}
\altaffiltext{10}{Department of Physics, Massachussettes Institute of Technology, 77 Massachusetts Avenue, Cambridge, MA 02139-4307, USA}
\altaffiltext{11}{Department of Physics, Florida Atlantic University, 777 Glades Road, Boca Raton FL 33431-0991, USA}

\begin{abstract}
We present optical lightcurves of variable stars consistent with the
positions of X-ray sources identified with the
Chandra X-ray Observatory for the Chandra Galactic Bulge Survey. Using data
from the Mosaic-II instrument on
the Blanco 4m Telescope at CTIO, we gathered time-resolved photometric
data on timescales from $\sim2$\,hr to 8\,days over the $\frac{3}{4}$
of the X-ray survey containing sources from the initial GBS
catalog. Among the lightcurve morphologies we 
identify are flickering in interacting binaries, eclipsing sources, dwarf 
nova outbursts, ellipsoidal variations, long period variables, spotted stars, 
and flare stars. $87\%$ of X-ray sources have at least one
potential optical counterpart.  $24\%$ of these candidate counterparts are 
detectably variable; a much greater fraction than expected for randomly selected field 
stars, which suggests that most of these variables are real counterparts. We
discuss individual sources of interest, provide variability
information on candidate counterparts, and discuss the characteristics
of the variable population.

\end{abstract}

\section{Introduction}

In the past, Galactic X-ray surveys of faint sources have focused on
the Galactic Center and Plane, which carries the advantage of high source
density \citep{Muno03}, but also the disadvantages of high crowding and large 
optical extinction up to $A_{V}=30$. Together, those
disadvantages make the determination of optical or infrared counterparts to 
X-ray sources very difficult \citep{DeWitt10,Mauerhan09}. Off-center multiwavelength surveys 
such as the ChaMPlane survey of bright X-ray sources \citep{Grindlay05} and the XMM-Newton
Galactic Plane Survey \citep[XGPS]{Hands04} have had success in 
identifying significant numbers of optical/NIR counterparts in low extinction windows, typically
finding large numbers of coronally active stars and AGN, with a few Cataclysmic Variables (CVs) 
\citep{Motch10,Hands04,Servillat12,vandenBerg12}. Narrow, deep 
surveys in the Galactic Plane or Bulge find even more coronal sources, as well as more CVs.
\citep[e.g.]{vandenBerg09}. 
Surveys of X-ray sources in globular clusters \citep{Heinke03,Heinke05,Lu09} 
avoid the problem of high 
extinction in the Galactic Plane, but crowding is even more severe, requiring observations with 
HST or adaptive optics. 
Also, because X-ray Binary formation is dominated by dynamical processes 
\citep{Pooley03,Pooley06,Peacock09}, they do not offer a probe of binary evolution in 
the field. Because knowledge of 
the counterpart is necessary for using such diagnostic tools as the ratio of X-ray 
to optical luminosities, ellipsoidal modulations of the companion, and
optical and infrared spectroscopy, identification of counterparts is
critical to the classification of the X-ray source, especially for faint X-ray sources with few 
detected counts. For systems
accreting through Roche Lobe overflow, the masses of each component in
a binary can be determined entirely by measuring the velocity amplitude of the 
donor star along the line of sight $K_{2}$ and the rotational broadening 
$v\sin i$ through spectroscopy to determine the mass ratio $q$, and the
inclination angle $i$ either through the modeling of the ellipsoidal
variations of the companion or the detection of eclipses.

\subsection{Galactic Bulge Survey Design, Expectations, and Goals}

\begin{figure}
\begin{center}
\includegraphics[width=0.4\textwidth,angle=90]{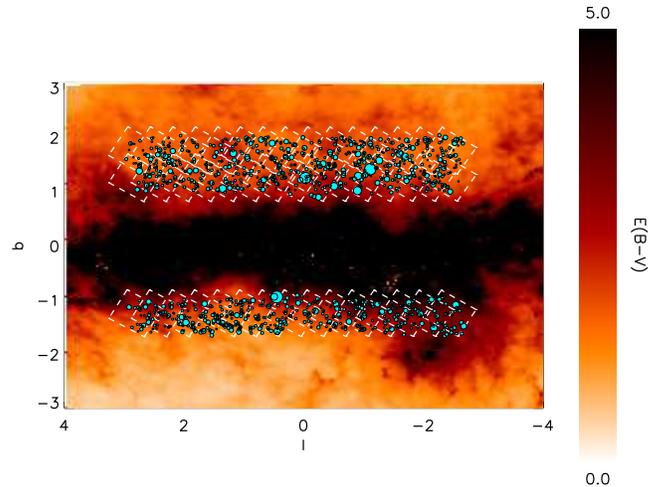}
\caption{GBS sources overlaid on the reddening map from
  \citet{Gonzalez12}. The size of each point is proportional to the
  brightness of the X-ray source at that position. The dashed lines
    are the outlines of Mosaic-II fields we used for the optical side
    of the survey. The southern edge of the Mosaic-II region also contains some
    CXB sources identified in \citet{Jonker14}.}
\label{fig:gbsmap}
\end{center}
\end{figure}

The Galactic Bulge Survey (GBS) \citep{Jonker11,Jonker14} is intended to avoid as much as
possible the problems of crowding and extinction present in previous
surveys of the Galactic Center, while giving up as little as possible
in the way of number of sources. The GBS makes use of both optical and
X-ray imaging of two $6\,^{\circ} \times 1\,^{\circ}$ strips located
$1.5\,^{\circ}$ above and below the Galactic Plane, cutting out the
region $b<1\,^{\circ}$ to avoid copious amounts of dust in the
Galactic plane. The GBS region and sources overlayed on the dust map from
\citet{Gonzalez12} can be seen in Figure \ref{fig:gbsmap}. The GBS
X-ray observations are short, only 2\,ks, in order to keep the
relative number of quiescent Low Mass X-ray Binaries (qLMXB) high
compared to CVs and to allow a wide survey area. 

There are multiple science goals to be achieved in such a census of
X-ray sources \citep{Jonker11}. We aim to expand greatly the known
number of Galactic X-ray binaries, including the likely discovery of the first
Galactic eclipsing black hole (BH) binary. The known population of BH and NS 
binaries is riddled with selection effects. For transient sources found in 
outburst, which includes most NS LMXBs and all dynamically confirmed BH LMXBs, 
the peak luminosity, outburst duration, and recurrence time depend strongly on
the orbital period \citep{Wu10}. There are no known dynamically confirmed BH LMXBs with 
$P_{orb} < 4\,$hours, which is a significant paucity \citep{Knevitt14}. There are, however, 3 
BHCs with $P_{orb}<4\,$hours, identified as BHCs based on their X-ray spectra 
\citep{Corral13,Kuulkers13,Zurita08}. Searching for LMXBs in 
quiescence avoids these selection effects against short period systems. By greatly increasing the
number of known LMXBs, we expect to increase
correspondingly the number of LMXBs for which mass
determinations are possible. Mass determinations of neutron stars
(NS) are useful in constraining the equation of state (EoS) of matter at 
super-nuclear density in ways that cannot be done on Earth. Identification 
of source class types is also valuable because it allows constraints to be 
placed on binary evolution models by comparing observed source class numbers 
to the predictions of population synthesis models. Such models vary widely in 
their predicted number of LMXBs in the Galaxy, from $10^{3}$ to $10^{5}$ 
\citep{Pfahl03,Zwart97,Kiel06}, depending on what assumptions are made. 
The common envelope stage of binary evolution typically dominates the uncertainties, 
though other important factors include the size of SN kicks, the initial binary fraction, the 
initial mass ratio distribution, and the initial orbital period distribution. 
Assuming a Galactic 
NS binary formation rate of 
$10^{-5}\,{\rm year^{-1}}$ and a typical lifetime of $10^{9}$ years, totalling 
$10^{4}$ qLMXBs in the Galactic population, 
\citet{Jonker11} predicted 71 qLMXBs with accessible optical
counterparts ($r'<23$)
in the GBS area, which corresponds to 53 qLMXBs in the region covered by the 
Mosaic-II optical observations we present here. We expect most of these to show 
ellipsoidal variations in their lightcurves, typically of $0.1-0.2$ magnitudes.

Before any of these science goals can be achieved, it is necessary to
identify properly the optical counterpart of the X-ray
source. Variability is a powerful tool for ensuring proper
counterpart identification. Using the OGLE III 
Catalog of Variable
Stars\footnote{http://ogledb.astrouw.edu.pl/\textasciitilde ogle/CVS/} 
\citep{Soszynski11a,Soszynski11b,Szymanski11}, $\sim98\%$ of field stars are 
non-variable to $\Delta I = 0.01$ in the range of $14<I<17$ 
photometry in the direction of the Bulge,
while many classes of sources that produce X-rays
should also have variable optical emission. Both \citet{Udalski12} and 
\citet{Hynes12} explore the variability of possible optical counterparts 
to GBS sources as well, using OGLE IV and ASAS data respectively. These 
papers focus on somewhat brighter optical sources than are considered here. 
The lightcurve morphology also can enable
determination of some system parameters, such as orbital period and
inclination angle. Variability searches also can reveal the presence
of high inclination systems through eclipses. Mass determinations of CVs and 
LMXBs are most accurate for eclipsing systems because the inclination angle is
well constrained given that $\sin{i} \simeq 1$ as the donor eclipses either the
disk or white dwarf. The derived masses for eclipsers are then also 
relatively insensitive to inclination angle, because they depend on $\sin^{3}i$
which is almost flat near $i=90^{\circ}$, varying by $<5\%$ between 
$80^{\circ}<i<90^{\circ}$.

\subsection{Expected X-ray Source Populations}
\label{sec:expected}

\citet{Jonker11} contains a summary of the population we expect in our 
survey area.  The total number of X-ray sources detected, 1640, very 
closely matches the predicted number of 1648. We expect many different 
source classes which typically show variability on the timescale of hours 
to days. Of these, many can share light curve morphologies. For
example, RS CVns and coronally active M-dwarfs can show sinusoidal
variations with a period of days due to star spots. These variations can be
difficult to distinguish from ellipsoidal modulations when the orbital period
and phasing have not been established. Although qLMXBs or CVs with main 
sequence (MS) donors should have periods of hours, periods greater than a day
are possible for evolved donors.

The CVs in our sample are X-ray selected, and so are likely either to be
magnetic systems or quiescent, as high accretion rates onto non-magnetic CVs
create an opaque boundary layer (BL) that quenches X-ray emission,
reprocessing it into UV wavelengths \citep{Patterson85,Warner03}. During most 
DN outbursts, the X-ray 
emission rises up until a critical value of $\dot{M}$, at which point the BL 
becomes opaque for the remainder of the optical outburst. There 
is at least one
non-magnetic CV, however, that brightens in the X-ray for the duration of the 
DN outbursts
\citep[i.e. U Gem][]{Swank78}. It is also important to note that while the X-ray emission during an outburst is typically supressed, it does not disappear entirely. To be clear, for the purposes of this paper we consider a ``quiescent'' CV to be a dwarf nova in the low state, without an ionized accretion disk. New LMXBs in our survey that have
optical counterparts are also likely to be in quiescence, as they are
too X-ray faint to have been detected by All Sky Monitors. 

Quiescent systems have a larger portion of their continuum light
contributed by the donor star. In systems where the donor
fills its Roche Lobe, the effective surface area of the donor changes
with phase due to tidal distortion. Because the donor contributes a
large portion of the continuum light in quiescent systems, we expect
to recover ellipsoidal modulations for these. Higher accretion rate systems 
have continuum emission dominated by the accretion disk. For systems with
brighter accretion disks, we expect aperiodic flickering to dominate over any
underlying periodic variations in our data set.

X-ray detected BH qLMXBs are likely fainter in the X-ray
than NS qLMXBs at the same period because energy is carried either through the event horizon by ADAFs \citep{Narayan97,Garcia01,Hameury03,Narayan08,Rea11} or away by jets \citep{Fender03}. The optical light, however, eminates from the accretion disk and donor star and is comparable to NS qLMXBs at the same periods. Therefore, low ratios of X-ray to optical light somewhat favor BH accretors over NS accretors. It is worth noting that $\frac{L_{X}}{F_{opt}}$ is a rough diagnostic, representing the balance of probability rather than a hard limit between NS and BH qLMXBs, especially in the absence of information about the orbital period. There is therefore some overlap between NS and BH qLMXBs in this metric. The population of qLMXBs with $\frac{L_{X}}{F_{opt}}\approx1$, however, should be composed primarily of NSs, and the population with $\frac{L_{X}}{F_{opt}}\approx\frac{1}{100}$ should host primarity BHs. We use this diagntostic to triage more intensive spectroscopic follow-up in the future.

\begin{figure}
\begin{center}
\includegraphics[width=0.4\textwidth,angle=-90]{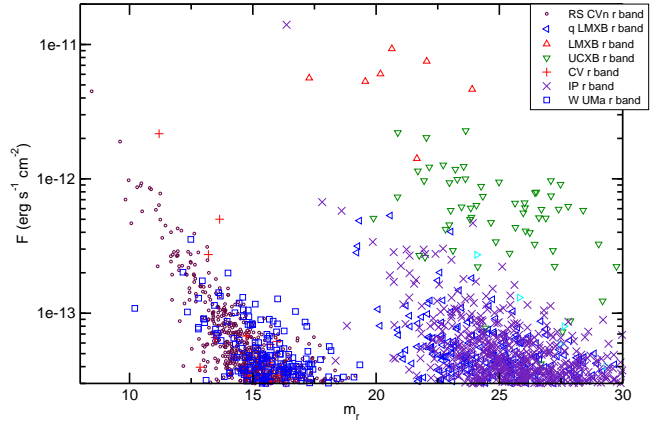}
\caption{Estimated $r'$ band magnitudes for the main population of GBS sources with assumptions in
  \citet{Jonker11}, who presented a similar figure for the Sloan $i'$ band. Changes between the expectations in the Sloan $i'$ band and the $r'$ band presented here are mainly due to the difference in extinction between these two optical bands.}
\label{fig:rpopsynth}
\end{center}
\end{figure}

This paper sets out to identify likely counterparts to X-ray sources
in the GBS based on variability characteristics, taking into account
the changing error in X-ray position with the off-axis angle of the
source detection and for a different number of X-ray counts detected in the source. We identify eclipsing sources as priorities for
detailed spectroscopic and X-ray follow up observations, measure the
orbital periods of systems where possible, and compare the
population estimates in \citet{Jonker11}, corrected for the change in
filter from Sloan $i'$ to Sloan $r'$ and shown in Figure \ref{fig:rpopsynth}, with possible population
numbers based on photometry.

\section{Observations and Data Reduction}
We acquired 8 nights of photometry, from July 8th-
15th 2010, with the Blanco 4.0 meter telescope at the
Cerro Tololo Inter-American Observatory (CTIO). Using
the Mosaic-II instrument, we observed the 9 square degree area
containing the X-ray sources identified by the first GBS X-ray observations
 \citep{Jonker11}. At the time of the optical observations, full X-ray
 coverage of the remaining 3 square degrees had not yet been approved, so we did
 not spend time observing the southernmost quarter of the survey area
 for which we had no X-ray source list. Those southern sources listed in
 \citet{Jonker14} were examined with DECam in June 2013 and
 photometry will appear in \citet{Johnson14}. Multiple Sloan $r'$ exposures with an
 integration time of 120\,s of 45 overlapping fields were taken to
cover the area. These fields are shown in Figure \ref{fig:gbsmap}. 
Positions for these 45 fields were dithered between
cycles to recover sources that fell on gaps between CCDs. Each field
was observed 19 times over the 8 night run, but because of overlap
between fields, many sources are observed at twice as many epochs, and 
some at half as many epochs in places with no overlap and chip gaps. The 
order in which the fields were cycled was randomized to minimize aliasing 
caused by regular sampling. Typical seeing for the run was around $1''$, 
though on nights 4, 7, and 8 of the observations the seeing was worse, 
peaking on night 8 around $2.5''$.

The data were reduced
via the NOAO Mosaic Pipeline \citep{Shaw09}, which also
added a World Coordinate System (WCS) to the images.
The NOAO pipeline searches for instrumental artifacts
in the image, corrects for cross talk between CCDs, applies a pupil 
ghost correction for light reflecting from
the filter to the back surface of the corrector then back
through the filter, applies bias and flat field corrections,
and calibrates WCS for each image based on USNO-B1
stars in the field. Dark current calibrations are unneccesary. A 
detailed explanation of each procedure can be
found in Chapter 2 of the NOAO Data Handbook \citep{Shaw09}. By
comparing our fields to the UCAC3 catalog \citep{Zacharias09}, we have 
found that the WCS
provided by the pipeline is typically accurate to within $0.2''$.

To determine the error in the X-ray position, we use the methods of
\citet{Hong05} for the 95\% confidence region. For sources at a
large off-axis angle and low number of counts, this can be as large as a
$10''$ radius. X-ray sources viewed close to an offset angle of zero have
significantly smaller errors in the position, which can be less than $1''$ and are dominated 
by the boresight correction. 

\subsection{Photometry}

Differential photometry was done using Alard's image subtraction routine ISIS,
described in detail in \citet{Alard98} and \citet{Alard00}. 
This yields changes in flux relative to the reference image. In order to 
convert this to magnitudes we used either aperture photometry 
or DAOPHOT-II \citep{Stetson87} to measure the zero-point flux in the 
reference image. In order to save computation
time, small cutouts of the full Mosaic images were taken around each object for
processing. These were typically $201 \times 201$ pixels ($52'' \times 52''$) 
or $401 \times 401$ pixels ($104'' \times 104''$), although in a few cases it 
was necessary to increase the field to $801 \times 801$ pixels if the field 
near the X-ray position contained many saturated stars.
A subtraction was deemed to be ``clean'' if it resulted in a variance image 
free of the vast majority of field stars and free of artifacts from the wings 
of saturated stars near the X-ray position.

We found that different fields required somewhat different keyword
values in ISIS to obtain a clean subtraction around the object of
interest. For most fields, we ran ISIS with a kernel composed of
3 Gaussians multiplied by polynomials of degree 6, 4, and 3 with
$\sigma=1.1,3.0, \&\ 5.5$ pixels, respectively. The kernel and
background were also
most often fit
with a 1st order polynomial to allow for spatial variation across the
field, though this was sometimes increased to a 2nd order polynomial if
the subtraction was not clean. 
Some fields
contain artifacts or very bright stars in one section; for these
fields we divided the image into separate parts for the image subtraction.

We adopt a definition of variability wherein an object is said to be
variable either if the standard deviation is at least three times the
typical relative photometric error or if there is at least one observation
greater than four times its relative photometric error away from the
mean. The only objects subjected to these tests are those that are
at all visible in the variance image produced by ISIS and those shown in other
observations to be emission line objects.

\subsection{Periodicities}

For all variable sources, periodograms were created using the
Lomb-Scargle statistic \citep{Lomb76,Scargle82} in an effort to search for 
periodicities. Since ellipsoidal variations have 2 maxima and
minima in a single orbital cycle, we also folded the lightcurve on periods
twice as long as prominent peaks on the periodogram. We
also consider both aliases and harmonics, as higher harmonics can 
sometimes show up at a higher power than the fundamental
frequency. For sources with multiple observations of an eclipse, we
also searched for periods with Phase Dispersion Minimization 
\citep{Stellingwerf78}. 
The significance of identified periods is checked through Monte Carlo 
simulations
in which the order of the observations is randomized. If the recovered
period is due to the overfitting of Gaussian noise, then reshuffling the
lightcurve should also provide similar fits to noise in a non-trivial
number of cases. We consider periods to be significant if these
searches result in less than 0.5\% of simulations with a higher
spectral power at any point in frequency space than the initial recovered 
period in the properly ordered data. Since we have \vble\ variable potential counterparts,
we expect of order 1 false period detection at this threshhold.

Systems with a power density spectrum of red or pink noise can easily yield
false periods that appear significant in white noise tests
\citep{Vaughan10}, though sparse sampling means that most flickering
should appear as white noise (i.e. nearby points are not correlated).
It is unclear that irregular variables in our data exhibit red noise
power spectra. In order to test the likelihood of spurious identification of 
periods in our data set, we used the existing OGLE-IV photometry for GBS sources
\citep{Udalski12}, which has a baseline $\approx100\times$ as long as our 
Mosaic-II data. Using irregular variables aligned with X-ray sources in this
data set and pulling out 8 days worth of data at a time, we ran period searches
and Monte Carlo simulations just as we do for the Mosaic-II data. We recover
spurious periods that appear to be significant under white noise simulations
approximately $6\%$ of the time. Furthermore, of the 11 systems which both 
appear to be long period variables in Mosaic-II data and appear in OGLE-IV
photometry, 10 are truly periodic in OGLE-IV data. We conclude that we have
$\lae 2$ sources with spurious periods between 1 and 4 days, and that the 
majority of periods recovered are likely real. This assumes that the power spectra for
the fainter objects is similar to that of the brighter objects.

\subsection{Magnitude Calibrations}

At present, all apparent magnitudes cited
here are scaled to nearby stars in the USNO-B1 catalog \citep{Monet03}
and are to be used with caution until secondary standards are
established for all Mosaic fields. The magnitude scaling, which is a 
pipeline calibration product,
carries an estimated uncertainty of $\pm 0.5$ magnitudes for
each source. This is adequate for estimating X-ray
to optical flux ratios, as the uncertainty is dominated by low photon
counts in the X-ray, uncertainty in the X-ray spectral shape, and 
uncertainty in the reddening and absorption for the vast majority of sources.

\section{Results}
\label{sec:results}

Of the 1234 X-ray sources identified with the northern three quarters of 
the Chandra observations, \duplicates\ sources are likely duplicates of 
the same X-ray source with positions separated by $>3''$ that were not 
removed when the catalog 
was made \citep{Hynes12} leaving 1216 unique sources. Of these, \offchip\ 
lie outside regions imaged by Mosaic-II, \saturated\ have 
likely counterparts saturated in the Mosaic-II data, and a further \chancesat\ 
sources are too near a saturated star or a bleed trail to do photometry 
on.  A sizable number of optically bright sources was anticipated, and many are covered by other
observations \citep[see e.g.]{Hynes12,Udalski12}, while the focus of this work was the optically
faint population. This leaves \ngood\ unique sources with useful Mosaic-II data. These 
comprise \noc\ sources with no detectable
counterpart in Sloan $r'$ inside the $95\%$ X-ray confidence region, \nv\ 
that have no counterpart that shows variability over the course of our 
observations in the $95\%$ confidence region, and \vble\
sources which have possible counterparts that show variability during our
observations.  The majority of variables show flickering, for which we
were unable to recover a period. 

\subsection{Consideration of Chance Alignments}

Inevitably when examining 1216 unique sources, variables
occasionally lie outside the $95\%$ confidence region. In order to estimate the rate of 
coincidence with variable interlopers, we first estimate typical crowding in our fields by counting
the number of stars in regions with a radius of $3''$ offset from a random selection of X-ray
positions by $15''$. We find 
a mean stellar density of $0.064\,{\rm arcsecond}^{-2}$. 
Approximately $61$
true counterparts should lie between the $95\%$ and $3\sigma$ confidence regions,
which means we should have $\approx9$ variable true counterparts
falling outside the error region given the observed rate of
variability within the $95\%$ confidence region of GBS sources. This is less than the $\sim40$
chance variables in the annulus between the $95\%$ and $3\sigma$ confidence
regions that one expects by assuming that the rate of variability for interlopers down to $r'<23$ 
is approximately that of the population measured by OGLE III ($I<17$), where the probable number of variable interlopers inside a given area is a Poisson distribution. Similarly, there should be 
$\sim20$ variable interlopers within $2''$ of the X-ray positions. The further from the center of the
 X-ray position a variable is, the more likely it is to be an interloper as the area of the sky being 
 searched increases. We compare the likelihood of finding a true variable counterpart at a certain 
 radius compared to the likelihood of finding a variable interloper within the same radius to assign 
 a False Alarm Probability to variables near X-ray positions, so that $FAP=[1-\frac{2}{\sigma_{X}\sqrt{2\pi}}e^{-(\frac{r}{\sigma_{X}/2})^{2}/2}][1-e^{-\pi r^{2}\times0.00128{\rm\,variables\,arcsecond^{-2}}}]$. For sources with more 
than one variable star aligned 
with the X-ray position, those that show light curve morphology unique to either CVs or 
LMXBs, such as DN outbursts, should be
considered the counterpart, while variables showing more ubiquitous 
morphologies, such as
sinusoidal or ellipsoidal modulations, should
be considered to be more likely to be a chance alignment without other 
compelling evidence to the contrary. The
number of sources we have in each category of variability is shown in
more detail in Table \ref{table:varsummary}.

\begin{table}
  \caption{The table below
  shows the total breakdown of sources in
  Mosaic-II data. The largest category is non-variables, with sources
  either saturated or obscured by bleed trails as the second
  largest.}
\label{table:varsummary}
\begin{center}
\begin{tabular}{l r}
\hline
Category & Number \\
\hline
Duplicates & \duplicates \\
Saturated & \saturated \\ 
Coincidentally Saturated & \chancesat \\
Off-chip & \offchip \\ 
Not Variable & \nv \\ 
No Counterpart ($r'\lae 23$)& \noc \\
Variable & \vble \\ 
\hline
\multicolumn{2}{r}{Types of Variability} \\
Flickering & \flicker \\
Outburst/flare & \ob \\ 
Eclipsing & \ecl \\ 
Ellipsoidal & \ellip \\ 
Sinusoidal & 38 \\ 
Long Period Variables & \lpv \\ 
\end{tabular}
\end{center}
\end{table}

Many sources (\nv) show no detectable optical
variability. For some of these, the true counterpart could be below
detection limits, leaving only field stars coincident with the X-ray
position. There are also \noc\ sources with no possible counterpart in
Sloan $r' < 23$. In \citet{Jonker11}, a simplistic population model predicted that the GBS
would find $\sim400$ sources in Mosaic-II fields without a
counterpart in Sloan $i'$ band, which increases in Sloan $r'$ because
of greater extinction. The predicted number is much higher than that
observed, suggesting that many possible non-variable counterparts are
in fact random field stars. Sources with large errors on the X-ray position
admit many possibilities as counterparts, but the true counterpart may
be below detection limits. There are 537 sources with an estimated 
error with a radius larger than $1.9''$, each of which admits probable 
chance alignments, assuming a Poisson distribution of stars with the measured density on the sky. Of these, 256 have at least one possible
non-variable counterpart
within detection limits. This suggests that hundreds of true counterparts 
are indeed non-variable. Some of these have been confirmed through 
spectroscopic studies, e.g. H$\alpha$ emitters CX561 and CX1004 
\citep{Britt13,Torres14}. We expect to be sensitive to intrinsic ellipsoidal 
modulations of $0.1$ magnitudes for more than $99\%$ of spatially isolated objects with 
$r'<19$, with sensitivity decreasing as photometric errors rise to the level of
the projected variations. Because binary systems are uniformly distributed 
in $\cos i$ on the sky while the projected variation scales with $\sin i$, 
we expect to maintain sensitivity to ellipsoidal intrinsic variations of $0.1$ magnitudes
for the majority of sources until $r' \approx21$, when photometric errors
become comparable to $1/3$ the RMS of the ellipsoidal modulations in our line of sight.

\subsection{Sensitivities of the Survey}

The X-ray to optical flux ratio is calculated in Table
\ref{table:varstats} for
sources with variable counterparts for both absorbed and unabsorbed
flux at Bulge distance using
assumptions in \citet{Jonker11}, the extinction law $R_{V}=3.1$ found in
\citet{Clayton89}, and the relation between optical extinction and
hydrogen column density $N_{H}=0.58\times 10^{22} \times E(B-V)$ from
\citet{Bohlin78}. $F_{Opt}=\nu F_{\nu}$ is calculated from apparent magnitudes using filter
properties for Sloan $r'$. For Bulge extinction values, we use the maps from
\citet{Gonzalez12}.  We transform these values to the 
Sloan $r'$ filter using filter properties given in \citet{Schlegel98}.  Many sources are closer than the Bulge distance, so
the absorbed and estimated unabsorbed flux ratios represent
approximate upper and
lower limits, respectively, to
$\frac{F_{X}}{F_{Opt}}$ though there is additional uncertainty in this ratio
because X-ray and optical observations are not simultaneous, as well as 
uncertanties in both
$F_{X}$ from photon noise and spectral shape and in $F_{opt}$ from the
calibration of the zero point magnitude.
We should not detect sources
with an X-ray luminosity below $10^{32}\,{\rm ergs\,s^{-1}}$ at Bulge
distance, so RS CVns, W UMas, CVs, and coronally active stars 
should tend to be in the foreground and therefore to have a value towards the upper end 
of a source's range of possible flux ratios. 

\begin{figure}
\begin{center}
\subfigure{\includegraphics[width=0.35\textwidth,angle=90]{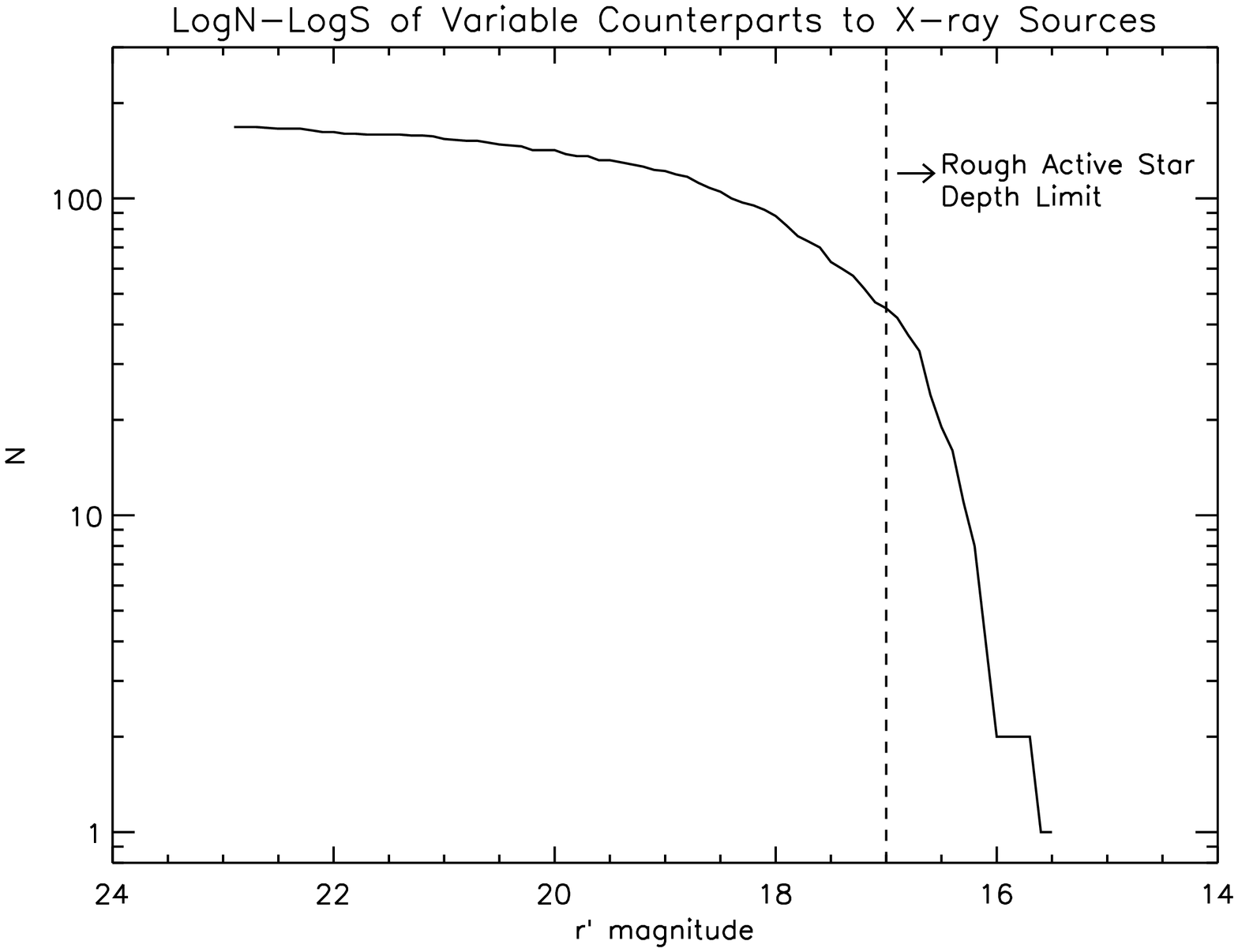}}
\subfigure{\includegraphics[width=0.35\textwidth,angle=90]{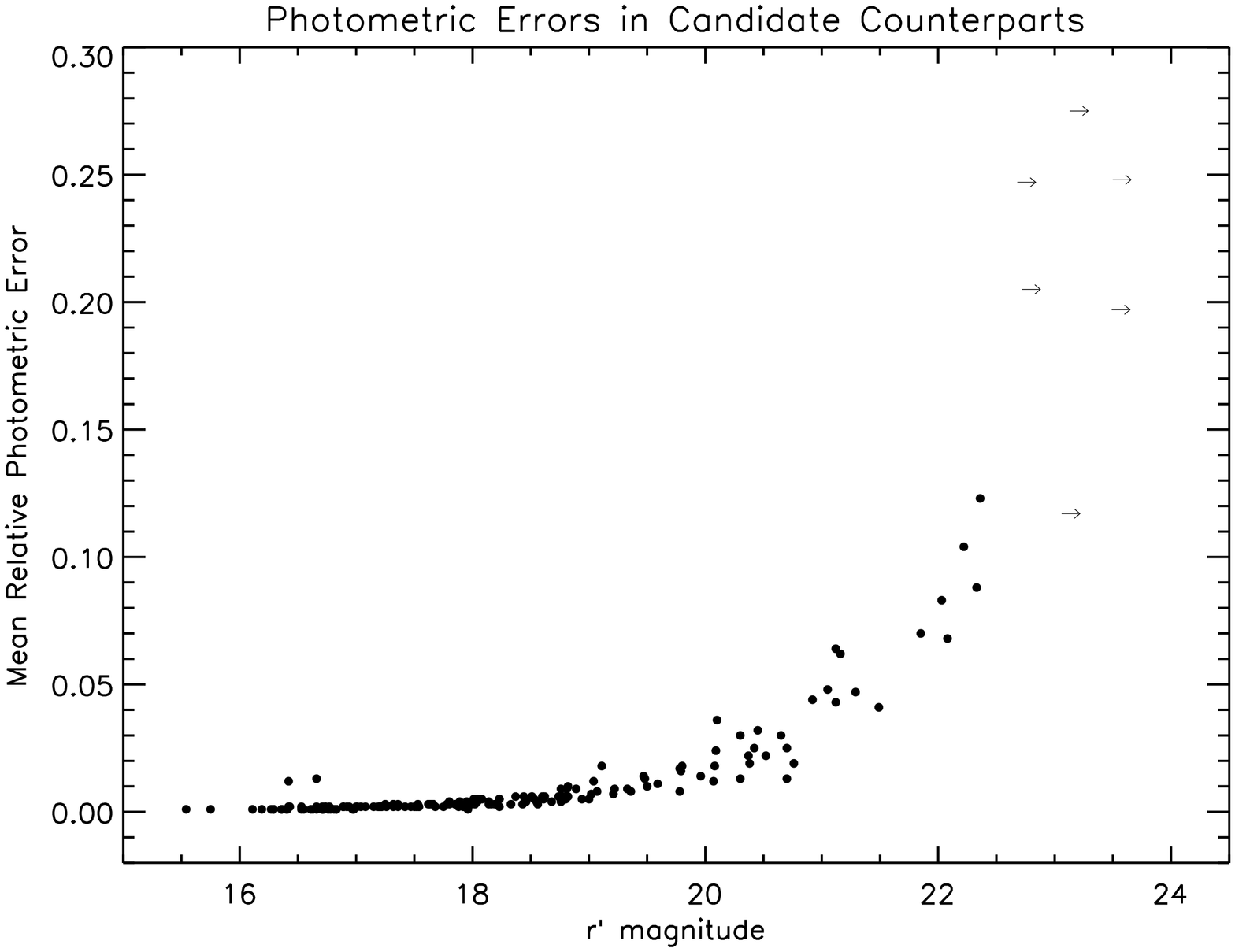}}
\caption{{\it Top:} A cumulative histogram of the mean magnitude of variable sources near 
X-ray positions in the GBS. The dashed line indicates the X-ray survey's detection limit of active stars assuming that they have 
$\frac{F_{X}}{F_{opt}}=\frac{1}{100}$ and $E(B-V)=0.5$.
This limit is uncertain due to factors discussed in \S\ref{sec:results}. The 
sudden change in slope at this point is expected because the
nature of the source population should change as one goes deeper into the 
optical, from systems with low $\frac{F_{X}}{F_{opt}}$ such as RS CVns, W UMas, 
and other coronally active stars, to systems with higher $\frac{F_{X}}{F_{opt}}$ 
such as qLMXBs and CVs.
{\it Bottom:} The mean photometric error in variable sources plotted against 
the mean magnitude. The relative photometric error does not become high enough
to mask the expected variability from ellipsoidal modulations in qLMXBs or CVs 
until $r'=21-22$, depending on the amplitude of the variations.}
\label{fig:lognlogs}
\end{center}
\end{figure}

The sources with potential variable counterparts are listed in Table
\ref{table:varstats}. Sources with the RMS scatter greater than 3
times the statistical error in the measurements are considered to be
variable. Sources with at least a single data point more than 4 sigma
from the mean magnitude are also considered to be variable. A cumulative 
histogram of the mean optical magnitudes of variable counterpart candidates is
shown in Figure \ref{fig:lognlogs}. There is a pronounced knee at roughly the
magnitude at which RS CVns and W UMa binaries should lie at a great enough 
distance to cease to be detected in the X-ray survey. The sudden change in 
slope in the $\log N - \log S$ distribution is attributable to the changing 
source population rather than a decreased discovery efficiency at that 
brightness. As shown in the bottom panel of Figure \ref{fig:lognlogs}, 
the average relative 
photometric error remains low well past the magnitude at which the break occurs.

\begin{figure}
\begin{center}
\includegraphics[width=0.35\textwidth,angle=90]{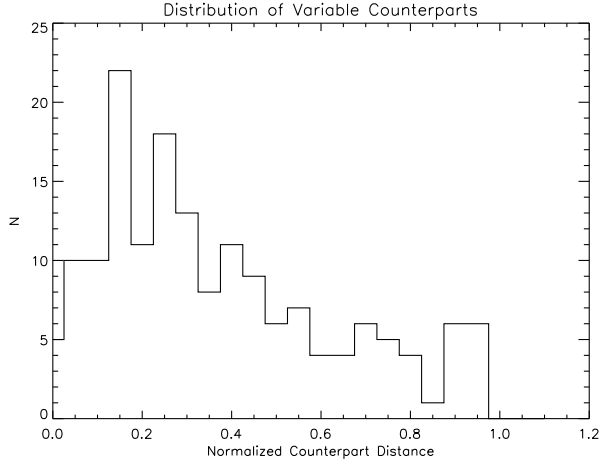}
\caption{A histogram of the distances between variable possible
  counterparts relative to the 95\% confidence radius of Chandra X-ray positions in
  \citet{Jonker11}. $72\%$ of variable counterparts are within half
  the $95\%$ confidence radius, which is a strong indicator that the
  vast majority of variables included here are likely true
  counterparts to the X-ray source. The number of interlopers should
  grow towards the outer edge of the error region as the area of the
  sky enclosed in each annulus increases.}
\label{fig:offsets}
\end{center}
\end{figure}

The majority of variables identified as candidate counterparts
are within $1''$ of the Chandra X-ray position. A histogram of all the
offsets between the optical counterpart and the Chandra position is
shown in Figure \ref{fig:offsets}, normalized to the uncertainty in the X-ray position. The 
fact that the variables are concentrated in the inner
half of the error region suggests that the 
vast majority are truly associated with the X-ray source,
as the effective area on the sky, and therefore the chance of variable
interlopers, increases with radius. 

\section{Selected Sources}

Finding charts and variance maps for all sources with possible
counterparts can be found
online\footnote{tigers.phys.lsu.edu/gbs/vardb/ and as figure sets on the ApJS online version of this paper.},
as can all lightcurves for sources with variable counterparts. Some
individual sources warrant more detailed discussion. Sources with
existing spectroscopy are covered in depth in
\citet{Britt13,Hynes14,Maccarone12,Ratti13,Torres14} and
\citet{Jianfeng14}. Most sources discussed here still require spectroscopy
for a full classification.

\subsection{CX11 - Magnetic CV or qLMXB}

\begin{figure*}[p!]
\begin{minipage}{0.9\textwidth}
\centering
\parbox{\textwidth}{
\subfigure{\includegraphics[width=0.4\textwidth,angle=90]{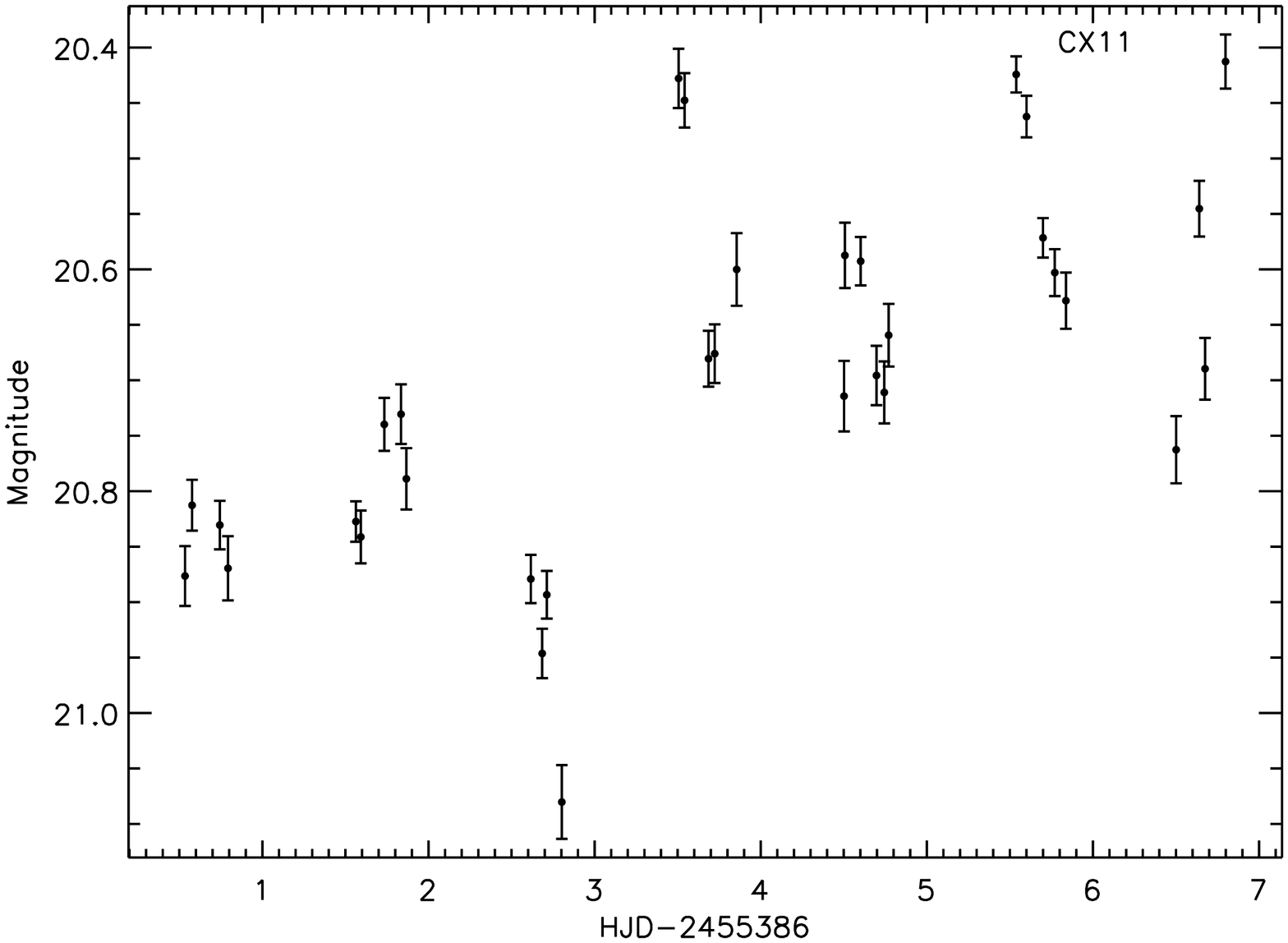}}\quad
\subfigure{\includegraphics[width=0.4\textwidth,angle=90]{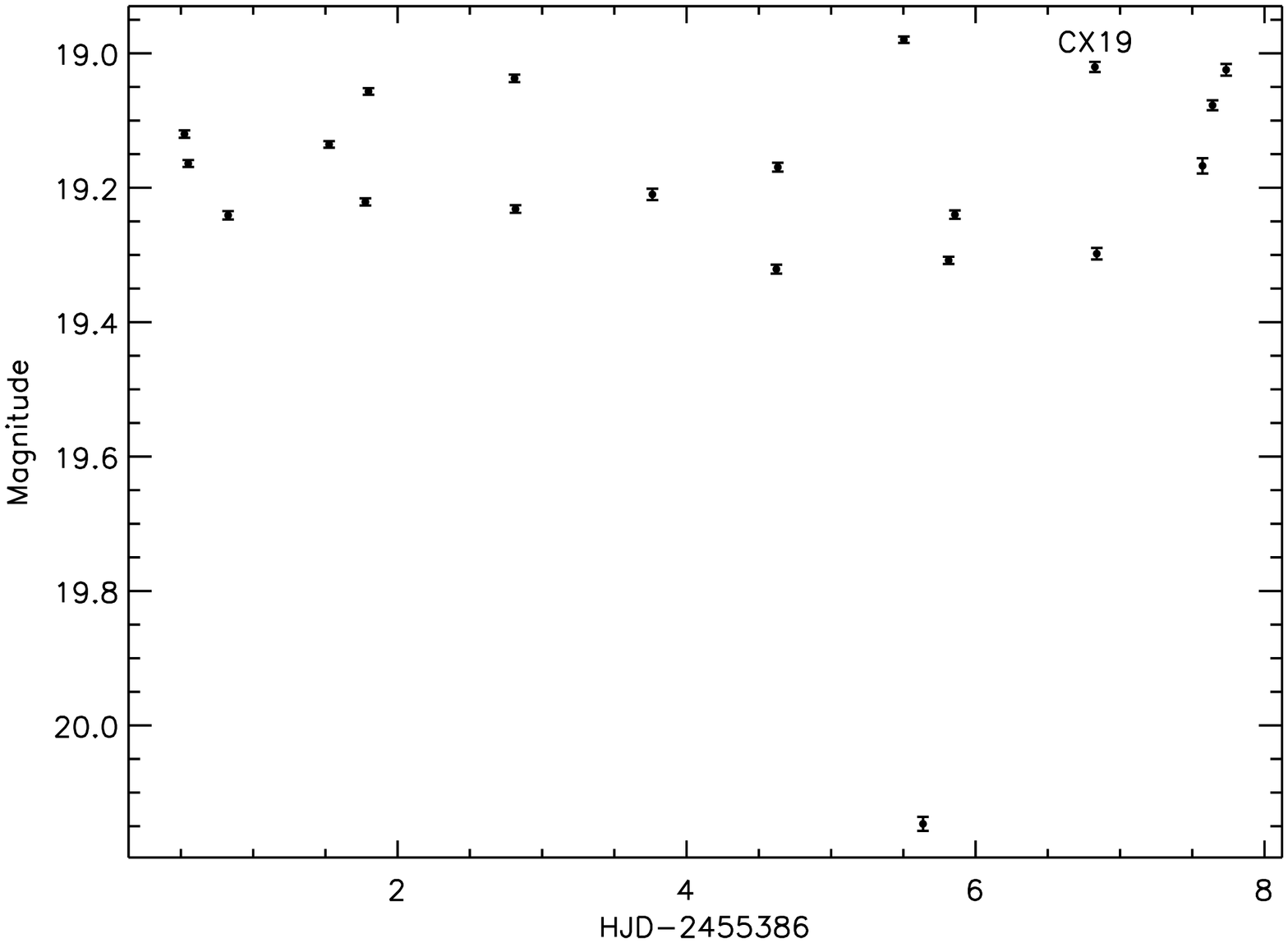}} \\
\subfigure{\includegraphics[width=0.4\textwidth,angle=90]{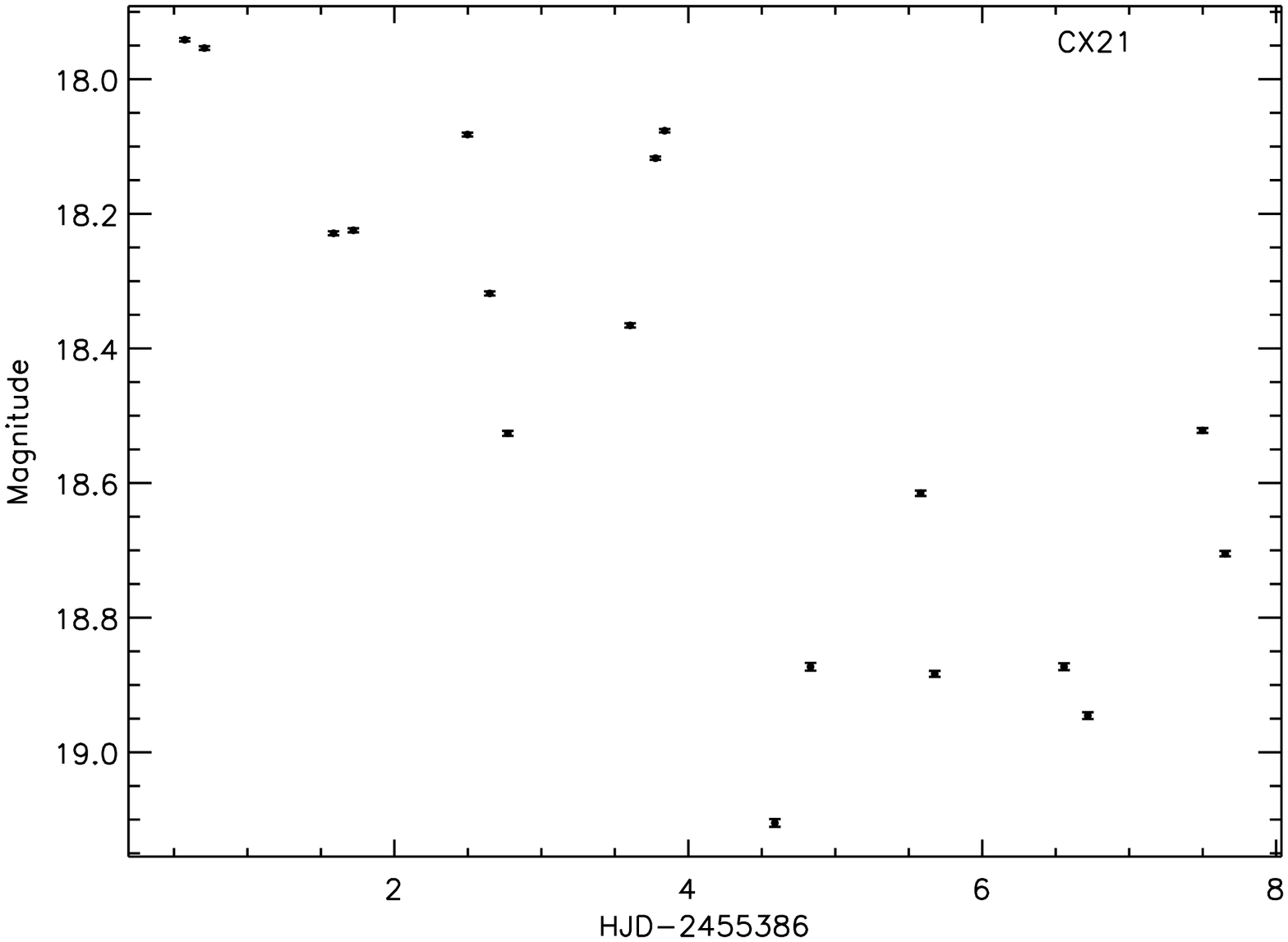}}\quad 
\subfigure{\includegraphics[width=0.4\textwidth,angle=90]{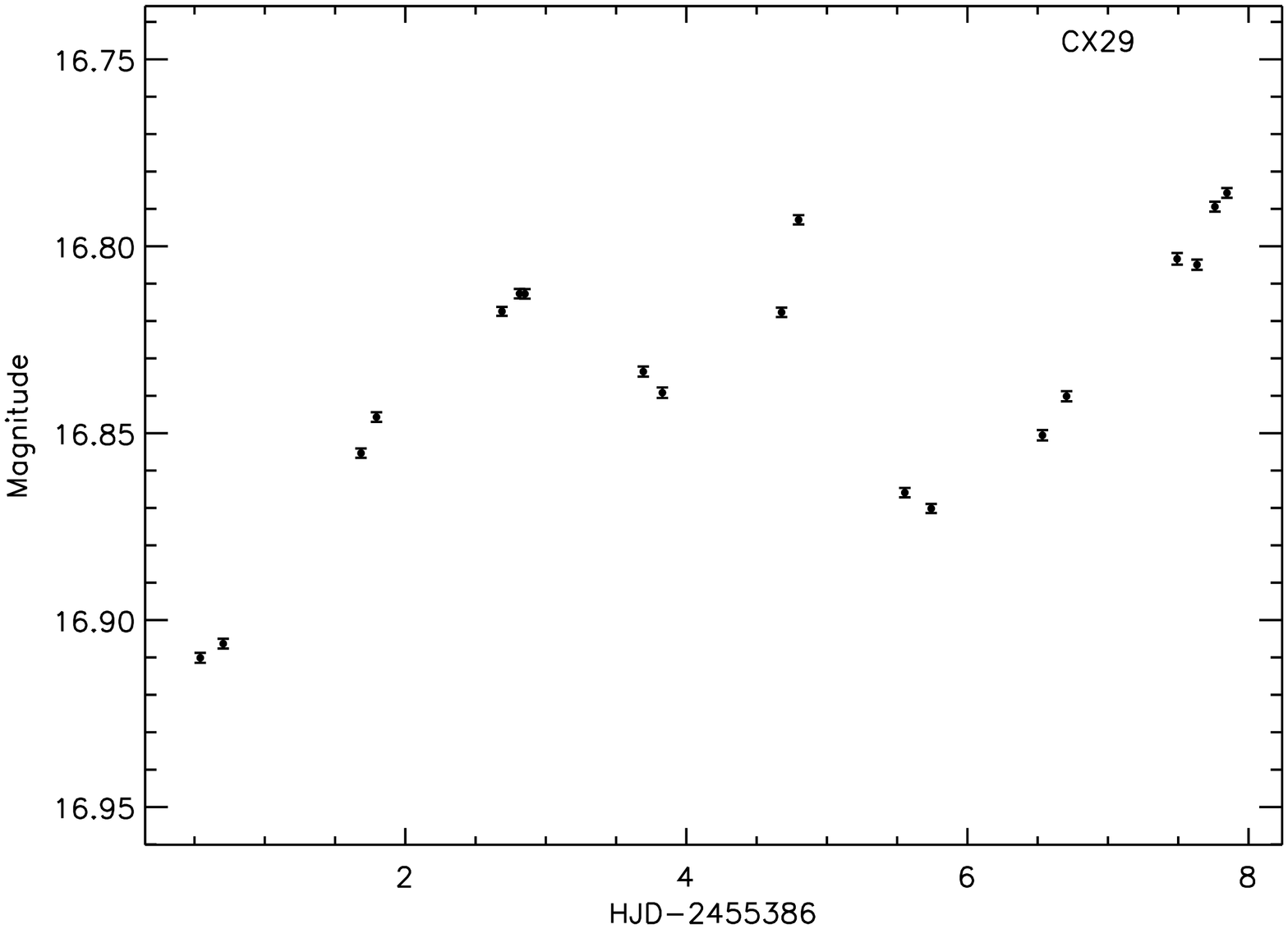}} \\
\subfigure{\includegraphics[width=0.4\textwidth,angle=90]{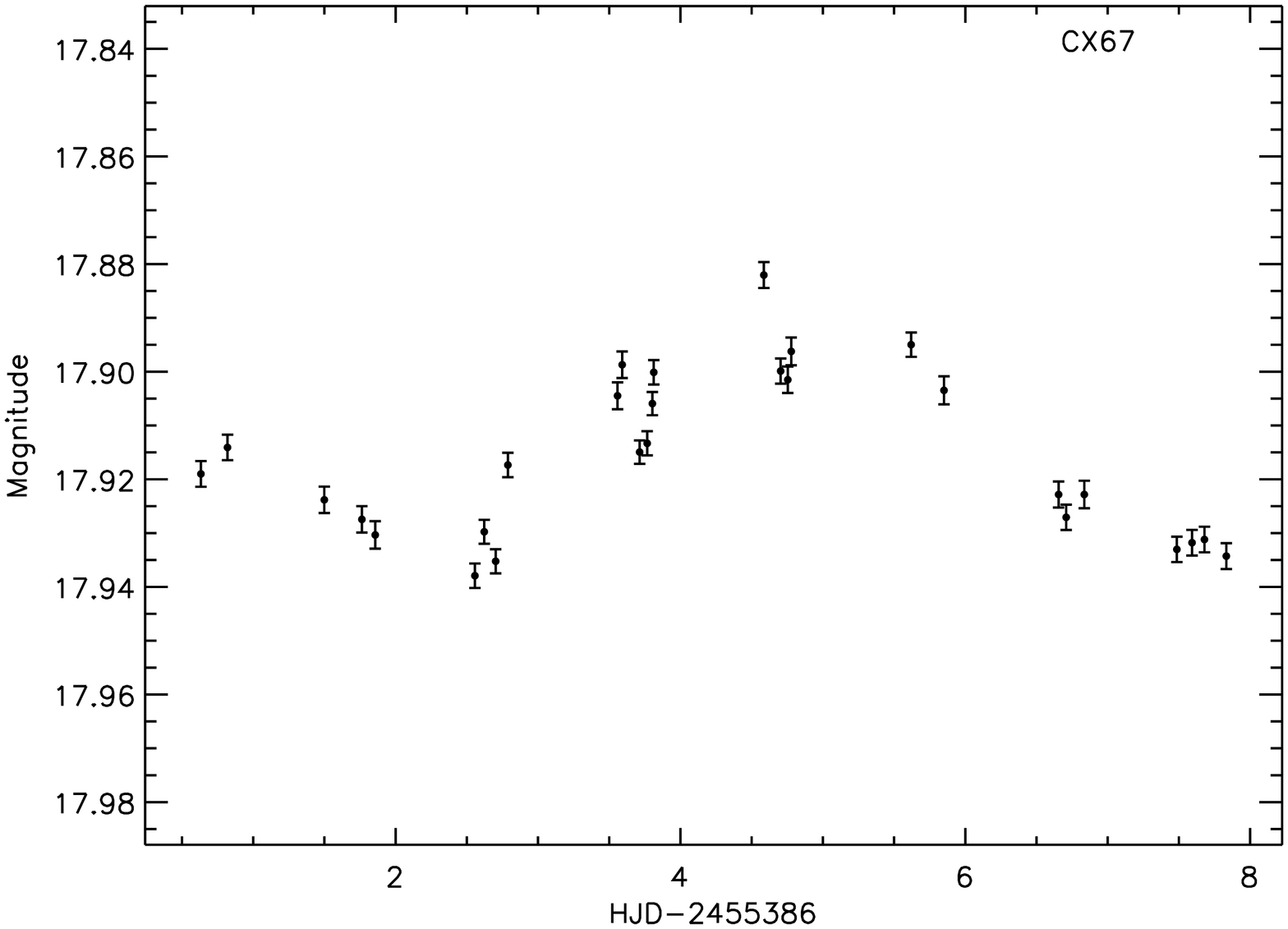}}\quad
\subfigure{\includegraphics[width=0.4\textwidth,angle=90]{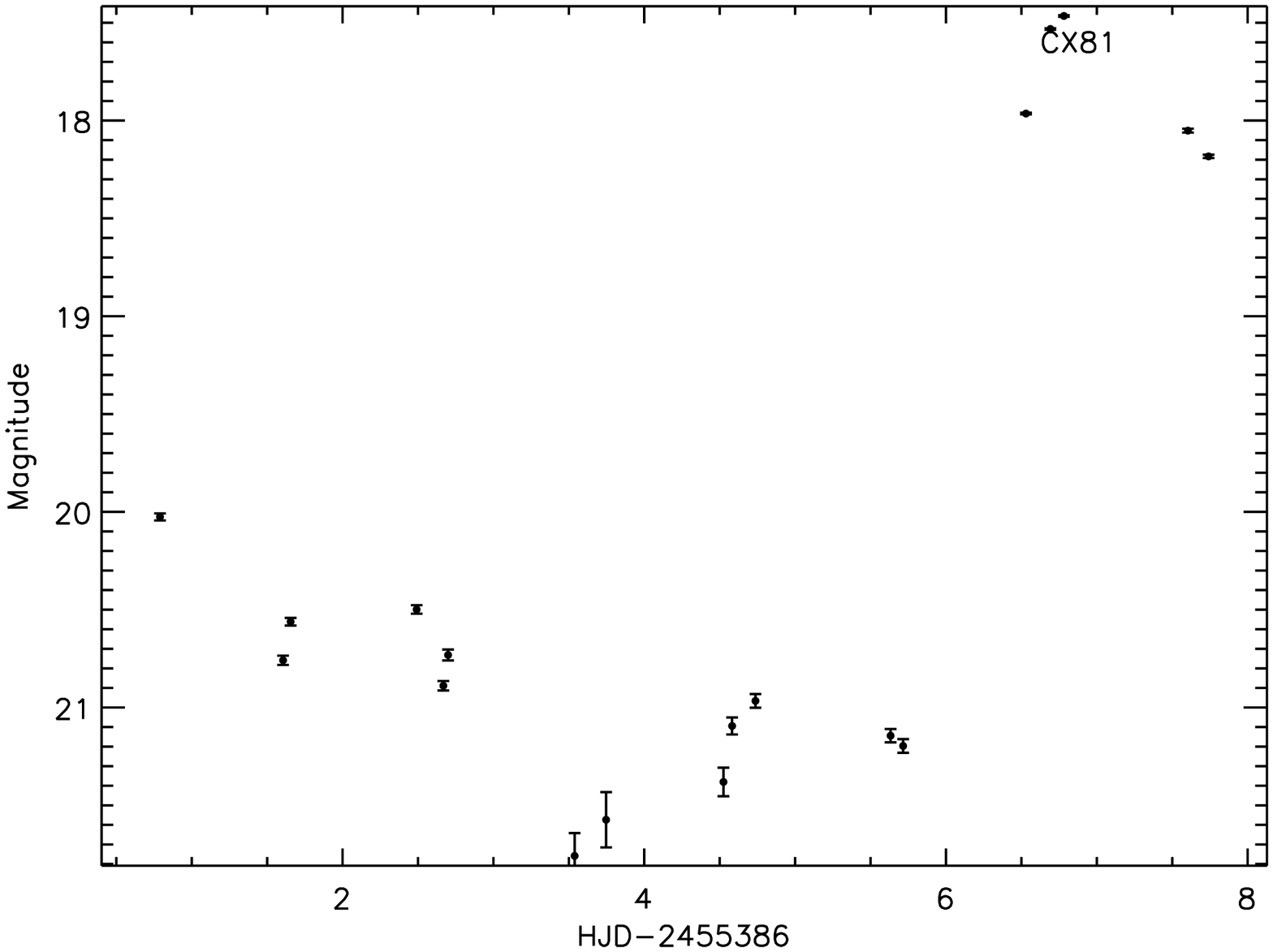}}
}
\caption{CX11, CX19, CX21, CX29, CX67, CX81 Mosaic-II Lightcurves}
\label{lcchunk1}
\end{minipage}
\end{figure*}

The variability of CX11 is irregular and on the order of $0.2-0.4$ 
magnitudes. This object has an X-ray to optical flux ratio of 40,
uncorrected for extinction, and shows
dramatic, aperiodic variations shown in Figure \ref{lcchunk1}. $L_{X}\approx
10^{34} \, (\frac{d}{8\,{\rm kpc}})^{2}\, {\rm ergs\,s^{-1}}$ and $\frac{F_{X}}{F_{opt}}=0.4$ with Bulge
reddening. $\frac{F_{X}}{F_{opt}}$ is high enough
to argue against a quiescent CV, which are not typically luminous enough in the X-ray
to be detected at this strength in the Galactic Bulge and thus could
not suffer the full amount of extinction predicted by the \citet{Gonzalez12} 
maps, though it is consistent with a closer magnetic CV or qLMXB. In addition, CX11 is X-ray bright 
enough that a meaningful hardness ratio can be calculated from the 2\,ks 
exposure [as described in \citet{Jonker11}], showing that the X-ray spectrum 
is fairly hard with $\frac{[2.5-8]-[0.3-2.5]}{[0.3-8]}=0.64$. This X-ray 
hardness further rules out a quiescent CV as well as arguing against thermal emission from a NS qLMXB, though some NS qLMXBs have strong powerlaw components that would result in a hard X-ray spectrum \citep[e.g. EXO1745-248]{Wijnands05}. Using the Web-PIMMS tool, we determine that the X-ray color is too hard for quiescent CVs assuming a 10\,keV thermal Brehmsstrahlung spectral shape. As discussed in Section \ref{sec:expected}, BH qLMXBs tend to be fainter in X-ray than NS qLMXBs at the same period, resulting in lower values of $\frac{F_{X}}{F_{opt}}$. We therefore prefer a magnetic CV interpretation for CX11, though a qLMXB interpretation cannot be firmly ruled out without further data.

\subsection{CX19 - CV or qLMXB}

CX19 also appears in OGLE-IV fields \citep{Udalski12}, where
it is observed to eclipse with a period of $P_{orb}=0.3587$ days. Our 
observations also show that CX19 eclipses, as shown in Figure \ref{lcchunk1}, 
in addition to having a fairly high X-ray
to optical flux ratio of 6 before correcting for absorption and $\frac{1}{10}$ 
assuming reddening for Bulge distance. Either of these is consistent with a
qLMXB or CV. It also shows large amplitude aperiodic
variability, up to 0.4 magnitudes, consistent with a large contribution from the
accretion disk to continuum light. The eclipse is deep, almost a
magnitude, implying a substantial temperature difference between bodies. The
eclipse lasts no more than 3 hours, and there
is only 1 observation in eclipse out of 20.  It is very likely a binary with a 
compact object.  This object is
$1.27''$ away from a star that is very red and which dominates over it in VVV data and in 2MASS
data. CX19 is a candidate eclipsing qLMXB or CV. \citep{Torres14} classify this object spectroscopically as a high accretion rate CV, which is consistent with our photometric classification.

\subsection{CX21 - CV}

CX21 is strongly variable, on timescales of hours, showing a
change in brightness of over 1 magnitude shown in Figure
\ref{lcchunk1}. There is no apparent periodicity to
these changes. Absorbed $\frac{F_{X}}{F_{opt}}=3$ which drops to $0.01$ for Bulge distance 
reddening, a range that is consistent with CVs,
qLMXBs, and AGN. $L_{X}\approx3\times10^{33} (\frac{d}{8\,{\rm kpc}})^{2}\, {\rm ergs\,s^{-1}}$. 
The high amplitude flickering suggests a
large contribution to the continuum light from the accreting material
compared to the donor
star, which argues weakly against a quiescent system.  
We can safely rule out an RS CVn, W UMa,
or active star because of the strong flickering. An AGN is possible,
but we consider it unlikely because of 
the short timescale of the variability. 

CX21 is a ROSAT source \citep{Jonker11}, which detected it in a
2418\,s observation and saw possible evidence of variability. It is
somewhat soft in the X-ray in GBS observations with a hardness ratio
of $-0.38$ in $\frac{[2.5-8]-[0.3-2.5]}{[0.3-8.0]}$\,keV, and somewhat hard in 
ROSAT observations with a 
hardness ratio of HR2$=0.28$ in $\frac{[0.9-2.0]-[0.5-0.9]}{[0.5-2.0]}$\,keV. Because these measures are of different parts of the spectrum, they do not suggest a change in the spectral shape between observations. The ROSAT observation is hard enough to argue against thermal emission from a NS qLMXB even with minimal extinction, while the hardness ratio from Chandra is too soft for a BH qLMXB even with maximal extinction assuming a powerlaw spectrum with $\Gamma=2$. Both hardness ratios are consistent with thermal Brehmsstrahlung from a CV.

\subsection{CX29 - Flare star}

CX29 shows smooth variations which both rise and fall, as shown in
Figure \ref{lcchunk1}. There is a
crest and a trough and our data are consistent with a period of
$\sim10.2$ days, but this is highly speculative. There is a small
flare, $\approx0.1$ magnitudes in amplitude, which lasts for
several hours on the fifth night of observations. Absorbed
$\frac{F_{X}}{F_{opt}}=0.5$, which is consistent with CVs,
qLMXBs and with
active M dwarfs since most the light they emit is in the infrared so that
$\frac{F_{X}}{F_{opt}} >> \frac{F_{X}}{F_{bol}}$. 
$L_{X}\approx2.4\times10^{33} (\frac{d}{8\,{\rm kpc}})^{2}\, {\rm ergs\,s^{-1}}$. If this were an
M dwarf with an absolute magnitude of $M_{r'}=15$, then the
distance would be $23\pm7\,{\rm pc}$ which would imply $L_{X}\approx
2\times10^{28} \, {\rm ergs\,s^{-1}}$ which is consistent with M dwarfs.  
In the 2MASS survey \citep{Skrutskie06}, this star has $J-K=1.15$ and 
$K=12.50$, which
is consistent with a late M dwarf at that distance. It is possible
that this is a flare star, with the complex multi-day variations arising
from a combination of star spots as the star rotates with $P_{spin} >
8$ days. CX29 appears in \citet{Udalski12} with a period of $12.77$ days, 
which is consistent with our observations. 
Differences among multiple star spots in phase and brightness, and even small
differences in period by latitude due to differential rotation, can lead to more
complex lightcurve morphologies such as in the case of CX29. This object is suggested to be a chromospheric active star or binary after spectroscopic observations in \citet{Torres14} pending further analysis, which agrees with our photometric classification.

\subsection{CX67 - CV or qLMXB}

CX67 shows variations of several hundredths of a magnitude, with a suspected period of 5.67 days. The Mosaic-II lightcurve is
shown in Figure \ref{lcchunk1}. Absorbed
$\frac{F_{X}}{F_{opt}} = 0.7$ which is consistent with CVs,
qLMXBs, and M dwarfs, but the flickering visible in the lightcurve
argues against an M dwarf interpretation. It is
possible that this is a longer period binary with a compact object and
a subgiant donor, such as V404 Cyg \citep{Hynes09}. 
If this is the case, $\frac{F_{X}}{F_{opt}}$ should
be lower for quiescent systems than for systems with a MS donor
because the optical contribution from the larger counterpart is much
higher. The amplitude of variations is quite low; if these variations
are due to tidal distortion of the donor, the inclination angle of the
system must be fairly low as well.

\subsection{CX81 - DN}

CX81 shows an outburst of $3.5$ magnitudes and lasts at least a few
days, as shown in Figure \ref{lcchunk1}, which is typical of
DNe. Absorbed $\frac{F_{X}}{F_{opt}} = 5$ in quiescence, which is consistent with
CVs. This source is very likely a CV undergoing DN outbursts.

\subsection{CX83 - CV or qLMXB}

\begin{figure*}[p!]
\begin{minipage}{0.9\textwidth}
\centering
\parbox{\textwidth}{
\subfigure{\includegraphics[width=0.4\textwidth,angle=90]{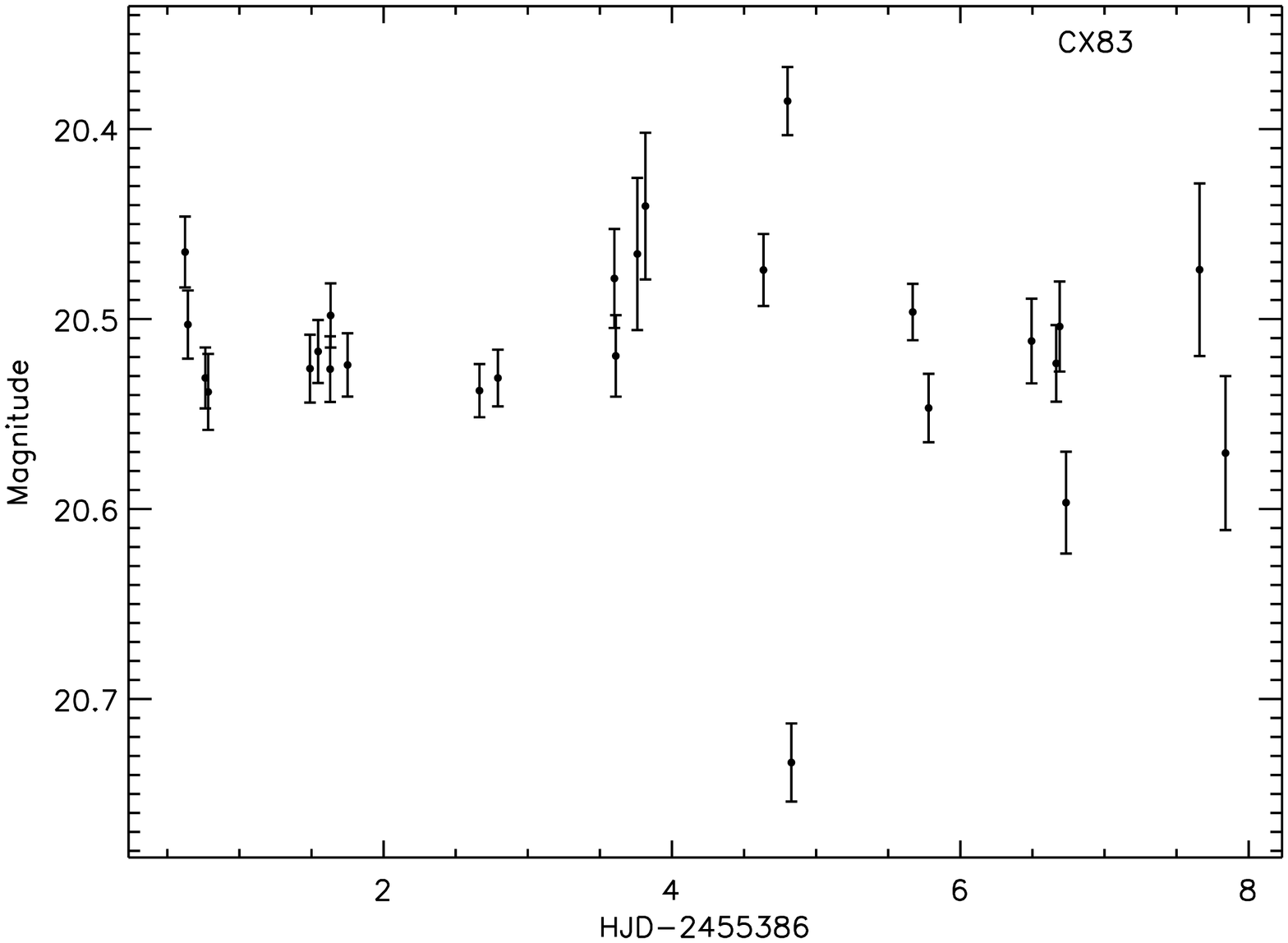}}\quad
\subfigure{\includegraphics[width=0.4\textwidth,angle=90]{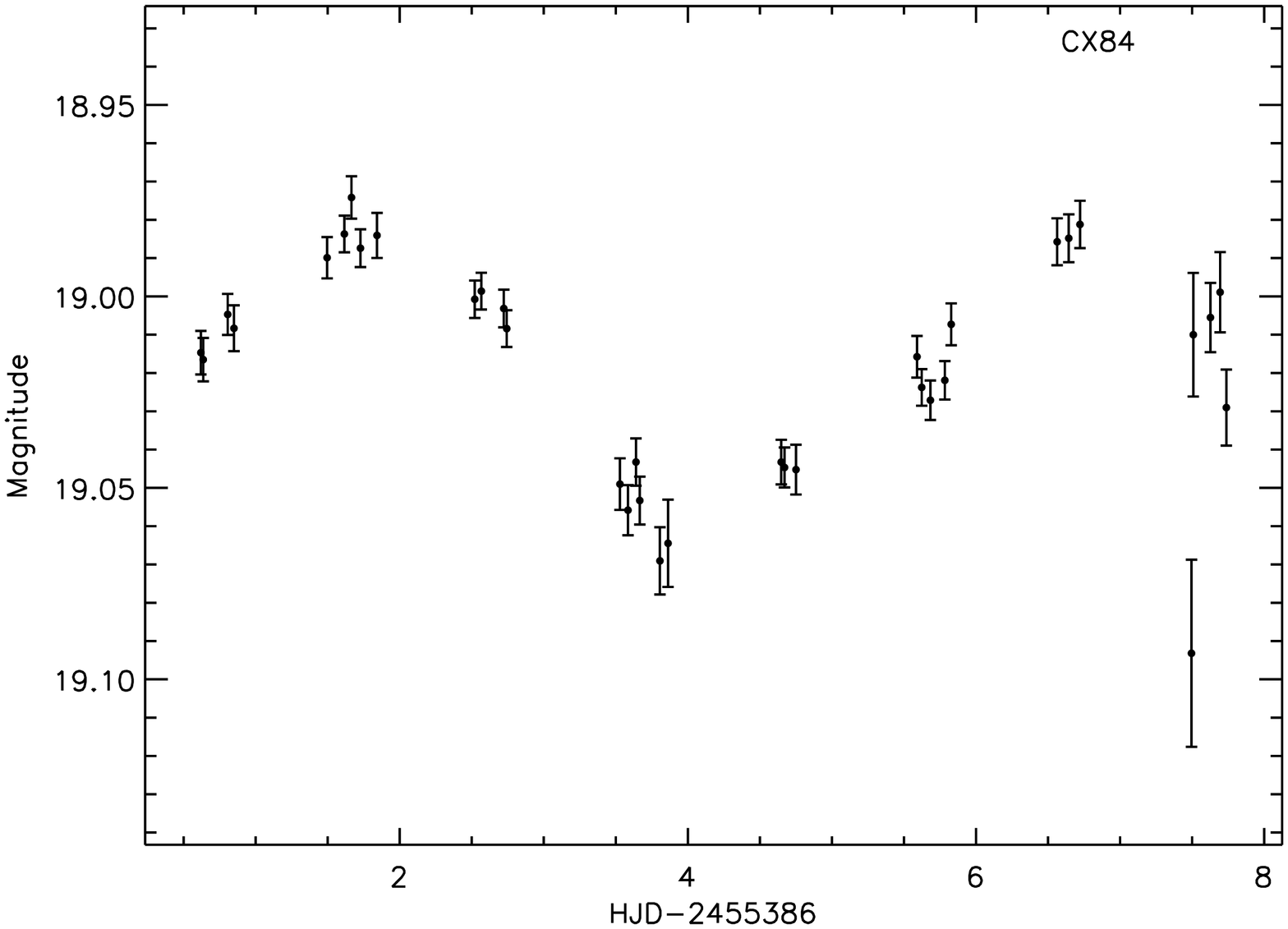}} \\
\subfigure{\includegraphics[width=0.4\textwidth,angle=90]{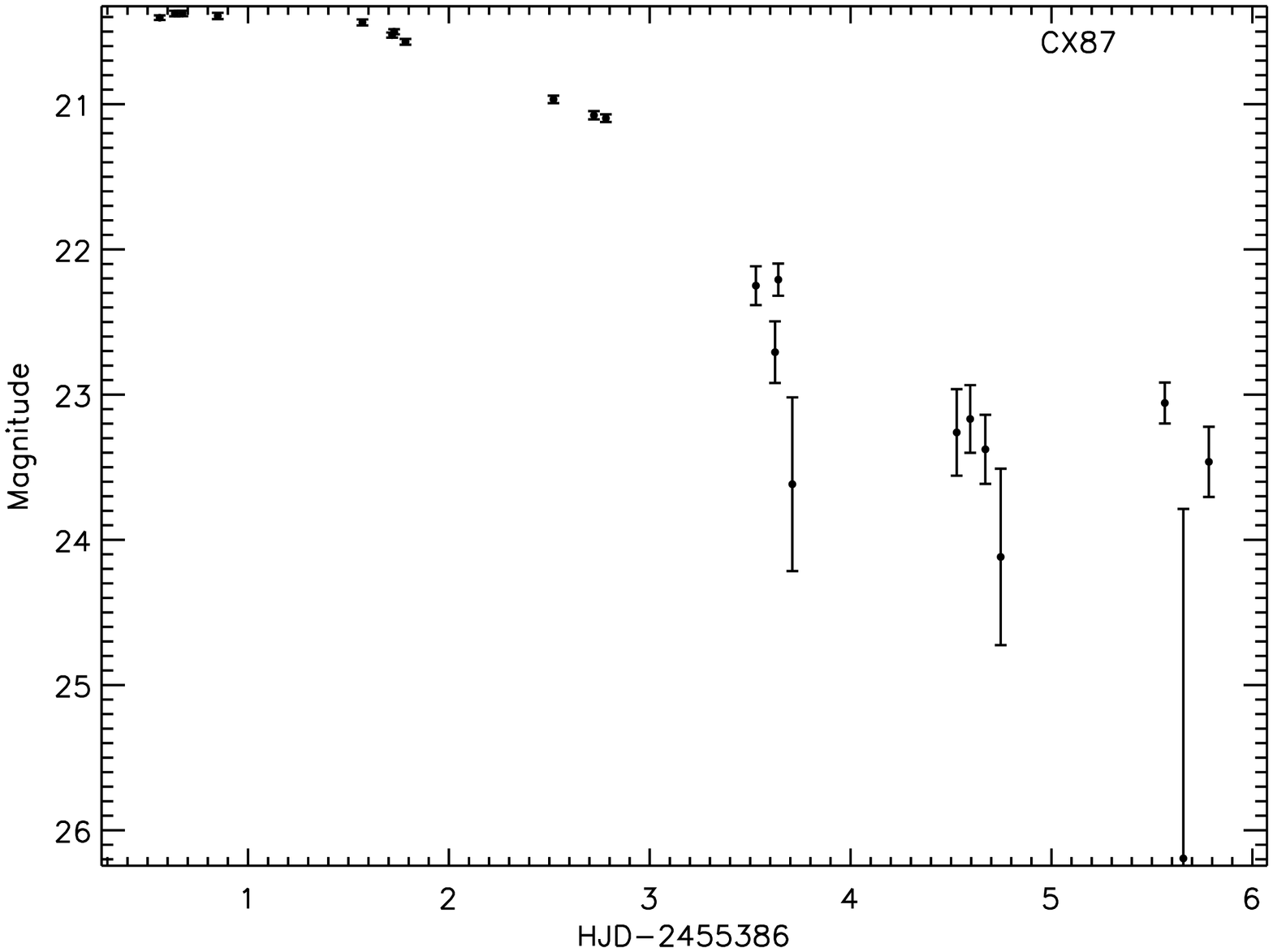}}\quad 
\subfigure{\includegraphics[width=0.4\textwidth,angle=90]{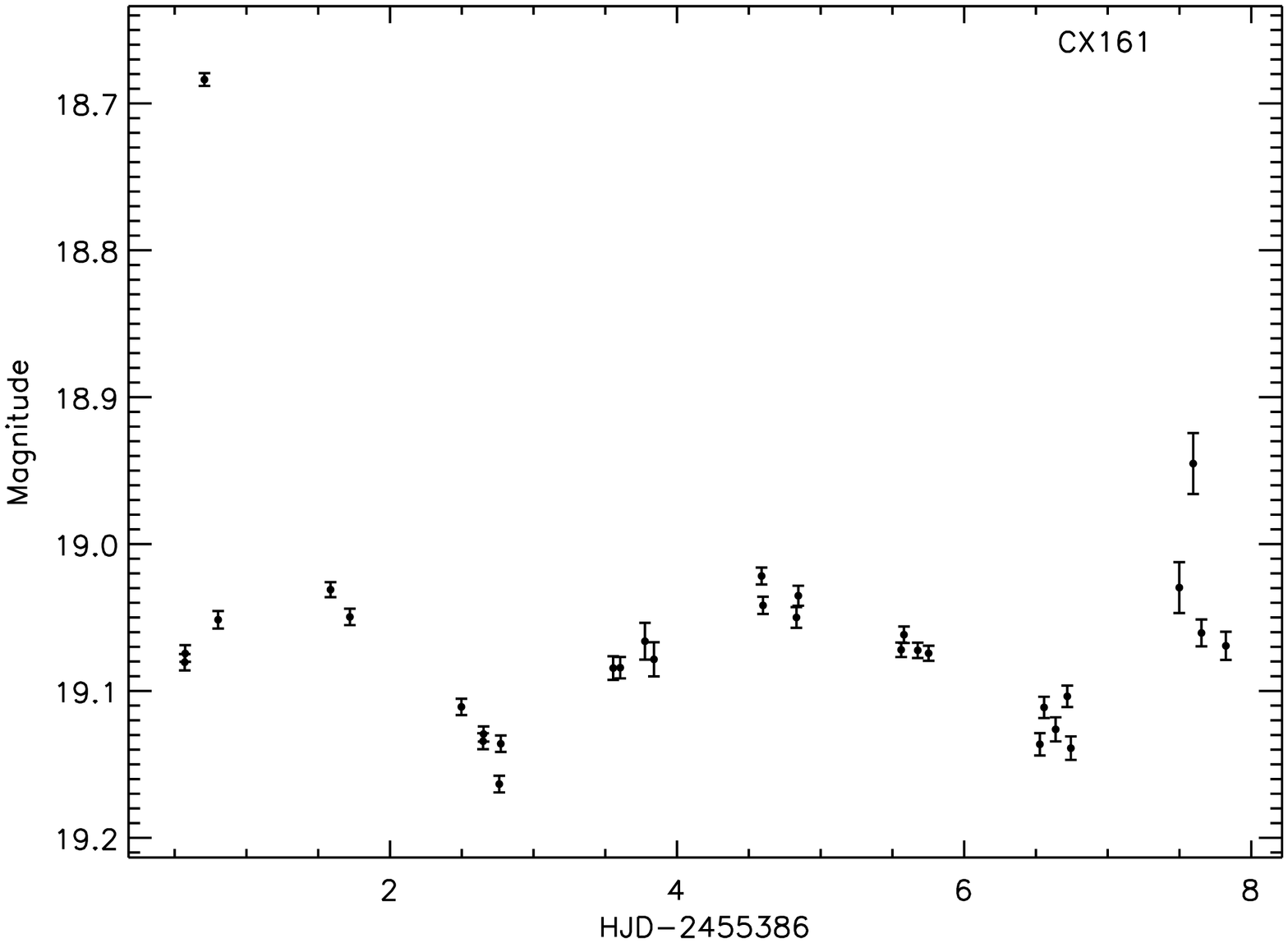}} \\
\subfigure{\includegraphics[width=0.4\textwidth,angle=90]{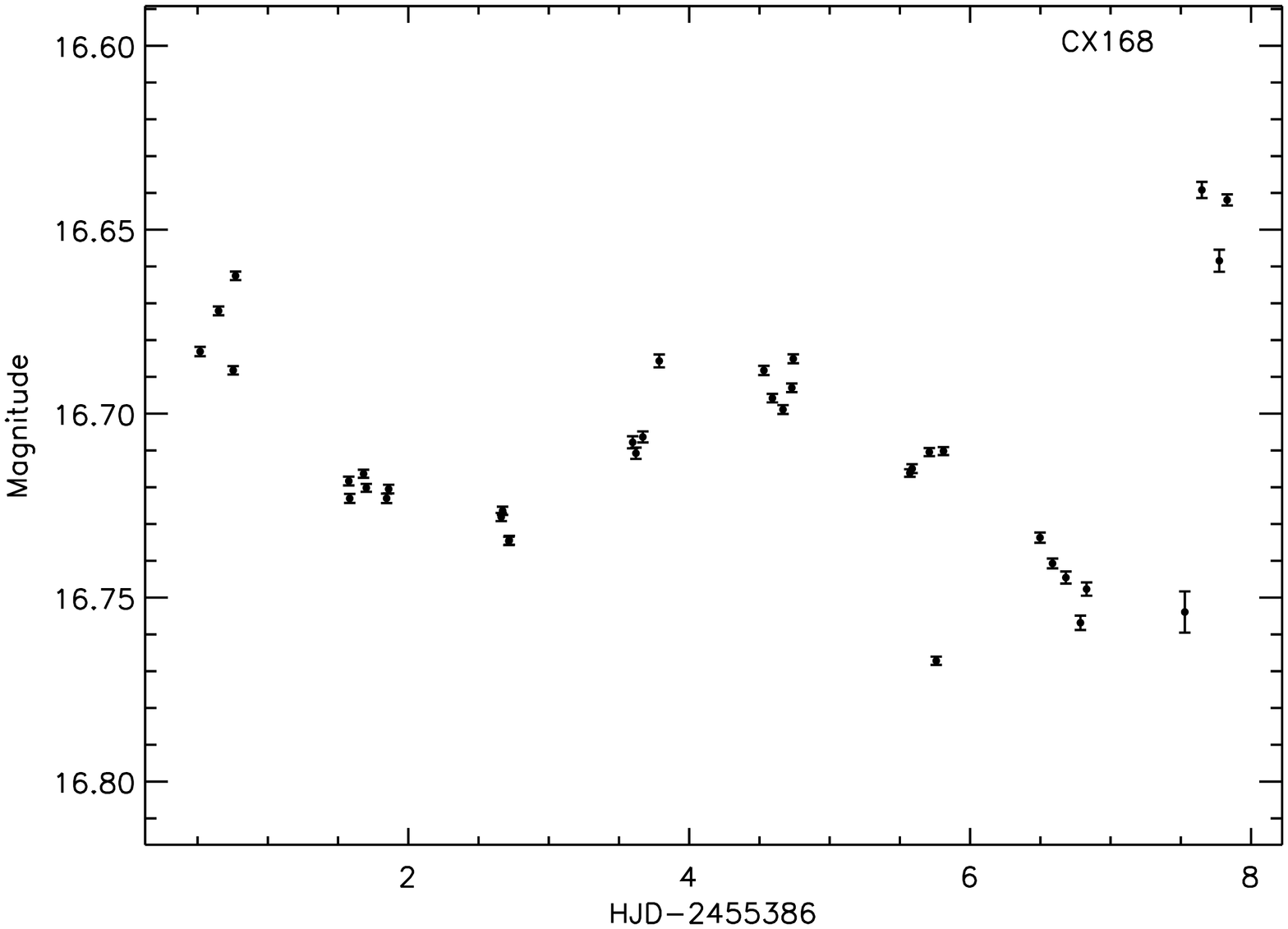}}\quad
\subfigure{\includegraphics[width=0.4\textwidth,angle=90]{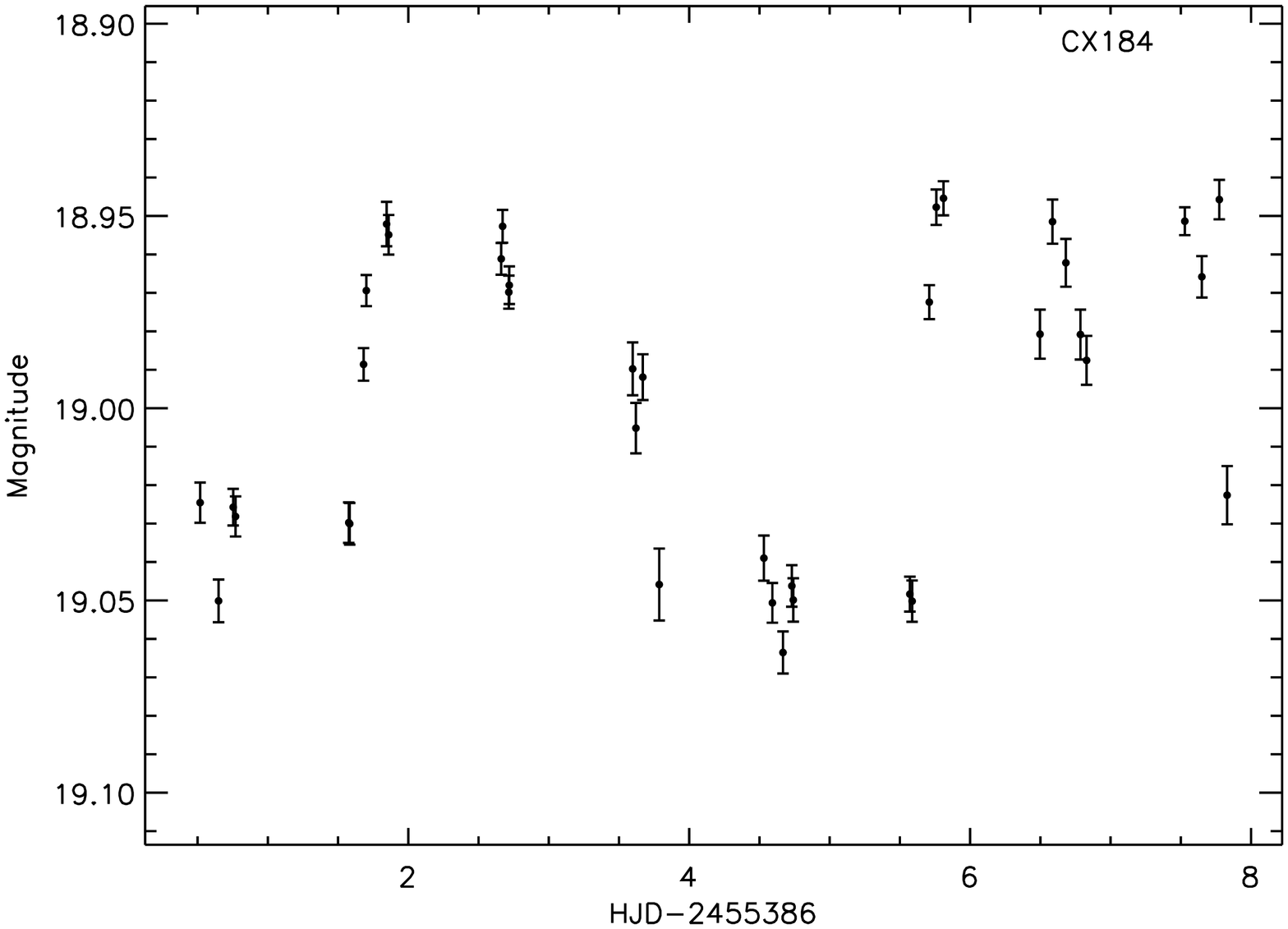}}
}
\caption{CX83, CX84, CX87, CX161, CX168, CX184 Mosaic-II Lightcurves}
\label{lcchunk2}
\end{minipage}
\end{figure*}

CX83 is not significantly variable except for a possible eclipse on the 5th
night of observations shown in Figure \ref{lcchunk2}. This eclipse is
$0.2$ magnitudes in depth and
lasts no more than 2 hours. Only 1 observation out of 26 points is in
eclipse which lasts only $4\pm4\%$ of the orbital phase, assuming that we evenly sample all phases. From
these constraints, we place an upper limit on the orbital period of 2
days. Absorbed
$\frac{F_{X}}{F_{opt}} = 7$, which drops to 0.02 with reddening at the
Bulge distance, which is a range consistent with both CVs and qLMXBs. The X-ray spectrum is very hard with a hardness ratio of 0.86, which is consistent with up-scattering from a disk corona at high inclination and not with the thermal emission expected for a quiescent CV, so a qLMXB interpretation is favored.

\subsection{CX84 - CV or qLMXB}

CX84 has a suspected period of 4.67 days shown in Figure \ref{lcchunk2}, and an absorbed
$\frac{F_{X}}{F_{opt}} = 2$. The amplitude of the variations is
$0.1$ magnitudes which is consistent with both ellipsoidal variations
from accreting binaries. Similarly to CX67, this object is
unlikely to be an RS CVn based on the high X-ray to optical flux
ratio, but it could be a qLMXB or CV with an evolved donor because of the 
multi-day orbital period.  In the
direction of the Bulge, $E(B-V)=1.98$, which implies 
$M_{r'}=-0.8$ if CX84 is in the Galactic Bulge. From VVV and UKIDSS
data \citep{Greiss13,Lucas08}, the counterpart to
CX84 has $J-K=1.41$ and $K_{S}=13.2$, yielding $M_{K}=-2$ and $J-K=0.38$
at Bulge distance after correcting for extinction. \citet{Jianfeng14}
classify this object as a G9 star. Assuming a subgiant G9 star with
$M_{r'}=2.5$ as the donor based on the multiday suspected orbital
period, and that reddening is roughly linear with distance, we derive
a distance of $D\approx3.4 \pm 0.7$\,kpc. At this distance, $L_{X}\approx2.4\times10^{33}{\rm ergs\,s^{-1}}$ which is consistent with qLMXBs. 

\subsection{CX87 - DN}

CX87 starts at close to $r'=20.4$ at the beginning of our
observations, and declines over the next few days to the limiting
magnitude of our observations at $r' \approx23$ as shown in Figure
\ref{lcchunk2}. This decline of 2-3
magnitudes over a few days is consistent with a DN
outburst. Measured from the magnitude when not in outburst,
$\frac{F_{X}}{F_{opt}} =100$, which is too high for a non-magnetic CV. The
faint magnitude in the optical and low
density of stars in the field
is consistent with some extinction. $E(B-V)=1.81$ in this field,
making $\frac{F_{X}}{F_{opt}} =2$ at Bulge distance, while
$L_{X}\approx1.3\times 10^{33} \, (\frac{d}{8\,{\rm kpc}})^{2}{\rm ergs\,s^{-1}}$. This is 
brighter than is consistent with a non-magnetic CV, suggesting it
would have to be closer, which means it likely cannot suffer enough
extinction to have $\frac{F_{X}}{F_{opt}}$ consistent with a non-magnetic CV. 
DN have been observed in
IPs, although rarely. It is possible that this is such a system at a distance 
of $2-3$\,kpc, which would give $L_{X}$ and $\frac{F_{X}}{F_{opt}}$ consistent
with CVs. Because the X-ray observations and optical observations are not
simultaneous, it is also possible that CX87 is a non-magnetic CV that 
was observed in the X-ray at the start of the outburst, while the X-ray luminosity was still increases as a result of increased mass accretion rate and before quenching begins during DNe 
outbursts \citep{Warner03}. If the X-ray 
observations was taken during such an outburst, then $\frac{F_{X}}{F_{opt}}$ 
would be much closer to $1-10$, consistent with non-magnetic CVs undergoing 
DN outbursts.  CX87 is classified spectroscopically by \citep{Torres14} as a CV.
Either scenario would make CX87 an interesting object for further study.

\subsection{CX161 - Flare star}

CX161 has a period of 3.32 days and shows a flare of $0.3$
magnitudes that fades back to the quiescent level before the next
observation $2.3$ hours later, as shown in Figure
\ref{lcchunk2}. Absorbed $\frac{F_{X}}{F_{opt}} = 1$, which
could be consistent with a flaring M dwarf between uncertainties in
reddening, flux levels, and the fact that most of bolometric luminosity 
for late M dwarfs is emitted in the IR rather than at optical wavelengths 
so that
$F_{opt}<<F_{Bol}$. The short timescale and magnitude of the flare is typical of
flare stars, while the period is attributable to stellar rotation and
star spots. The IR colors $H=13.885$ and $K_{S}=14.075$ from VVV data are also consistent 
with an M-dwarf \citep{Greiss13}. This object is suggested to be a chromospheric active star or binary after spectroscopic observations in \citet{Torres14} pending further analysis, which agrees with our photometric classification.

\subsection{CX168 - CV, qLMXB, or M dwarf?}

CX168 shows variations with a possible period of 3.8 days as
shown in Figure \ref{lcchunk2}, but
there is also some flickering superposed on these variations and a
brightening of $0.1$ magnitudes lasting at least a few hours on night
8. Absorbed $\frac{F_{X}}{F_{opt}} = 0.1$, which is consistent with
CVs, qLMXBs, and active M dwarfs. The absolute magnitude of this object if it 
were in the
Galactic Bulge is $M_{r'}=-6.1$ which too bright for CVs or qLMXBs, and would 
suggest a supergiant companion if this object were in the Bulge rather than 
in the foreground. In
addition, with Bulge reddening, $\frac{F_{X}}{F_{opt}} \approx
10^{-4}$ which is too low for CVs or qLMXBs. Assuming a subgiant
companion for a CV or qLMXB with an absolute magnitude of
$M_{r'}=2.5$ and that reddening is linear with distance, the colors
and magnitudes are consistent with a
distance of $2.3$\,kpc.  Therefore, the true
X-ray to optical flux ratio is likely towards the high end
of the given
range with the distance of the source substantially closer than the
Bulge.  From 2MASS,
this object has $J-K=1.14$, which is consistent with a nearby M
dwarf, though it could also be a qLMXB with a larger, cool companion at
a greater distance. The small flare on night 8 of observations could be flaring in an 
active M dwarf, while the absorbed $J-K$ values in 2MASS are consistent with
an M dwarf in the foreground. In
addition to the flare on night 8, there is a small dip of $0.07$
magnitudes on night 6 that lasts less than 2 hours. Spectroscopy can
quickly differentiate between an isolated M dwarf and an accreting CV
or qLMXB with a subgiant donor.

\subsection{CX184 - CV, qLMXB, or M dwarf?}

\begin{figure*}[p!]
\begin{minipage}{0.9\textwidth}
\centering
\parbox{\textwidth}{
\subfigure{\includegraphics[width=0.4\textwidth,angle=90]{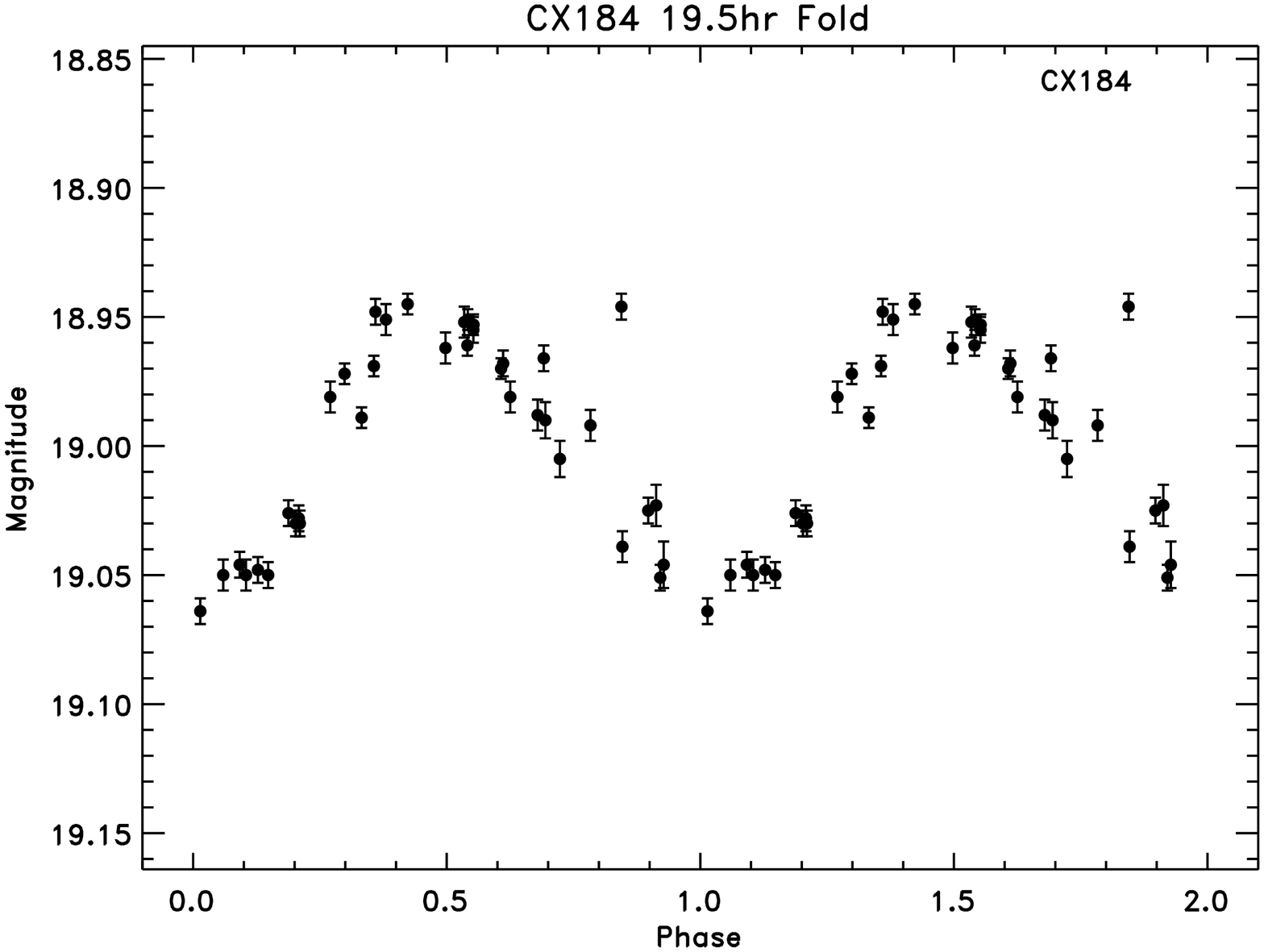}}\quad
\subfigure{\includegraphics[width=0.4\textwidth,angle=90]{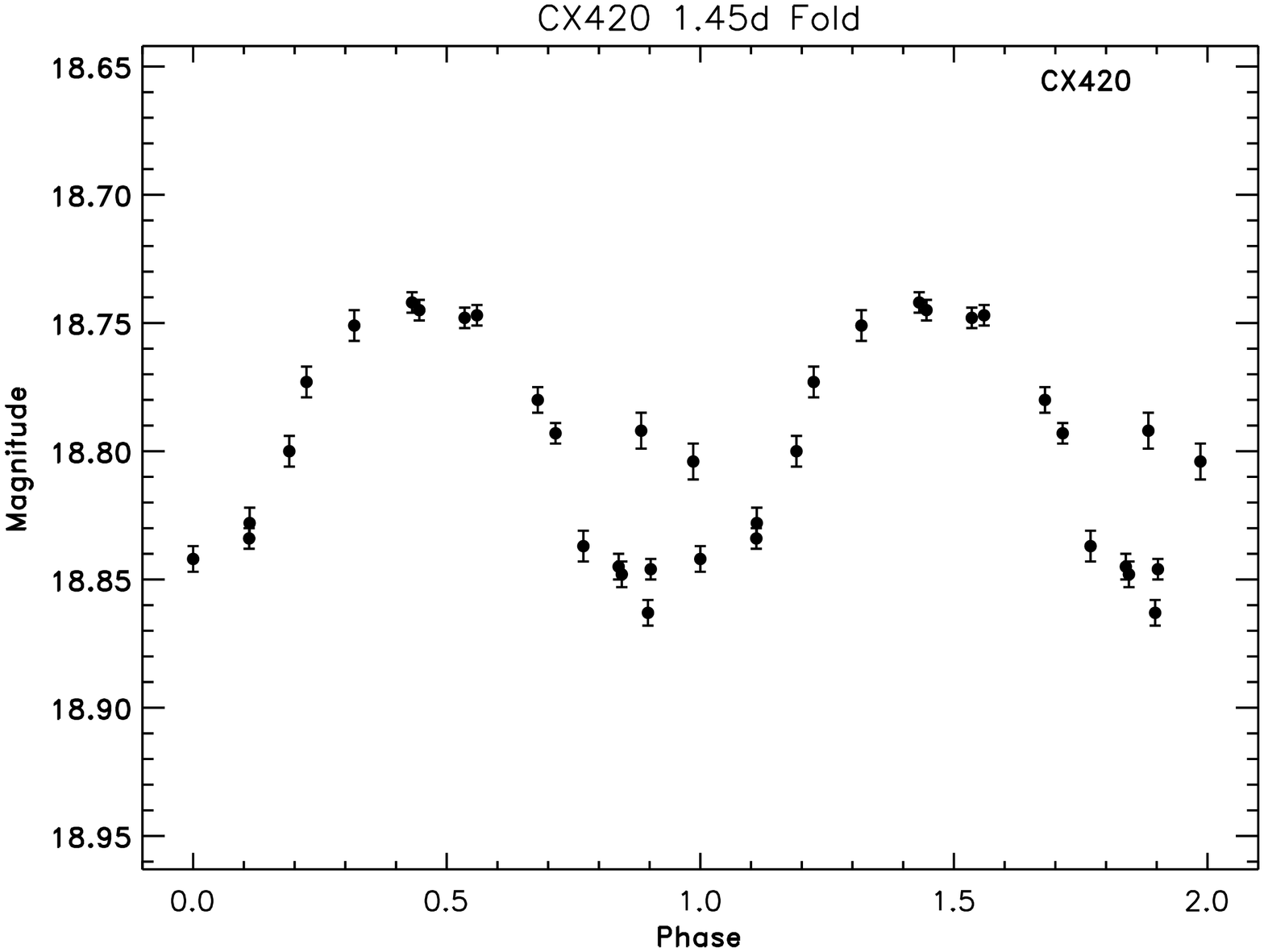}} \\
\subfigure{\includegraphics[width=0.4\textwidth,angle=90]{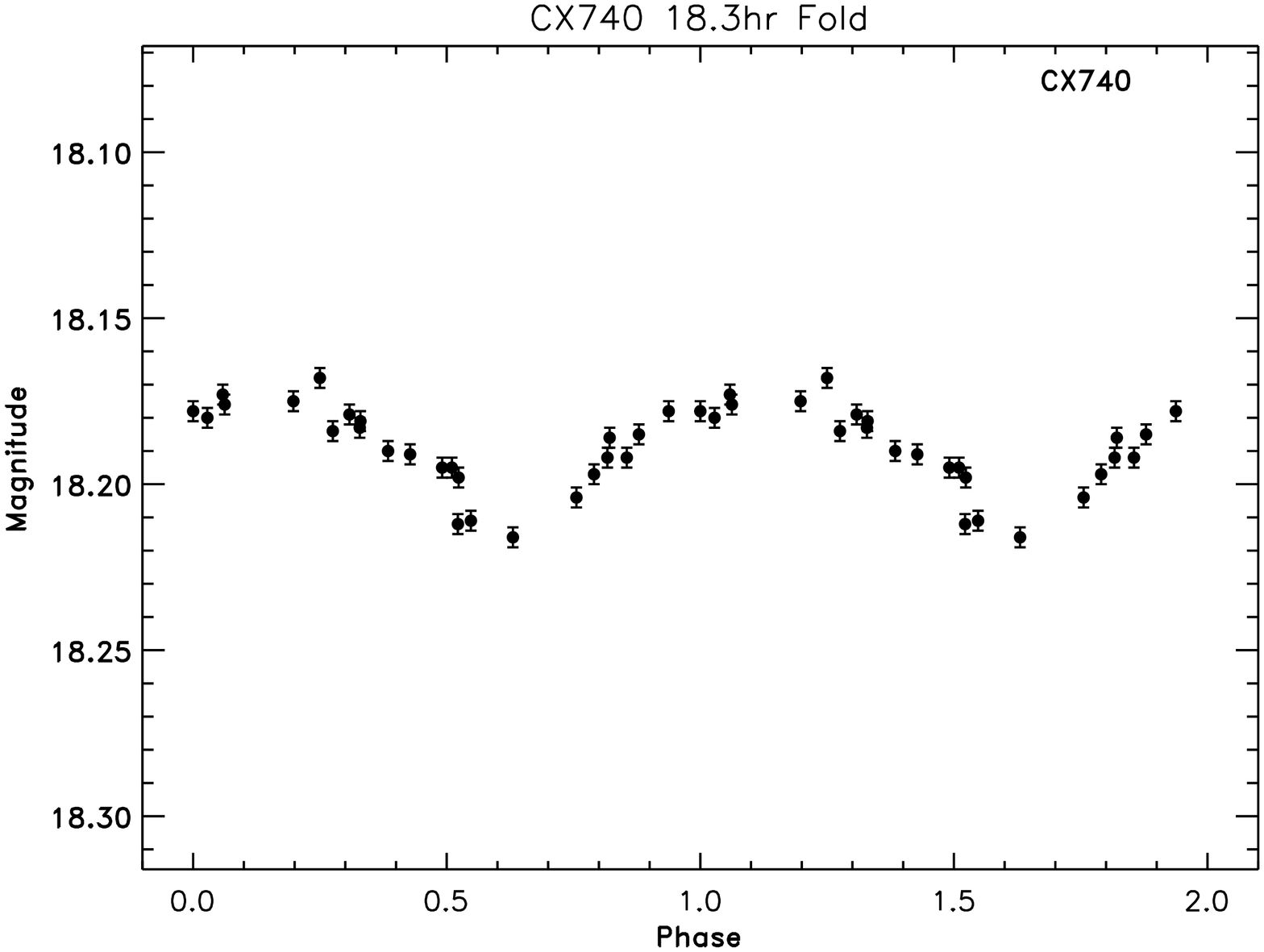}}\quad 
\subfigure{\includegraphics[width=0.4\textwidth,angle=90]{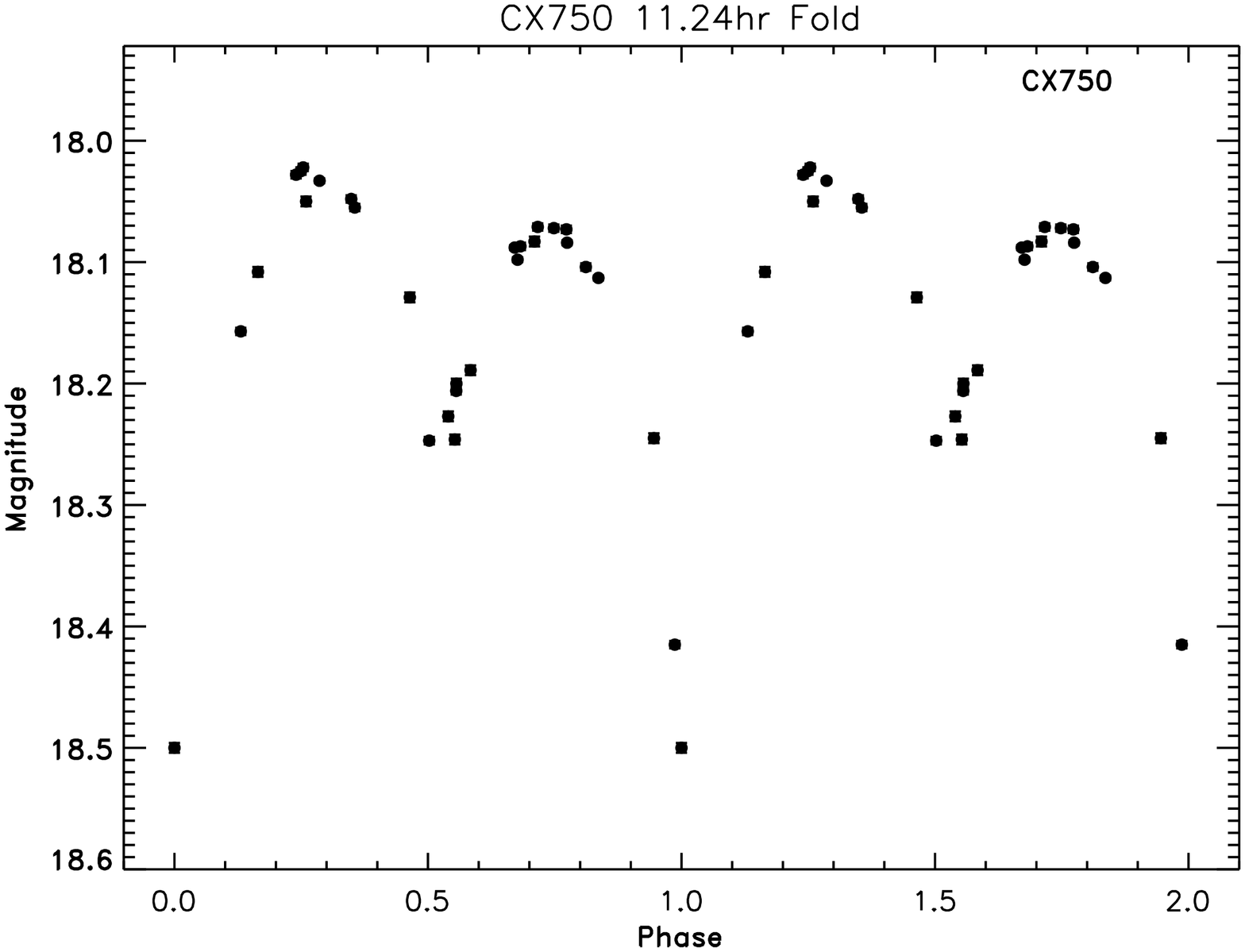}} \\
\subfigure{\includegraphics[width=0.4\textwidth,angle=90]{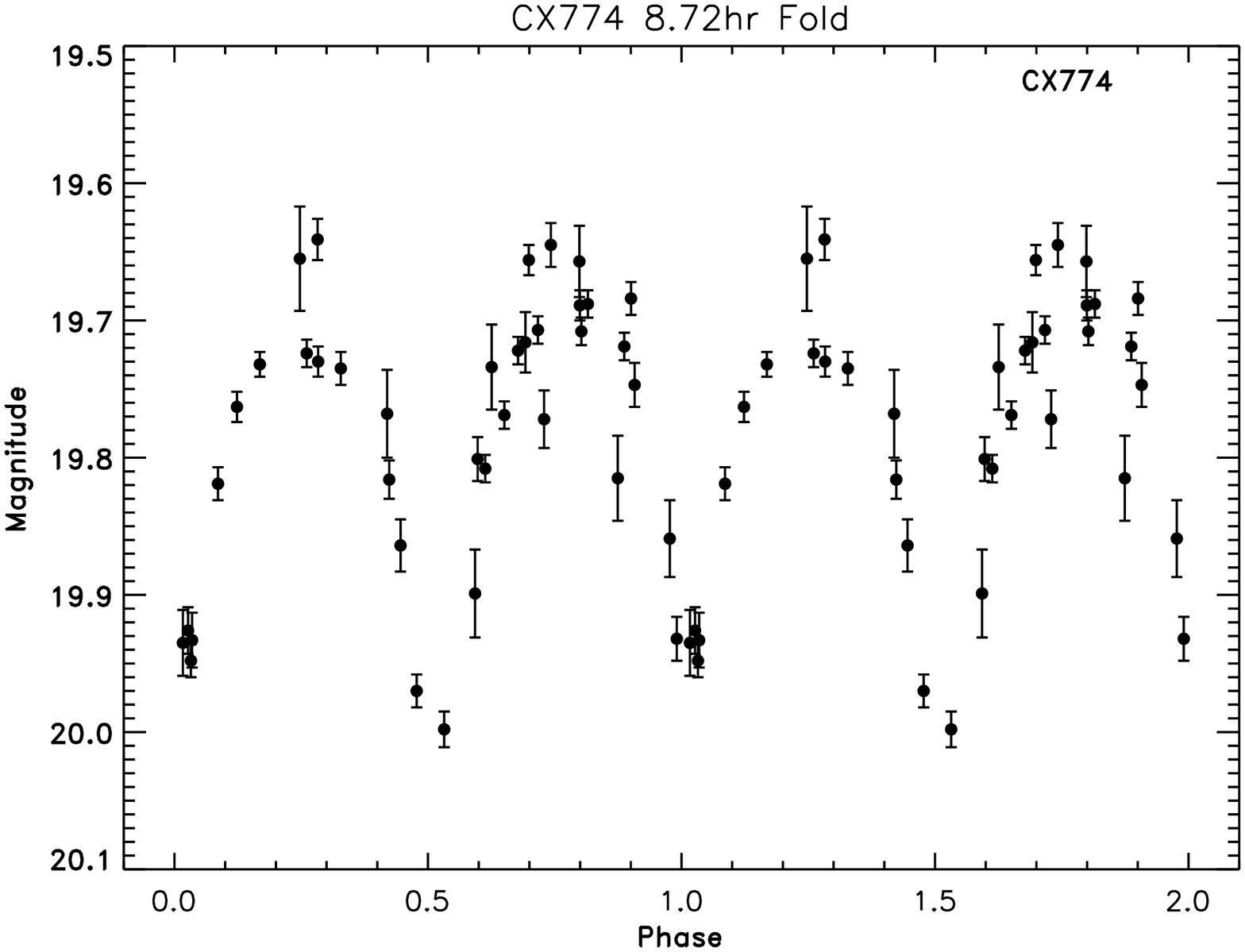}}\quad
\subfigure{\includegraphics[width=0.4\textwidth,angle=90]{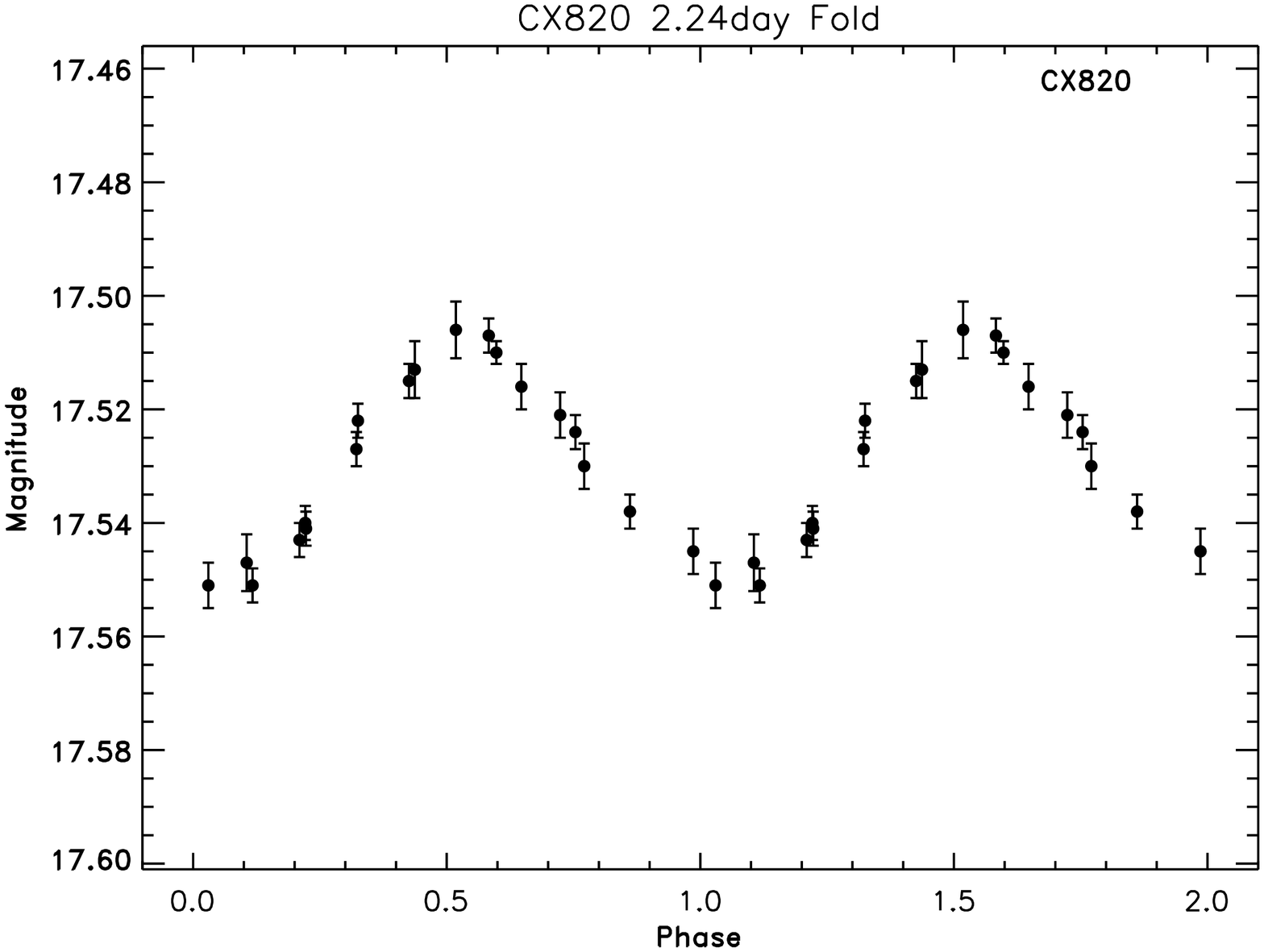}}
}
\caption{{\it Top Left:} Mosaic-II lightcurve of CX184 folded
on a 19.5 hour orbital period. 
{\it Top Right:} Mosaic-II lightcurve of CX420 folded
on a 1.45 day orbital period.
{\it Middle Left:} Mosaic-II lightcurve of CX740 folded
on an 18.3 hour period.
{\it Middle Right:} Lightcurve of one of the possible
  counterparts to CX750
  folded on an 11.24 hour period.
{\it Bottom Left:} Lightcurve of one of the possible counterparts to CX774 folded on a 8.72 hour period.
{\it Bottom Right:} Lightcurve of CX820 folded on a 2.242 day period.}
\label{foldchunk1}
\end{minipage}
\end{figure*}

CX184 has a photometric period of 0.811 days, or 19.5 hours. The
unfolded lightcurve is shown in Figure \ref{lcchunk2}, while the folded
lightcurve is shown in
Figure \ref{foldchunk1}. Absorbed
$\frac{F_{X}}{F_{opt}} = 1$, which is consistent with qLMXBs,
CVs, or M dwarfs. The variations
are single humped, with a steeper rise than decline, and have an amplitude of
$0.1$ magnitudes. It is also possible that they are ellipsoidal
variations with roughly equal minima, but the data are insufficient to
differentiate the two. The higher X-ray to optical flux ratio in
combination with a possible orbital period below a day makes this a
candidate qLMXB or CV, though spectroscopic follow up is necessary to
differentiate those two possibilities both from one another, which is 
non-trivial, and from a fast rotating active M dwarf. From VVV data, $J-K=0.74$, which could be consistent with either a foreground early M dwarf or a CV or qLMXB. 

\subsection{CX251 - M dwarf, possibly binary}

\begin{figure*}[p!]
\begin{minipage}{0.9\textwidth}
\centering
\parbox{\textwidth}{
\subfigure{\includegraphics[width=0.4\textwidth,angle=90]{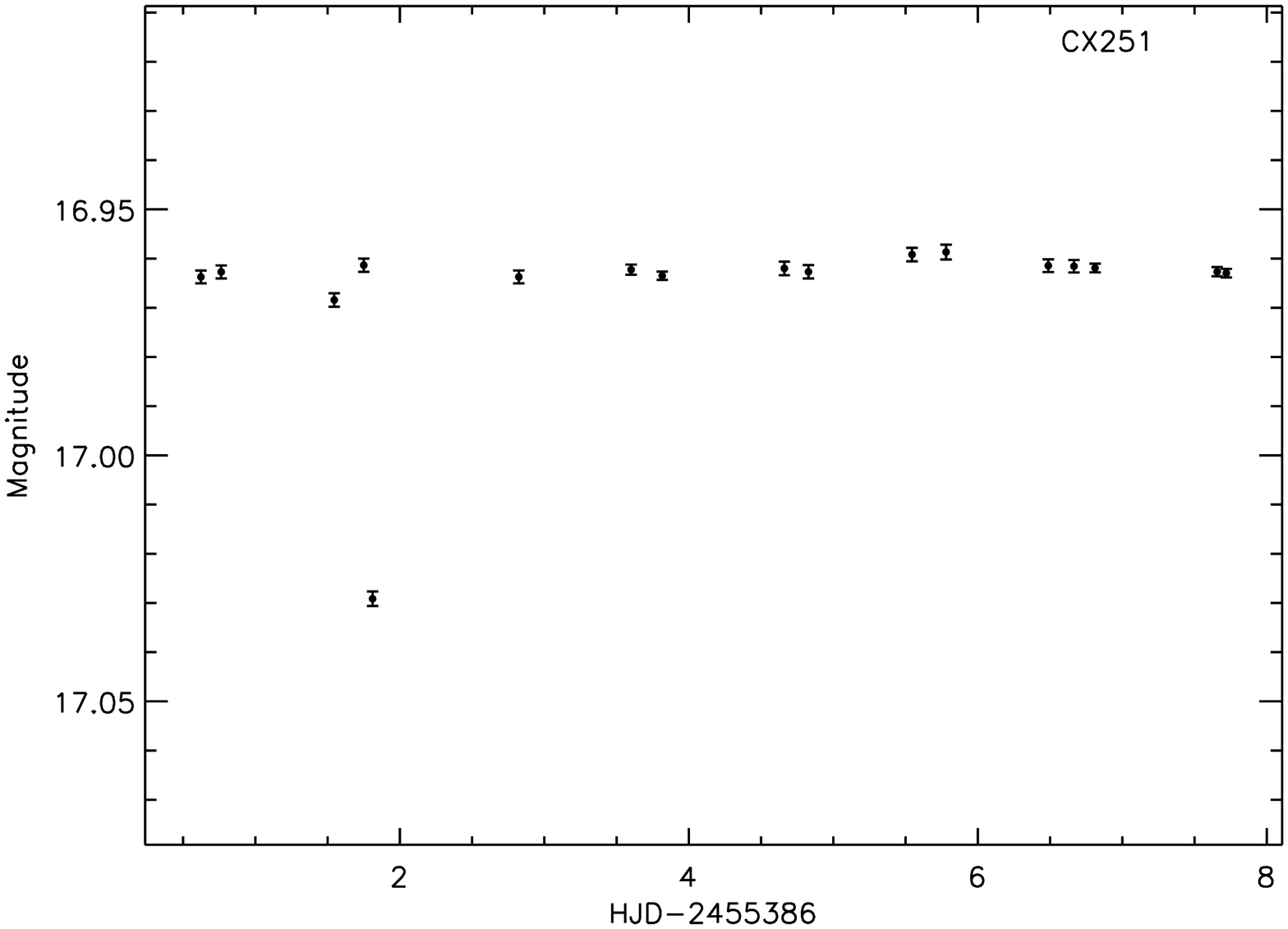}}\quad
\subfigure{\includegraphics[width=0.4\textwidth,angle=90]{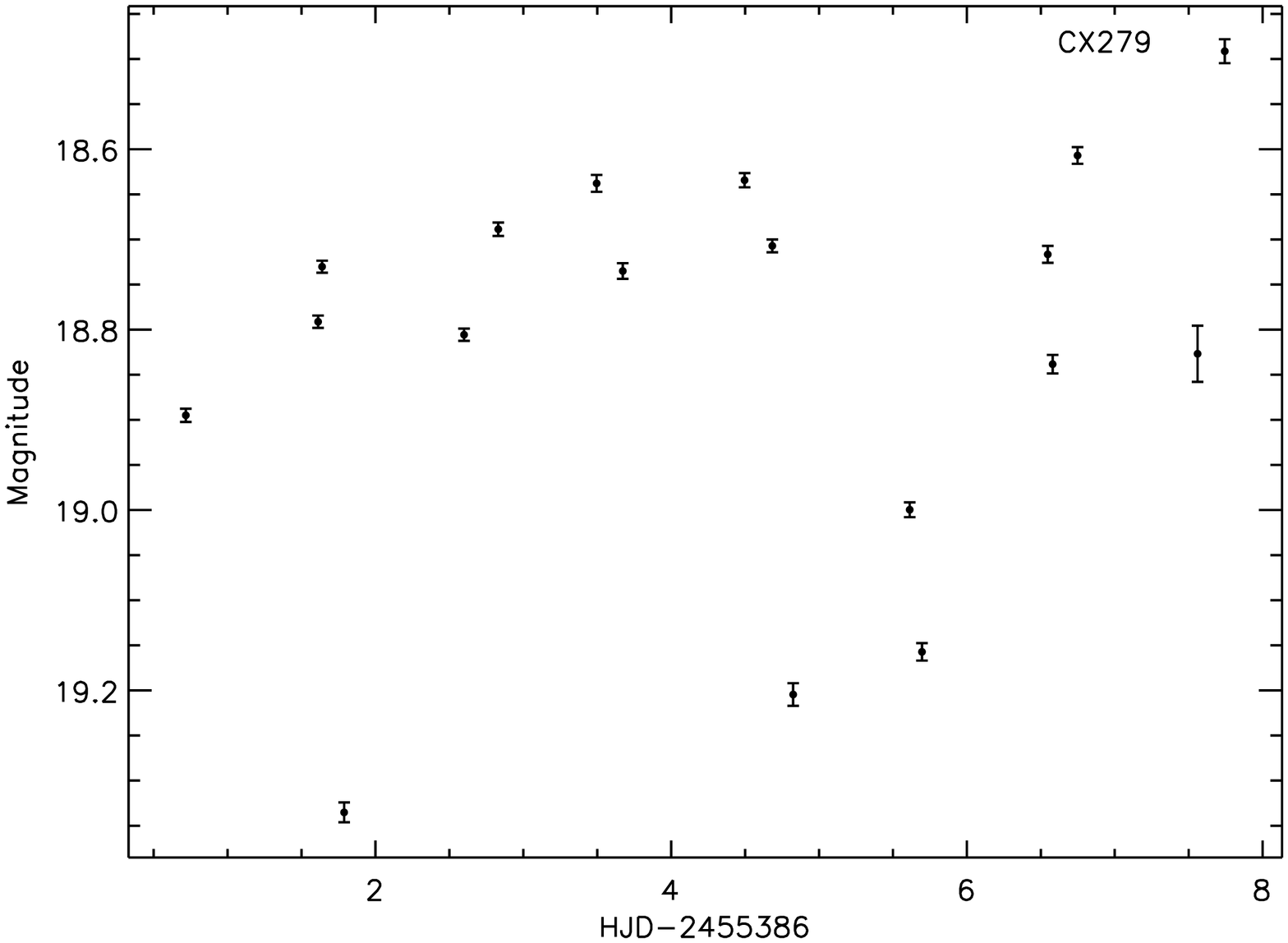}} \\
\subfigure{\includegraphics[width=0.4\textwidth,angle=90]{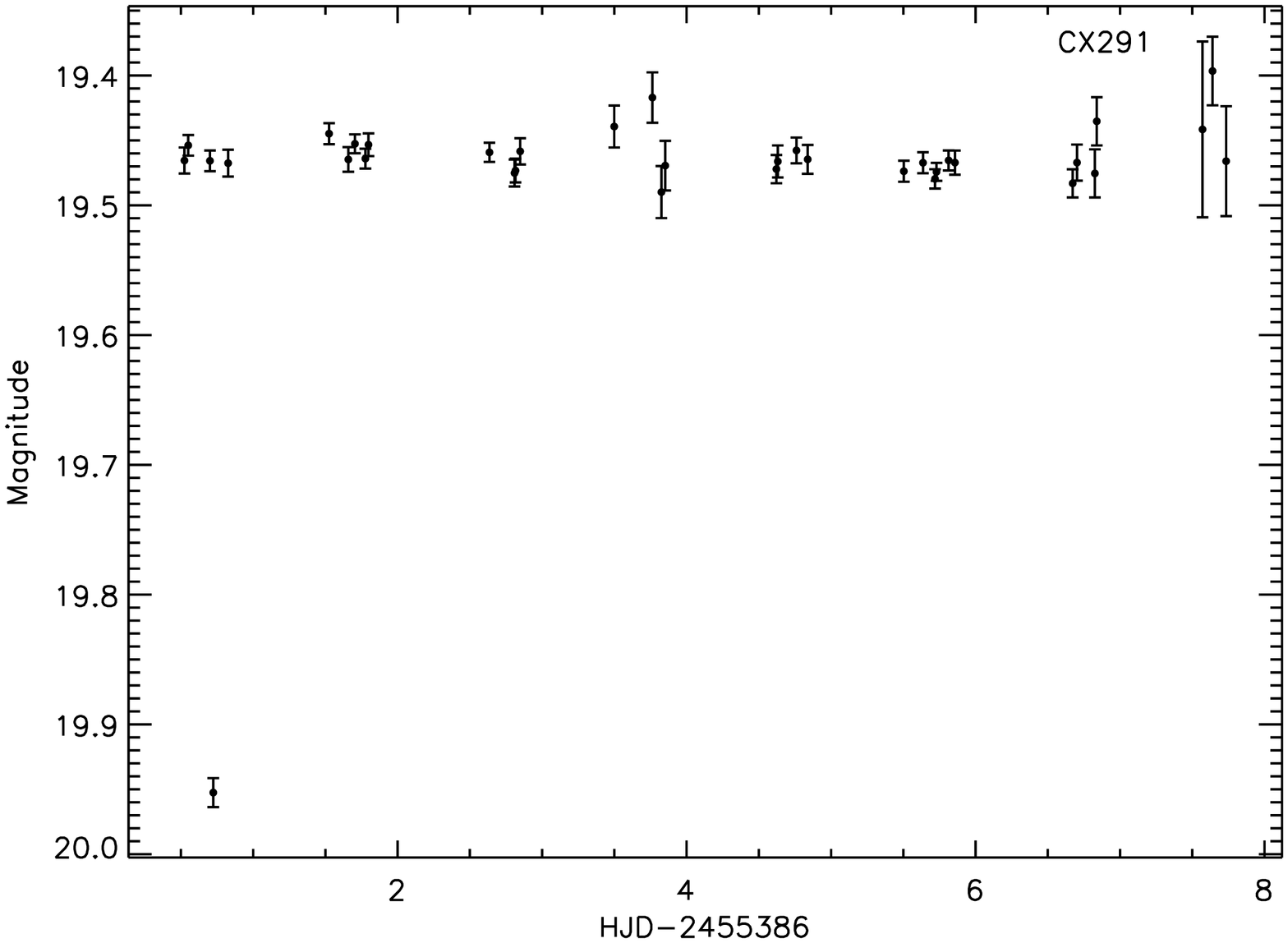}}\quad 
\subfigure{\includegraphics[width=0.4\textwidth,angle=90]{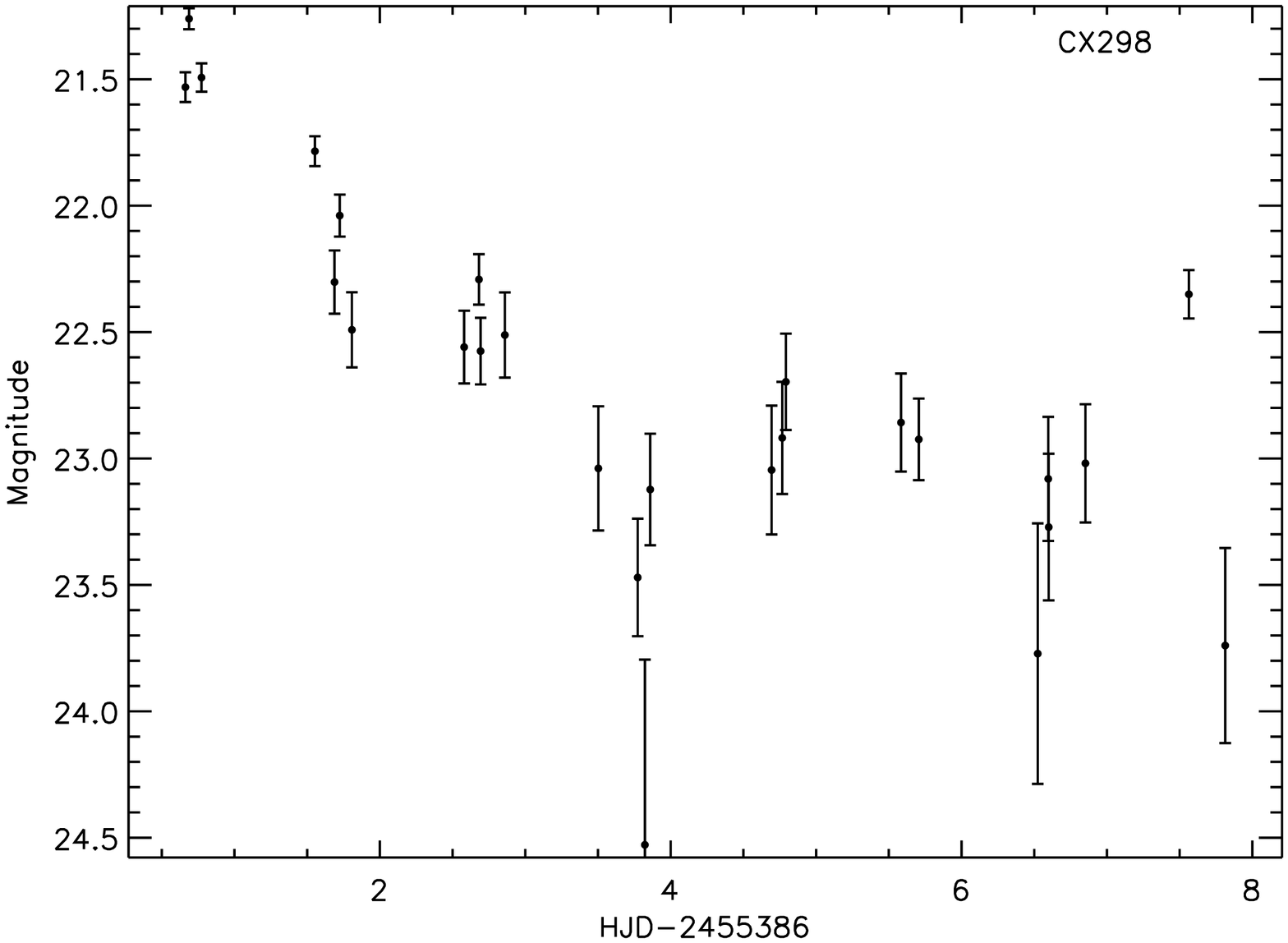}} \\
\subfigure{\includegraphics[width=0.4\textwidth,angle=90]{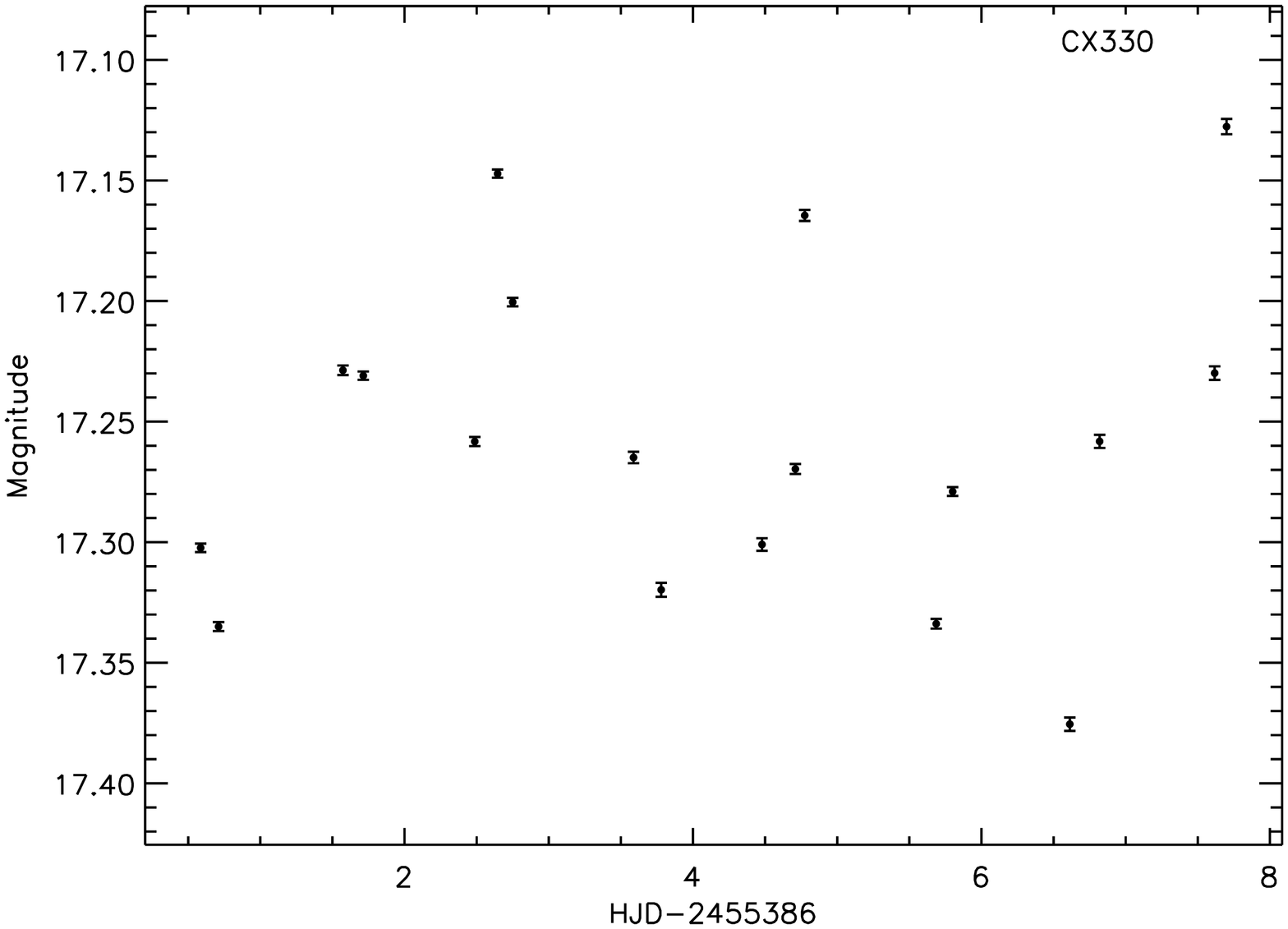}}\quad
\subfigure{\includegraphics[width=0.4\textwidth,angle=90]{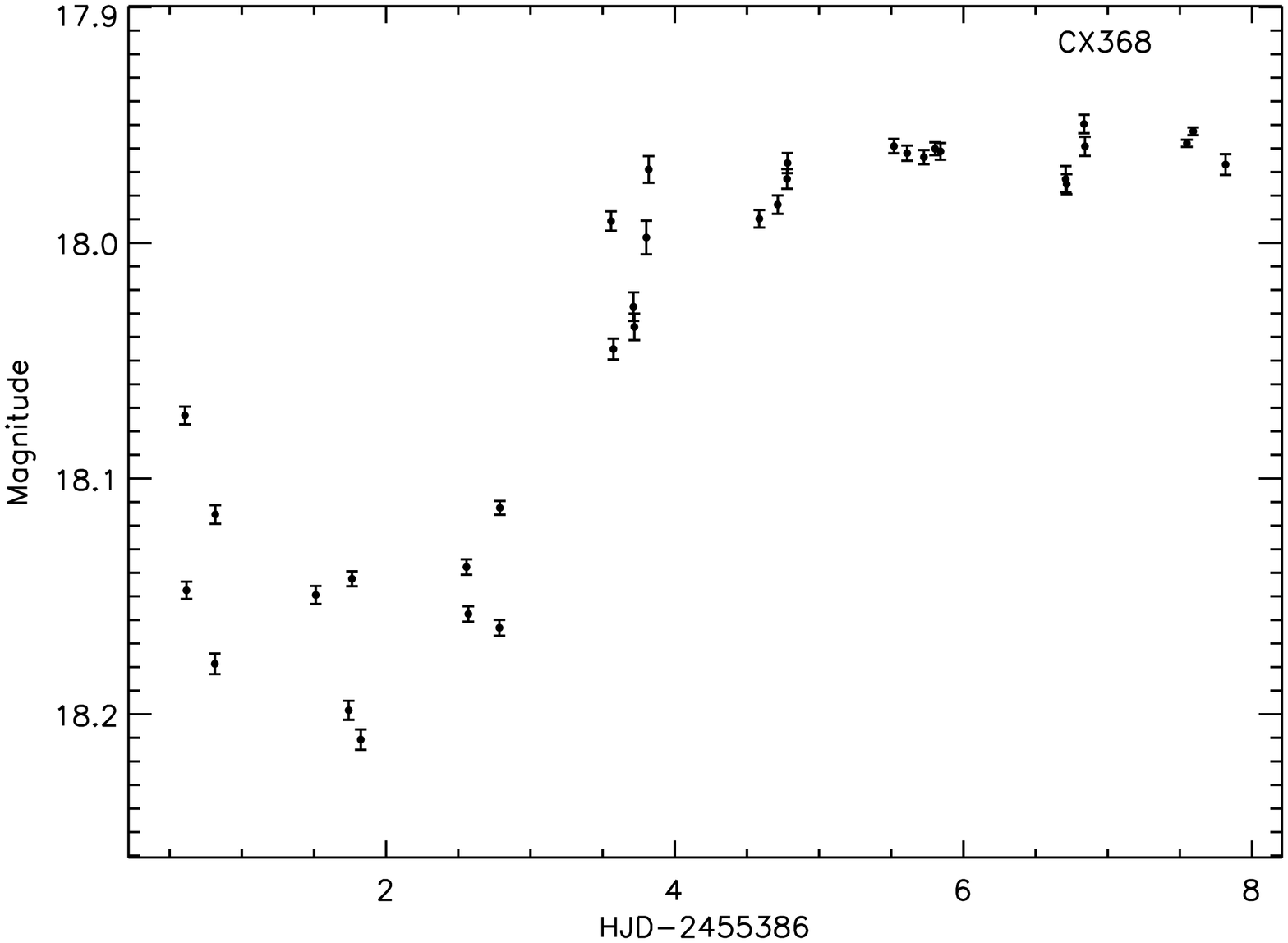}}
}
\caption{CX251, CX279, CX291, CX298, CX330, CX368 Mosaic-II Lightcurves}
\label{lcchunk3}
\end{minipage}
\end{figure*}

CX251 is very steady except for a single dip at least $0.07$
magnitudes deep, shown in Figure \ref{lcchunk3}, which could be part of
an eclipse. The eclipse depth and duration are not well constrained as the
last observation in night 2 is the only point in eclipse, so many different 
depths and durations are consistent with the data. Absorbed
$\frac{F_{X}}{F_{opt}} = \frac{1}{10}$ which is consistent with qLMXBs, CVs, and M dwarfs, though the lack of ellipsoidal variations, even at high inclination, 
argues against a system accreting through Roche-Lobe overflow.  
It is possible that this object is simply an eclipsing 
binary in the field coincident with the X-ray position. $\frac{F_{X}}{F_{opt}}$ 
is too high for the X-ray emission to be the result of coronal activity in 
this object unless it is an M dwarf with most of the bolometric luminosity 
in the infrared. If the dip is part of an eclipse, the high $\frac{F_{X}}{F_{opt}}$ would suggest that both binary members are M dwarfs. $K=12.0$ and $J-K=1.63$ in VVV \citep{Greiss13}, which is consistent with 
an M dwarf. This object is suggested to be a chromospheric active star or binary after spectroscopic observations in \citet{Torres14} pending further analysis, which agrees with our photometric classification.

\subsection{CX279 - CV or qLMXB}

We recovered no significant period for CX279 as determined by Monte
Carlo simulations with the Lomb-Scargle statistic. This source appears
to be eclipsing, shown in Figure \ref{lcchunk3}, with eclipses of $0.6$ magnitudes which are
relatively broad, taking 4 out of 18 observations, implying they last for 
$22\pm11\%$ of the orbital phase. Absorbed $\frac{F_{X}}{F_{opt}} = 0.6$ which is 
consistent with qLMXBs and CVs. This object is also very 
faint in the infrared, at $K_{S}\approx16.5$ \citep{Greiss13}, which favors 
bluer sources such as CVs.

\subsection{CX291 - Binary M dwarfs or Interloper?}

CX291 has a single point in eclipse in the first night of
observations, with a depth of $0.5$ magnitudes, shown in Figure \ref{lcchunk3}.
There is no variability apart from this eclipse; the weighted average as a model for the brightness
has $\chi^{2}/\nu=1.4$ with the eclipse excluded. As with CX251, the only
observation in eclipse comes at the end of the night, leaving its
duration nearly unconstrained. The absorbed X-ray to optical flux ratio is
1, which is consistent with CVs, qLMXBs, and M dwarfs, though the lack of ellipsoidal 
modulations from tidal distortion is problematic for a compact binary interpretation.
Also, from VVV data \citep{Greiss13}, $K_{s}=12.35$, which makes this
object quite red. From 2MASS, $J-K=0.84$, which  is also consistent
with nearby M dwarfs, though $r'-K_{S}=7.1$ could also point to
significant reddening. Like CX251, this object could also simply be an
interloping eclipsing binary.

\subsection{CX298 - DN}

CX298 drops from $r'=21.4$ on the first night of observations to the
limiting magnitude of $r'=23$ by night 4, where it remains
steady as shown in Figure \ref{lcchunk3}. This large, steady decline could 
be the end of a DN outburst. The X-rays are likely emitted in quiescence, as 
the high accretion rate during the DN outburst quenches X-ray emission 
substantially \citep{Collins10,Patterson85}. 
If the limiting magnitude is taken as the 
brightness in the quiescent state, absorbed $\frac{F_{X}}{F_{opt}} = 20$, 
though it could be lower with reddening. At the bulge, 
$\frac{F_{X}}{F_{opt}} = 1$, which is consistent with CVs. If it is a
CV undergoing DN outbursts, however,
then the distance should be substantially closer than the Bulge as
their X-ray luminosity is below what we are likely to detect at that
distance. With a moderate amount of extinction, however, this ratio
of X-ray to optical light is consistent with a CV undergoing DNe
outbursts. This is a similar source to CX87, which also appears to be
the end of a DN outburst with a very high X-ray to optical flux ratio,
fading to or below our limiting magnitude. There is no infrared counterpart to
this object in VVV or 2MASS, which further supports a DN interpretation.

\subsection{CX330}

CX330 appears in OGLE-IV data \citep{Udalski12} as an irregular variable. In 
our data, it shows large amplitude aperiodic flickering covering a range of
$0.3$ magnitudes shown in Figure \ref{lcchunk3}.  The X-ray to optical flux
ratio is fairly low at $0.1$ before correcting for extinction. The
brightness varies on a timescale of hours, which makes it unlikely to be an 
AGN. The flickering is consistent with an accreting source. 

\subsection{CX368 - CV, qLMXB, or Symbiotic Binary?}

This source shows a dip of $0.2$ magnitudes before holding plateauing at
the peak brightness, as shown in Figure \ref{lcchunk3}. The dip lasts for
at least 3 days, though it
begins before the start of our observations. The flickering present
throughout the lightcurve favors an accreting binary instead of star
spots on an M dwarf or RS CVn. 
The magnitude of
the counterpart is $r'=17.96$ which is consistent with a giant star at
the Bulge distance. This object appears in
the 2MASS catalog at $J=14.185$ and $K=12.692$. For Bulge reddening,
$E(J-K) = 0.91, A_{K}=0.63$. This implies $M_{K}=-2.46$ at the Bulge which is also
consistent with a giant star. In Symbiotic Binaries,
X-rays are produced when
winds from a giant star accrete onto a WD or NS. We have already
identified a possible Carbon star Symbiotic XRB in CX332
\citep{Hynes14}, which are much more rare than ordinary symbiotics. 
Absorbed $\frac{F_{X}}{F_{opt}} = 0.2$, which is consistent with CVs
or qLMXBs, which cannot be ruled out. At Bulge
reddening, $\frac{F_{X}}{F_{opt}} = 0.006$ which is also consistent
with a Symbiotic Binary.  $L_{X}\simeq 10^{32}\,(\frac{d}{8\,{\rm kpc}})^{2}\,{\rm
  ergs\, s^{-1}}$, which is also consistent with Symbiotic X-ray Binaries at the Bulge.

\subsection{CX420 - CV or qLMXB}

\begin{figure*}[p!]
\begin{minipage}{0.9\textwidth}
\centering
\parbox{\textwidth}{
\subfigure{\includegraphics[width=0.4\textwidth,angle=90]{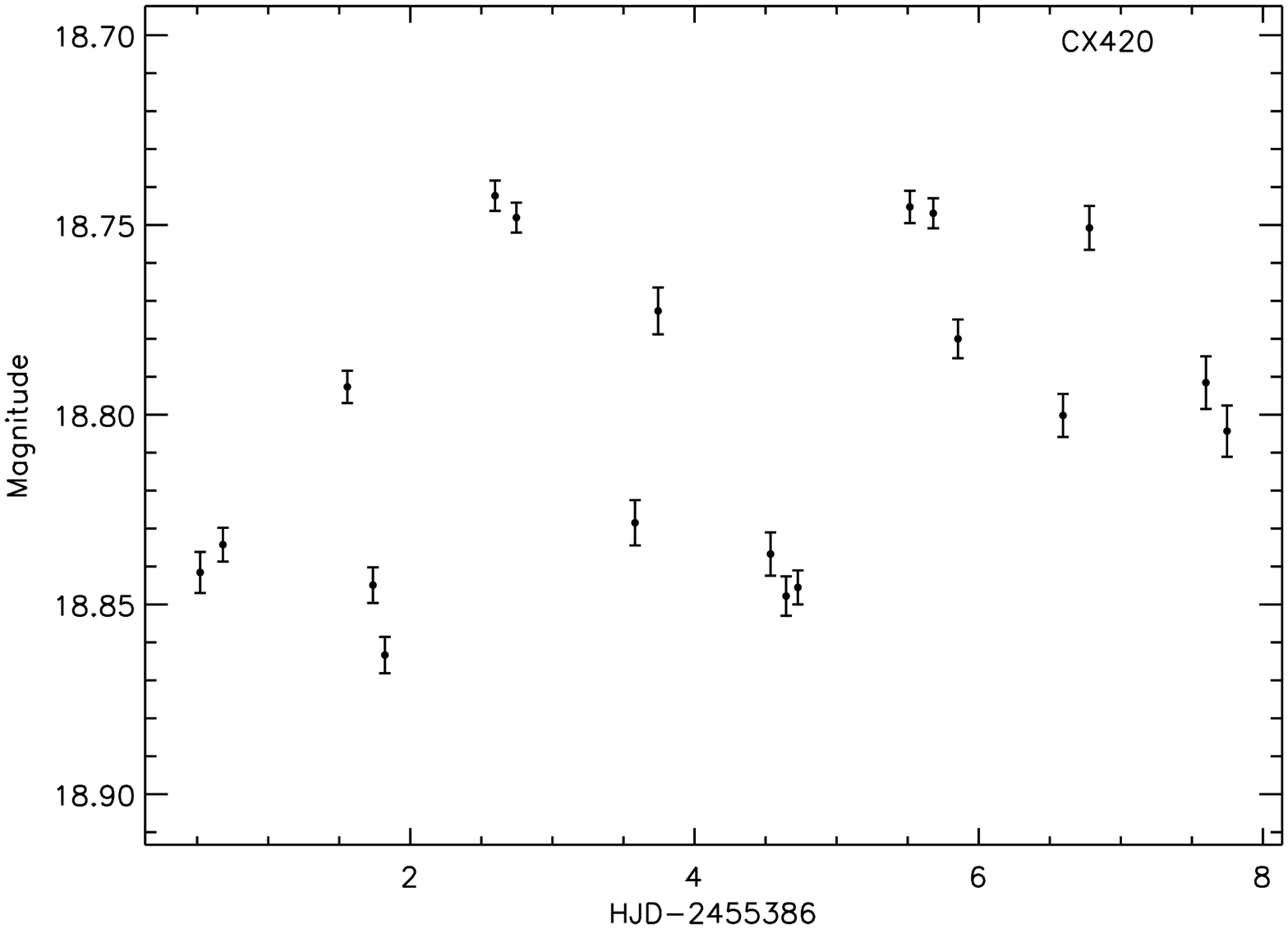}}\quad
\subfigure{\includegraphics[width=0.4\textwidth,angle=90]{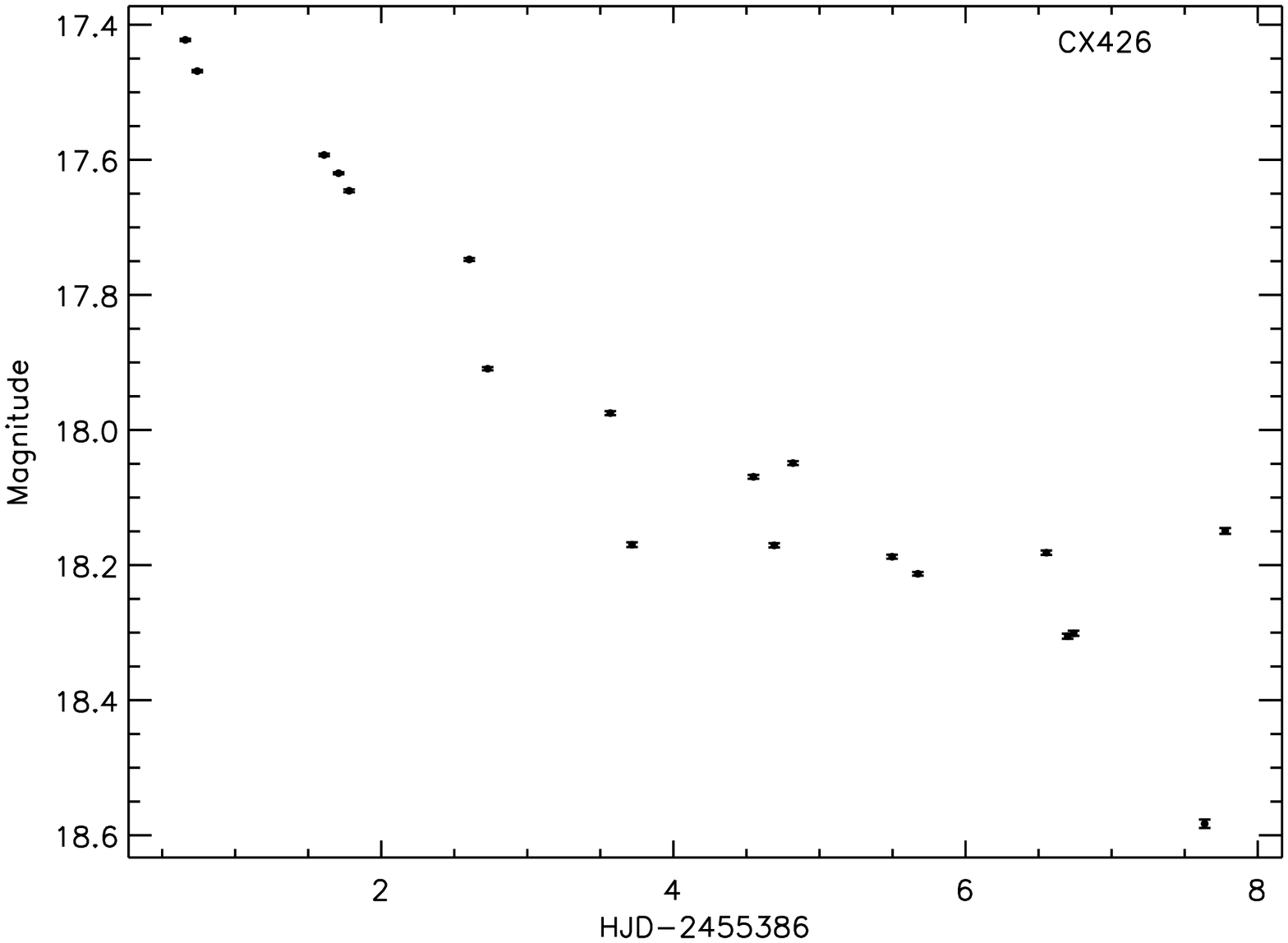}} \\
\subfigure{\includegraphics[width=0.4\textwidth,angle=90]{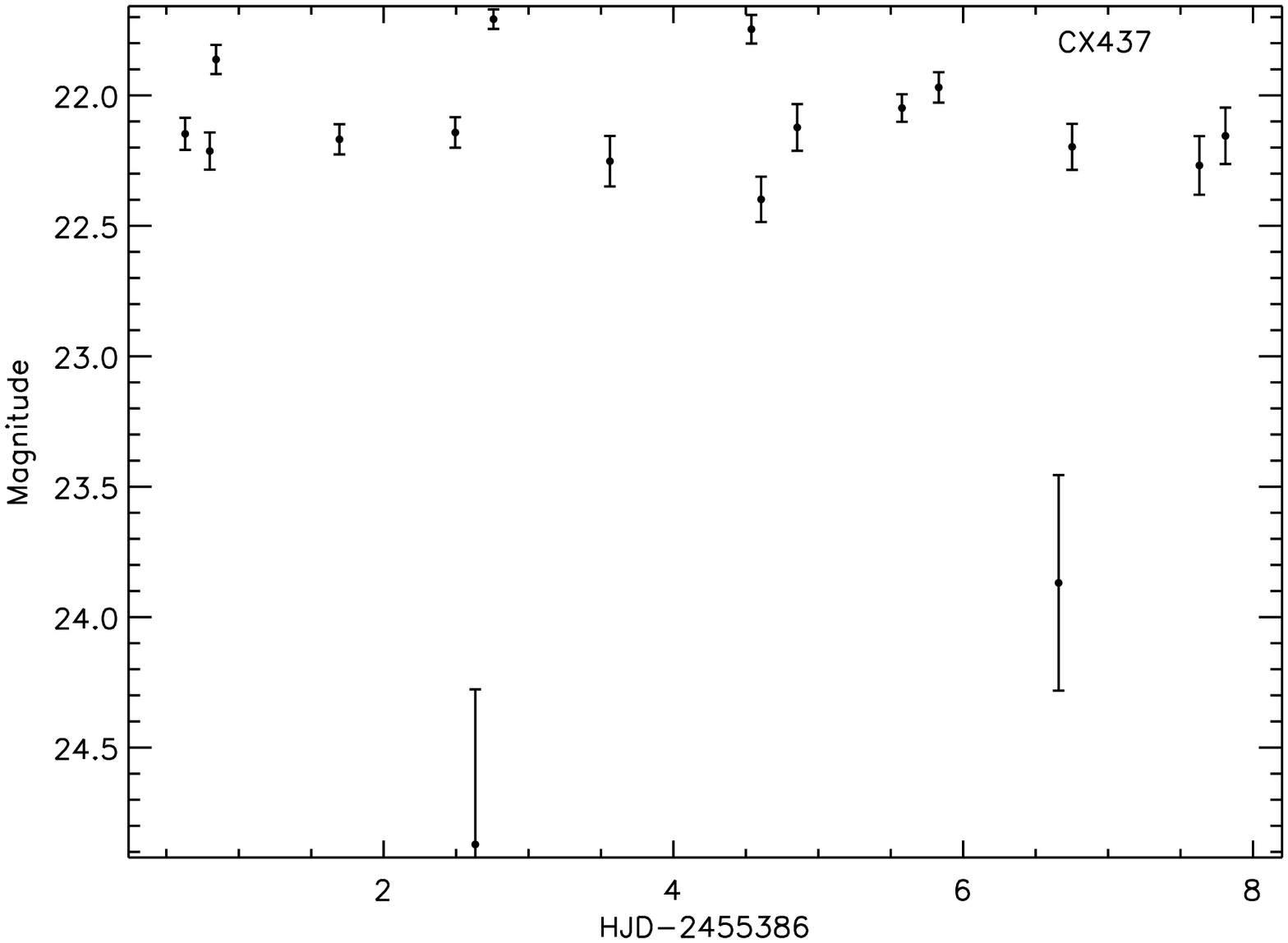}}\quad 
\subfigure{\includegraphics[width=0.4\textwidth,angle=90]{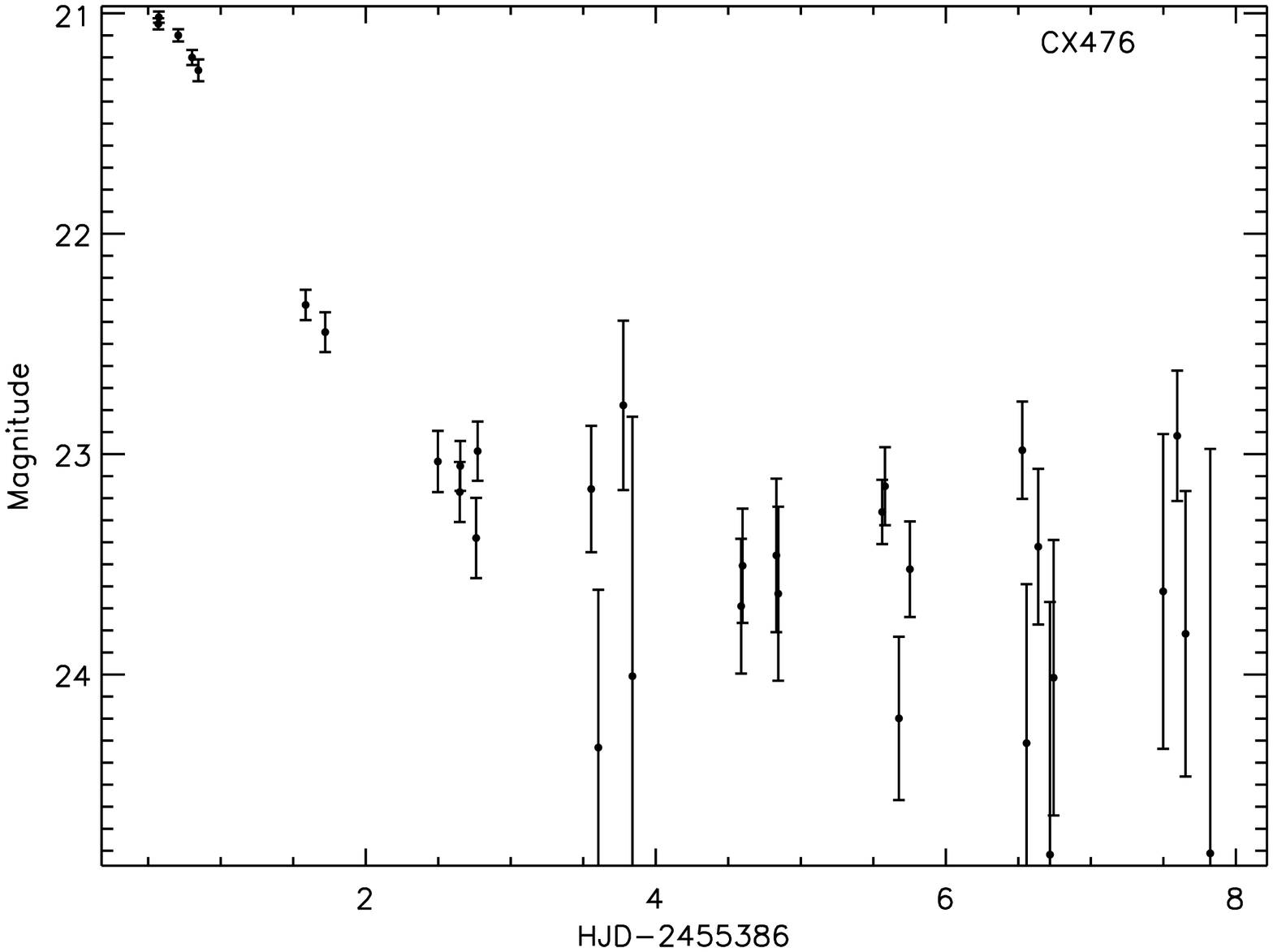}} \\
\subfigure{\includegraphics[width=0.4\textwidth,angle=90]{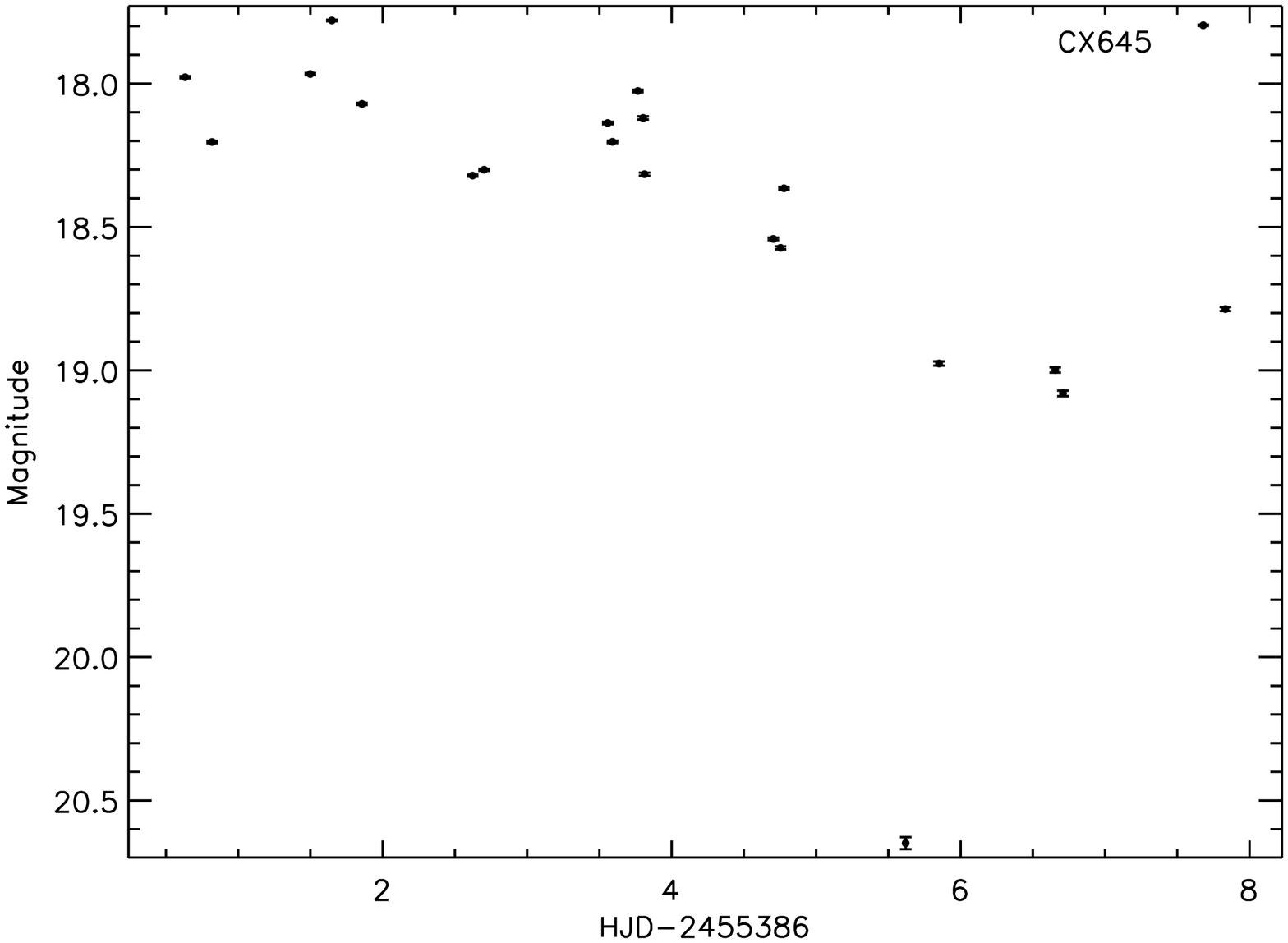}}\quad
\subfigure{\includegraphics[width=0.4\textwidth,angle=90]{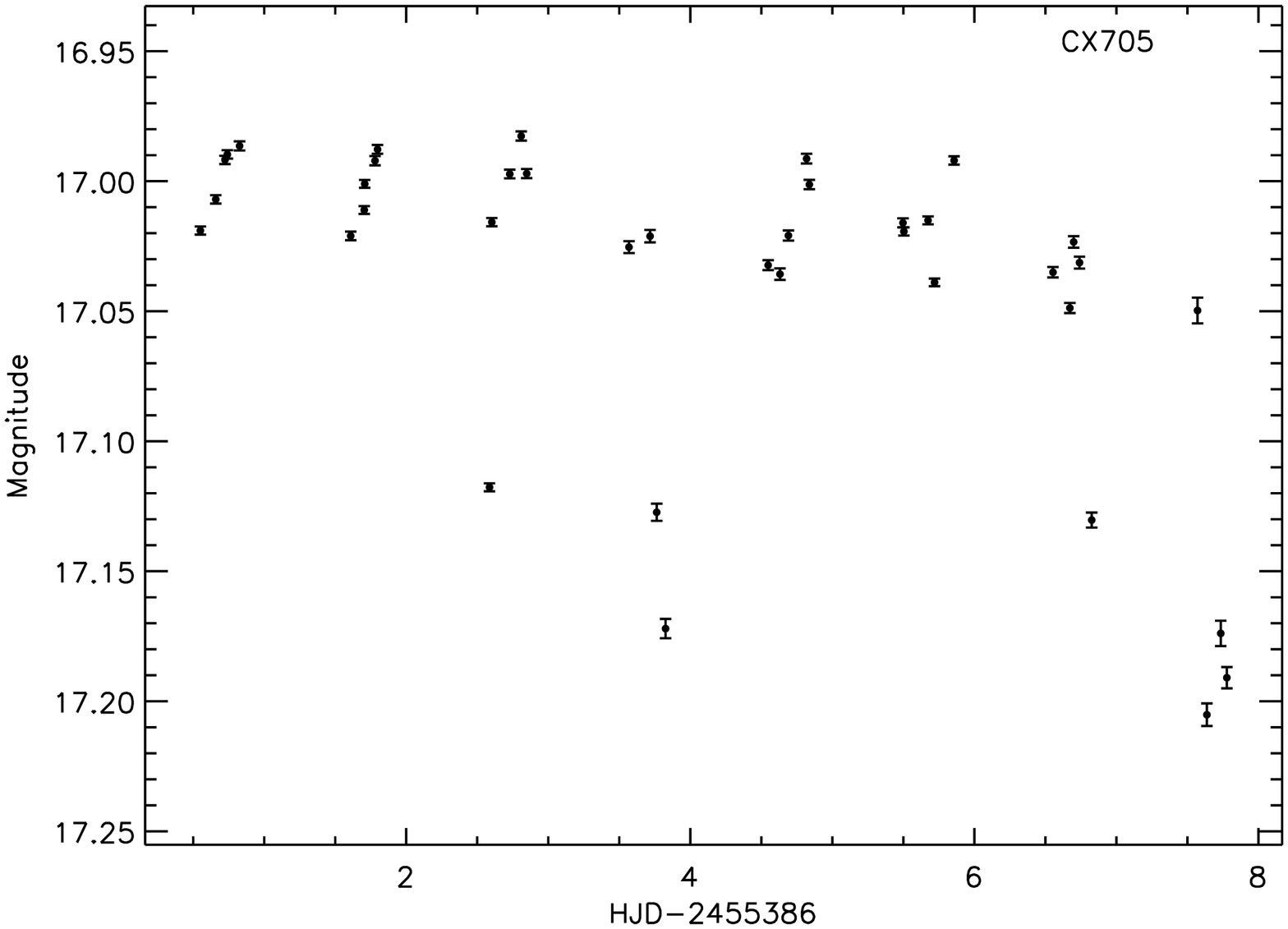}}
}
\caption{CX420, CX426, CX437, CX476, CX645, CX705 Mosaic-II Lightcurves}
\label{lcchunk4}
\end{minipage}
\end{figure*}

CX420 shows a period of 1.45 days. The Mosaic-II lightcurve is shown
in Figure \ref{lcchunk4}, while the folded lightcurve is displayed
in Figure \ref{foldchunk1}. It has an absorbed
$\frac{F_{X}}{F_{opt}} = 0.4$ which is consistent with qLMXBs,
CVs, and active M dwarfs. With the Bulge distance reddening in this line of 
sight, $\frac{F_{X}}{F_{opt}} = 0.002$ is consistent
with coronally active stars as well, but those are too faint in the X-ray to be
detected at Bulge distance. This object is not very red, appearing in 
VVV data with $K_{S}=15.75$ and $J-K_{S}=0.63$, which rules out an active 
M dwarf. There is also some indication of
flickering, which would also rule out an RS CVn or M dwarf. Long
period qLMXBs or CVs indicate an evolved companion, which should
have lower X-ray to optical flux ratios than those with MS
donors due to the brighter donor. CX420 is consistent with such an object.

\subsection{CX426 - DN}

CX426 appears in OGLE-IV data 
\citep{Udalski12}, where it is observed to undergo multiple DN outbursts.
In our data, it shows a steady decline of $0.9$ magnitudes over the course of
observations, shown in Figure \ref{lcchunk4}, consistent with a
DN outburst. The X-ray to 
optical flux ratio is $0.3$ using the faintest observations from
Mosaic-II data. This is consistent with a CV 
undergoing DNe outburst. There is some flickering around a best fit line as
well with an amplitude $0.1-0.2$ magnitudes which is much higher than
expected from simple photon noise or instrumental errors as well,
which argues in favor of a CV interpretation as ordinary long period variables
should not flicker. In VVV data \citep{Greiss13}, this object is not firmly
detected, though it can be seen by eye in some of the images. It is therefore
quite blue, further supporting the interpretation of a CV undergoing DN 
outbursts. We independently strengthen the conclusion of \citet{Udalski12}, 
that this is a CV undergoing DN outbursts.

\subsection{CX437 - CV or qLMXB}

CX437 shows 2 eclipses of at least a magnitude, shown in Figure
\ref{lcchunk4}. The counterpart is
very faint, $r'\approx22.1$, and the eclipses drop below the limiting
magnitude of our observations. There are 2 eclipses seen in 17
observations, suggesting an eclipse duration of $12\pm 8\%$ of
the orbital phase, which is typical of CVs and qLMXBs. The eclipses
last no more than 5 hours, which places an upper
limit on the orbital period of $2$ days. Absorbed
$\frac{F_{X}}{F_{opt}} = 10$ which is consistent with either qLMXBs or
CVs with moderate extinction. Spectroscopic follow 
up is needed to differentiate between the CV and qLMXB interpretations, which 
will likely be possible because the inclination angle is strongly constrained 
by the fact that the system is eclipsing.

\subsection{CX476 - DN}

CX476 starts out near $r'=21$ before dropping below the limiting
magnitude after a few days, shown in Figure \ref{lcchunk4}. This
appears to be a CV undergoing DN
outbursts. Absorbed $\frac{F_{X}}{F_{opt}} = 20$ which is a little
high for a CV, but with some extinction and uncertainties in both
X-ray and optical fluxes, it is consistent. This object is not present in VVV
\citep{Greiss13}, which is unsurprising given its faintness in quiescence.

\subsection{CX645 - CV or qLMXB}

CX645 shows large scale aperiodic variability, changing 
over a magnitude in brightness on a timescale of days as shown in
Figure \ref{lcchunk4}. One data point
$1.7$ magnitudes below the nearest observation $5.5$ hours later
appears to be an eclipse. There is also a large flare of 1 magnitude in 
brightness that lasts no more than 3.7 hours seen on the 8th night of 
observations. Absorbed $\frac{F_{X}}{F_{opt}} =\frac{1}{4}$, while with
Bulge distance reddening, unabsorbed $\frac{F_{X}}{F_{opt}}
= 0.003$. This range is consistent with a nearby CV or
qLMXB. The large amount of aperiodic variability suggests a larger
contribution of continuum light from the 
disk rather than the donor star. 
This object does not appear in the VVV catalog \citep{Greiss13}, meaning
it is quite blue ($r'-J \lae 2$), which is consistent with continuum emission from an 
accretion disk. Spectroscopy is needed to further support any classification. 

\subsection{CX705 - Flare star}

CX705 is in the OGLE-IV catalog \citep{Udalski12}, which lists it as an 
irregular variable. It is also next to a Mira variable in OGLE-IV, which is 
at minimum light and much fainter than the counterpart to CX705 in our 
Mosaic-II data, though they 
are of comparable brightness in OGLE-IV $I$ band photometry at the same time. 
Indeed, after examining the public OGLE-IV photometry of these two sources, it
is evident that the counterpart to CX705 is being slightly contaminated by 
the light from the nearby Mira in OGLE-IV data, resulting in a slight rise in 
the average brightness of the counterpart to CX705 when the Mira variable is 
at maximum brightness. If these few observations are ignored, a strong period
very near 1 day appears in the OGLE-IV data along with some flaring. In 
Mosaic-II data, CX705 has peaks in the periodogram at periods near 1 day and an 
integer fraction of 1 day, showing heavy aliasing between 1 day, $\frac{1}{2}$ 
of a day, and $\frac{1}{3}$ of a day. None of these periods is significant in 
our data, though Monte Carlo simulations show that with any amount of 
flickering it is unlikely we would recover real modulations at such a period. 
This period and morphology from OGLE-IV data seems to shift very slightly 
over the length of the 
OGLE-IV observations, which could be attributable to star spots and 
differential rotation. CX705 is faint in the X-ray for its
optical brightness, with absorbed $\frac{F_{X}}{F_{opt}} =0.05$, dropping
to $\frac{1}{1000}$ at Bulge reddening. This is consistent with coronal
activity from a flare star, with some star spots appearing and disappearing
over the course of OGLE-IV observations. This object appears in the VVV 
survey \citep{Greiss13} with $J-K_{s}=1.53$ and $K_{S}=12.48$. This is 
consistent with a flare star.

\subsection{CX740 - CV, qLMXB, or M dwarf}

\begin{figure*}[p!]
\begin{minipage}{0.9\textwidth}
\centering
\parbox{\textwidth}{
\subfigure{\includegraphics[width=0.4\textwidth,angle=90]{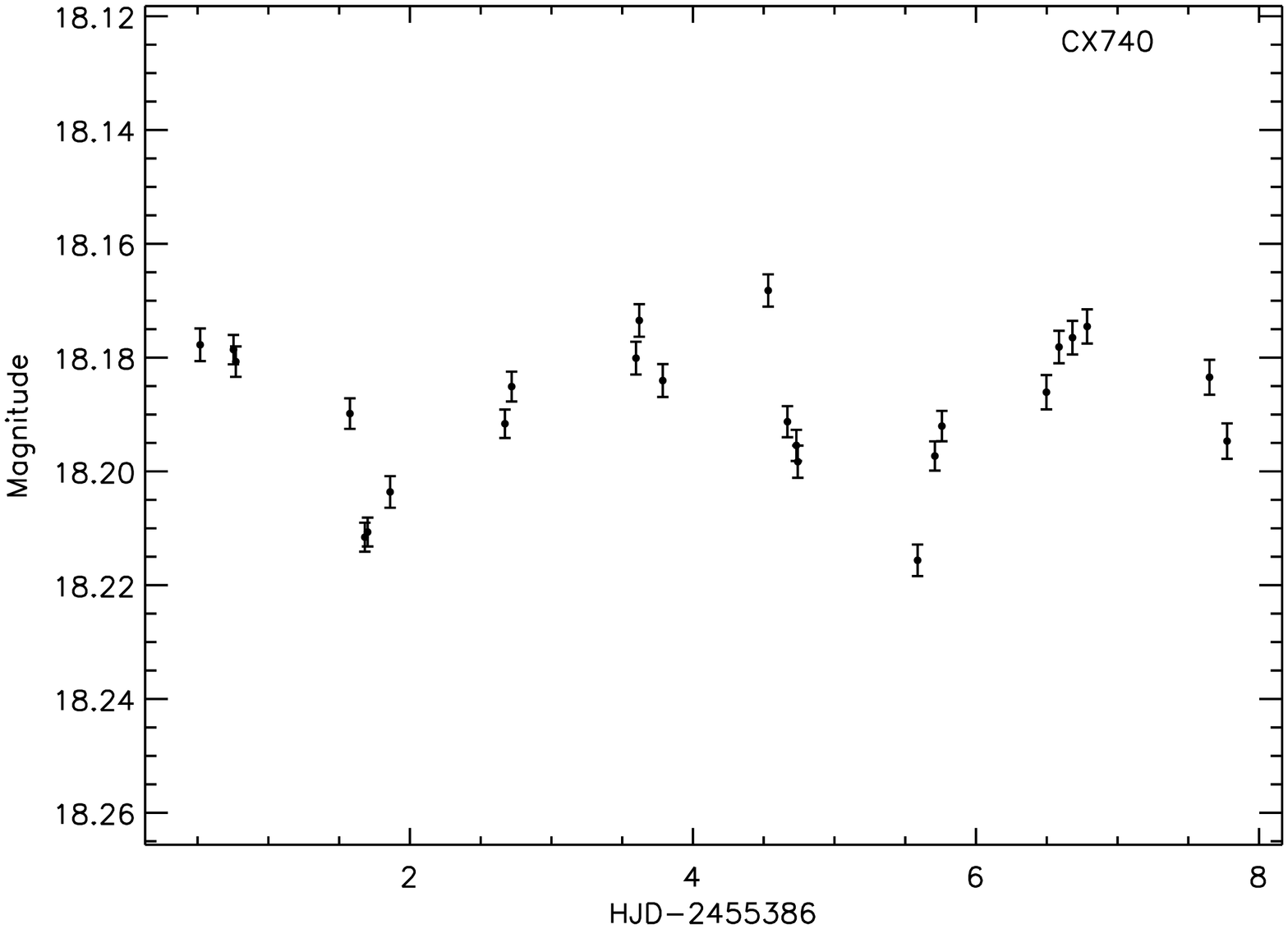}}\quad
\subfigure{\includegraphics[width=0.4\textwidth,angle=90]{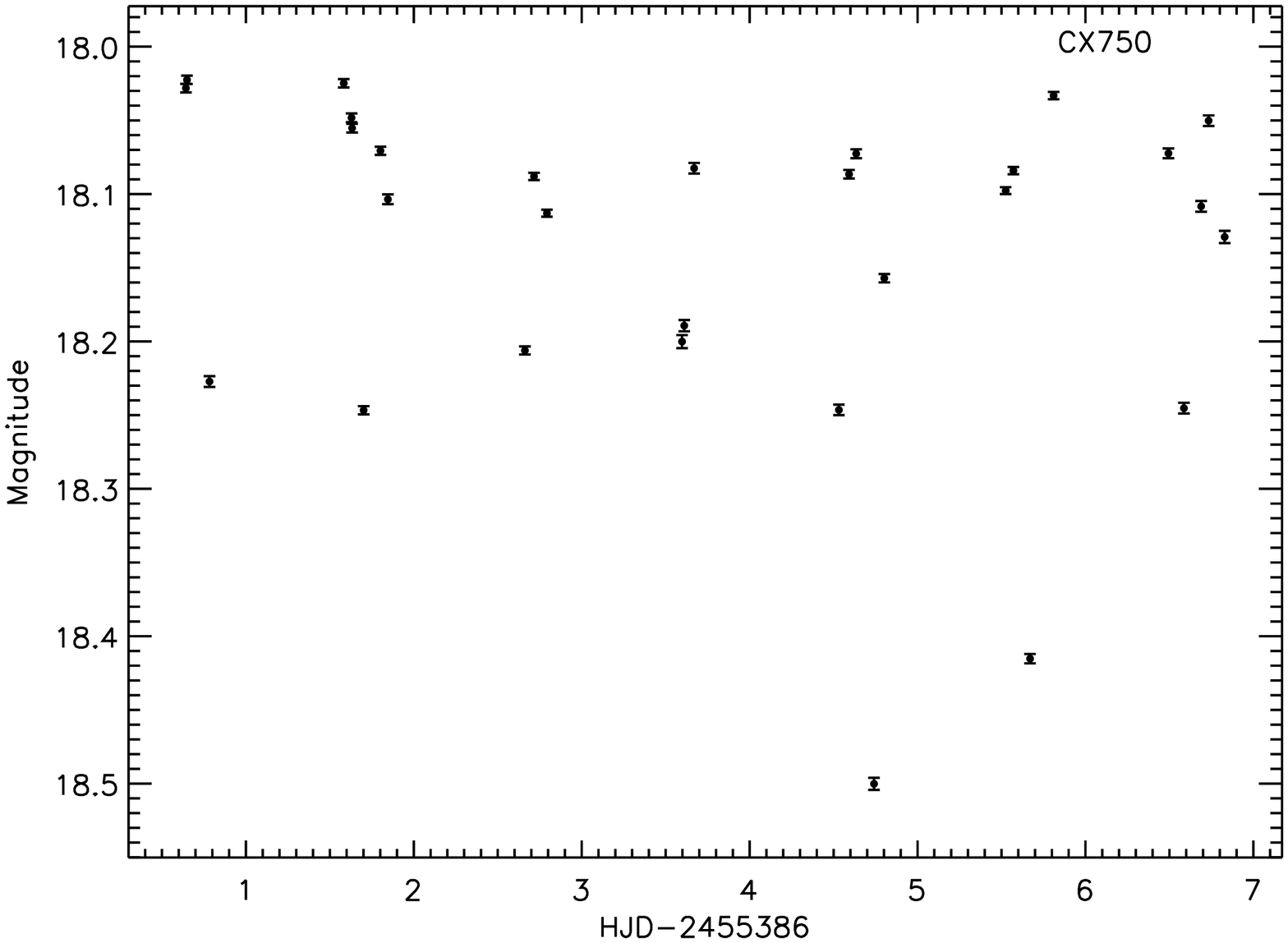}} \\
\subfigure{\includegraphics[width=0.4\textwidth,angle=90]{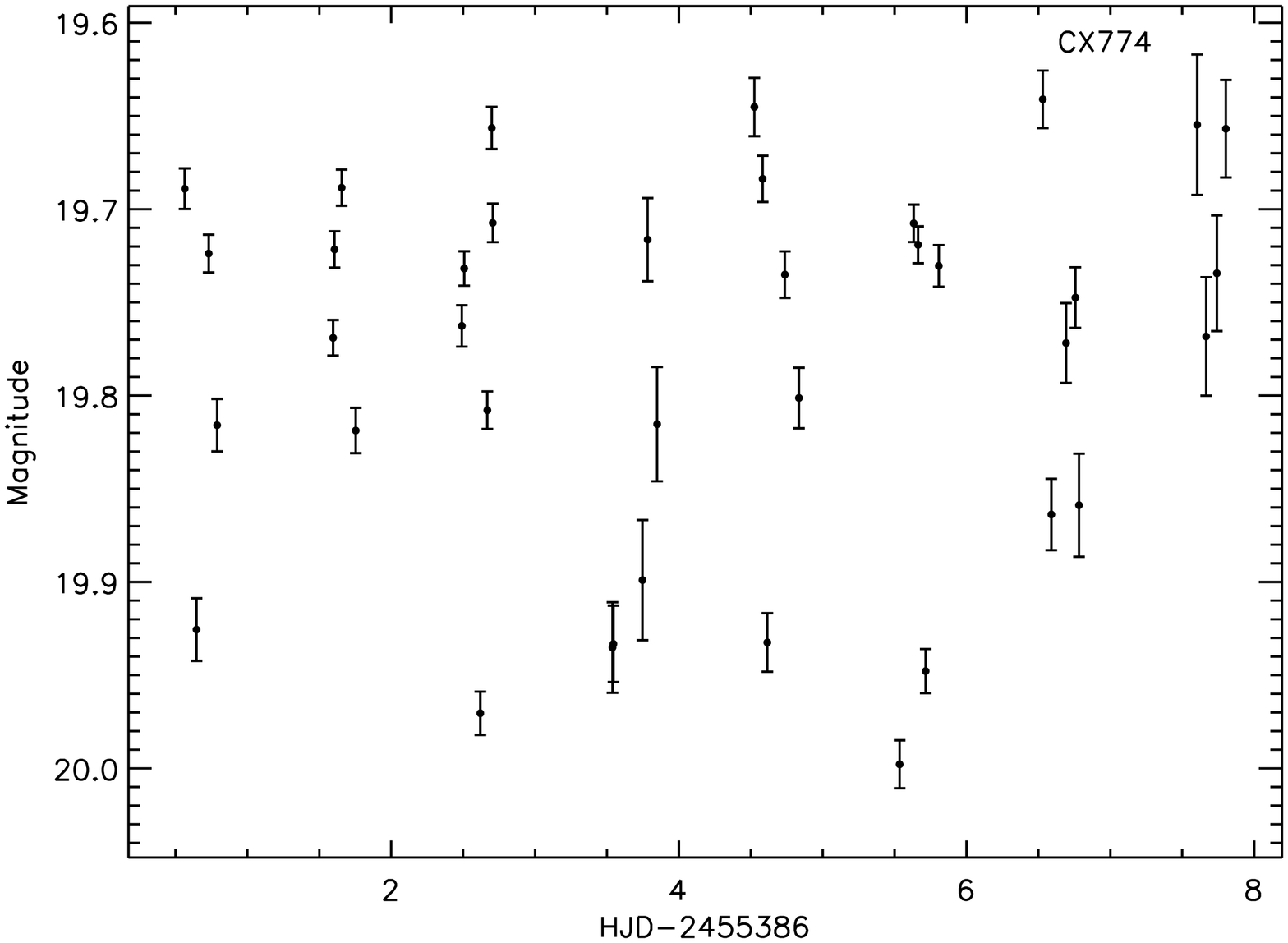}}\quad 
\subfigure{\includegraphics[width=0.4\textwidth,angle=90]{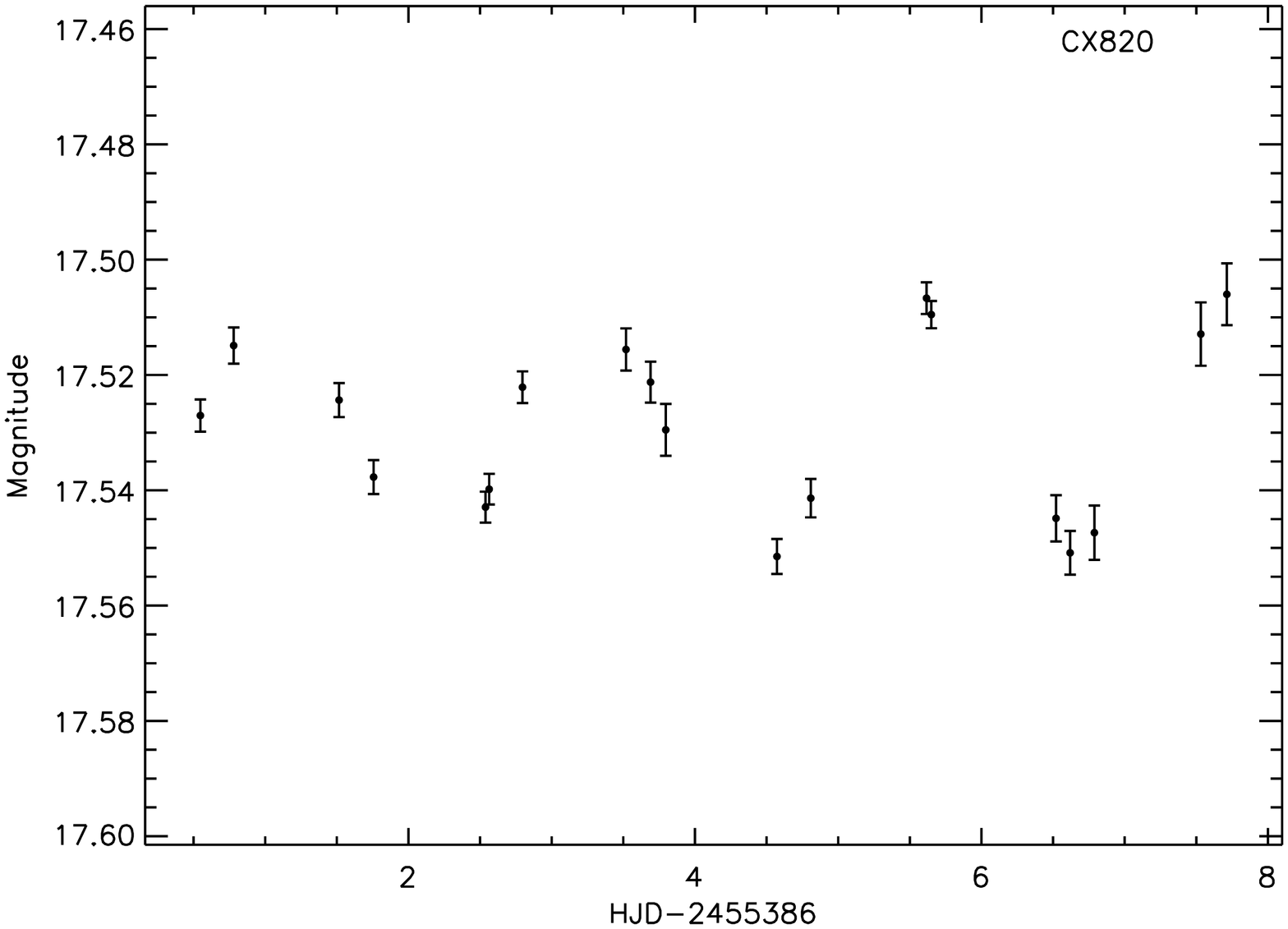}} \\
\subfigure{\includegraphics[width=0.4\textwidth,angle=90]{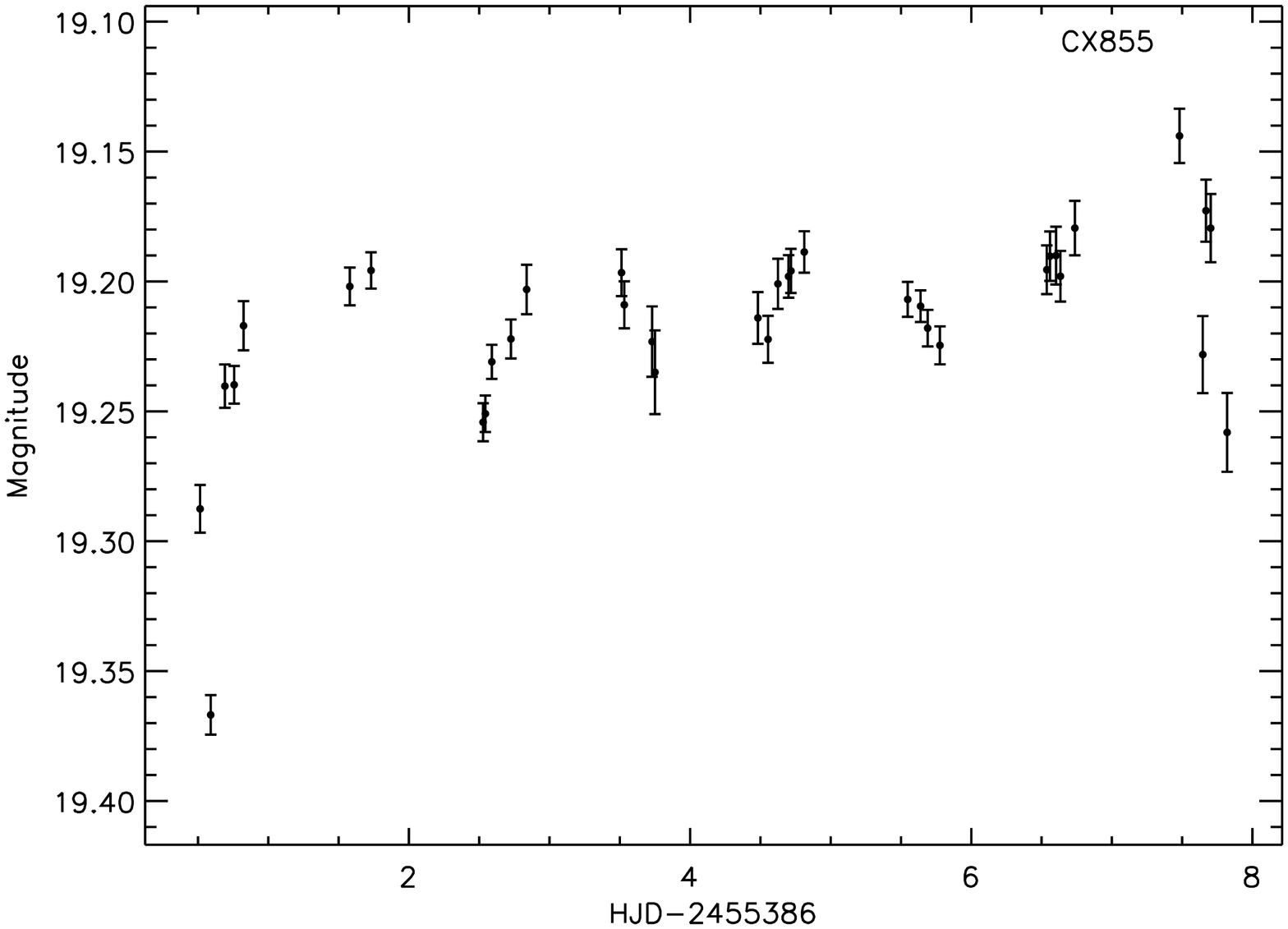}}\quad
\subfigure{\includegraphics[width=0.4\textwidth,angle=90]{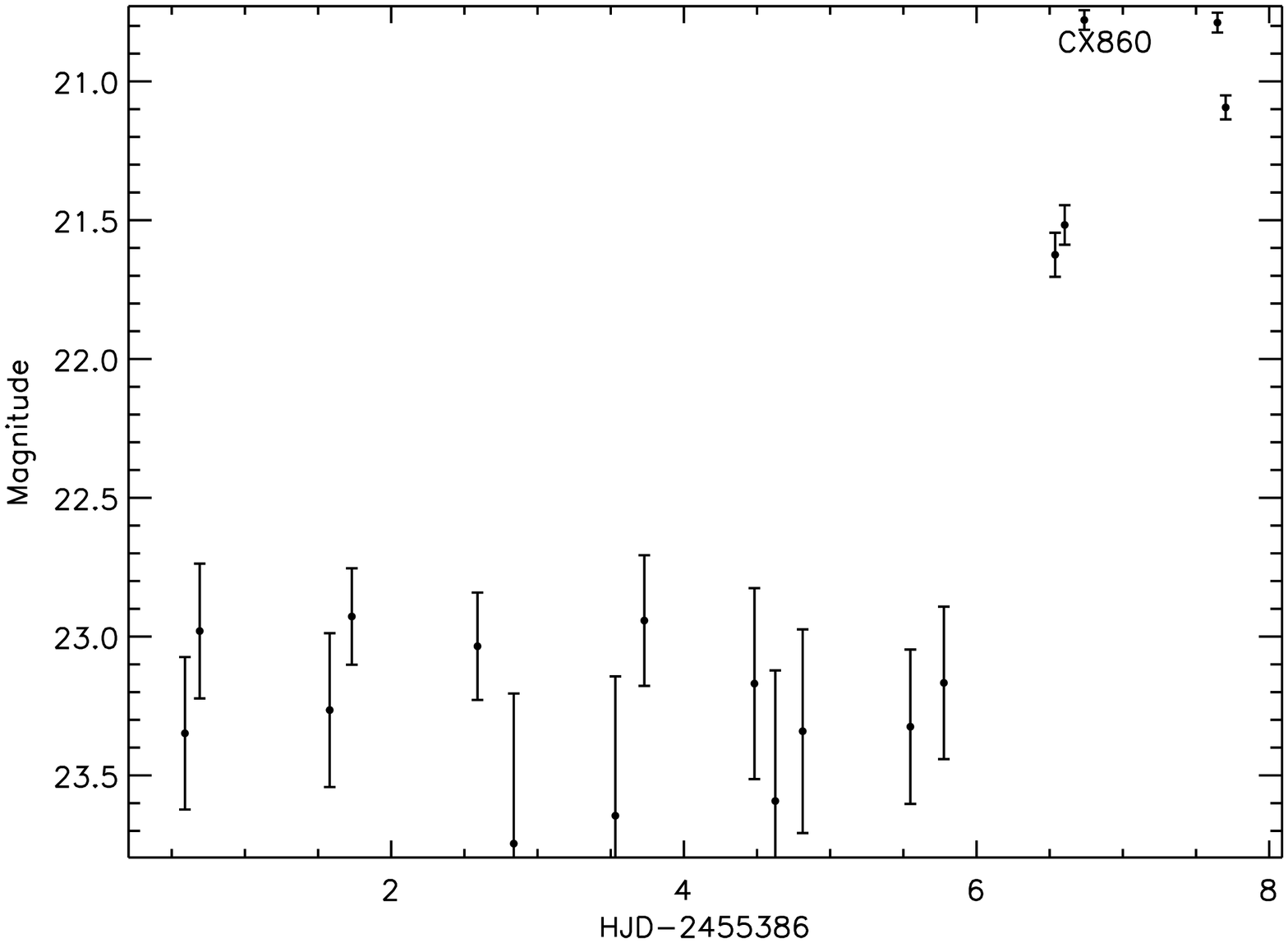}}
}
\caption{CX740, CX750, CX774, CX820, CX855, CX860 Mosaic-II Lightcurves}
\label{lcchunk5}
\end{minipage}
\end{figure*}

CX740 shows modulations of $0.06$ magnitudes, shown in Figure
\ref{lcchunk5}, and with a period of $0.765$
days, and is shown in Figure \ref{foldchunk1}. Absorbed
$\frac{F_{X}}{F_{opt}} =\frac{1}{6}$, though extinction
in this line of sight is quite high and at Bulge distance this drops
to $2\times10^{-5}$. This range is consistent with qLMXBs,
CVs, or active M dwarfs which have been observed to rotate this
quickly, though the short period with the low X-ray to optical flux
ratio for Bulge reddening 
strongly implies that this object is significantly closer than Bulge
distance. The changes could 
also be ellipsoidal with a period of 1.53
days, but this is so close to $\frac{3}{2}$ of a day that we
cannot distinguish between the periods because of gaps in phase
coverage. The lower values of the X-ray to optical flux ratio
are favored by larger distances, but it is unlikely that we would
detect a W UMa system much further than several hundred parsecs
because they are not luminous enough in the X-ray to be detected at
the Bulge in our short X-ray exposures, and W UMas are unlikely to
have a period as long as 1.5 days. CX740 is a candidate qLMXB or CV
based on the possibilty of ellipsoidal variations, but some
active M dwarfs have rotation periods below a
day as well. From VVV data \citep{Greiss13}, $J-K_{S}=1.18$ without correcting 
for reddening, which changes to $J-K_{S}=-0.88$ for a Bulge distance reddening 
of $E(J-K)=2.06$. The color is consistent with a foreground M dwarf, and we 
cannot rule this possibility out either without
spectroscopy, which should quickly differentiate between these cases.

\subsection{CX750 - CV or qLMXB plus a field W UMa?}

There are 2 variable stars within the X-ray error circle for
CX750. Both show what appear to be ellipsoidal variations, one also
shows an eclipse. The Chandra observation that detected this source
was made at a large off-axis angle, and the $95\%$ confidence region has a radius
of $11.9''$. The variable closest to the center of this region shows
ellipsoidal modulations with $P_{orb}=1.12\pm0.03$ days and a mean
magnitude of $r'=19.4$. It is possibly a field W UMa, which would imply
an X-ray brightness, given
typical values for W UMa stars, of $\sim10^{-15}  \, {\rm
  ergs\,cm^{-2}\,s^{-1}}$ which is well below the X-ray detection
limit. If it is the true counterpart to the X-ray source, it has an
absorbed $\frac{F_{X}}{F_{opt}} =\frac{1}{20}$, which could be
consistent with CVs or qLMXBs, while the period above a day would
suggest a somewhat evolved donor star in either scenario if this were
to be the true counterpart, though a MS donor could be possible for a BH qLMXB following the
period-mass relation for MS donors in \citet{Frank02}. This object appears faintly in the 
VVV survey 
\citep{Greiss13} at $K_{S}=15.94$ and $J-K_{S}=0.59$ or $J-K_{S}=-1.17$ for Bulge
distance reddening, for which $\frac{F_{X}}{F_{opt}} =
3\times10^{-5}$. This strongly implies that this object is closer than Bulge distance. 

The other object has a period of
$0.468\pm0.005$ days, or $11.24$
hours, with a minimum $0.25$ magnitudes deeper at phase $\phi=0$ than at
phase $\phi=0.5$ (with the phase set arbitrarily to zero at the deeper minimum),
which is consistent with qLMXBs and CVs, as is the observed absorbed X-ray to
optical flux ratio of $\frac{1}{8}$. The Mosaic-II lightcurve is
shown in Figure \ref{lcchunk5}, and the folded lightcurve is shown
in Figure \ref{foldchunk1}. The period is consistent with W
UMas, but the asymmetry in the minima depths means that the
temperatures of the two bodies are substantially different which is not the case for
a W UMa in which the two stars share a common envelope. Indeed, the
dip at $\phi=0.5$ is consistent with a minima from ellipsoidal
variations without an eclipse, while the eclipse at $\phi=0$ is incompatible 
with ellipsoidal modulations, which means this is unlikely to be a W
UMa. The maxima of the ellipsoidal modulations are also asymmetric, though the
brighter maximum trails the deep eclipse in phase, while a hotspot from an
accretion stream impact point on the disk should lead donor star in phase; 
this behavior is observed in the LMXB 4U 1822-37 \citep{Mason80} and the 
White Dwarf CAL 87 \citep{Schandl97} which was also modeled as a spray of 
colder matter obscuring the inner disk region before the eclipse, causing 
the asymmetry. These objects, however, are intrinsically much more
luminous than CX750 would be at Bulge distance. 
This object appears in VVV data with $J-K_{S}=0.83$ and
$K=13.89$. Because the period is only 11 hours, we can place a limit
on the spectral typle of the donor. Using the Mass-Period relation in \citet{Frank02},
this implies a donor mass of $\approx1.2\,M_{\odot}$ for a main
sequence donor, which would mean
this object has an intrinsic color of $J-K_{S}=0.24$
\citep{Ducati01}. Assuming that reddening is linear with distance,
$E(J-K)\approx0.6$ implies $d=2.7\,$kpc and $A_{r'}=3.2$. This allows
a tighter lower limit $\frac{F_{X}}{F_{opt}}>0.012$, as more evolved
donors which fit in this period are redder and optically fainter, requiring less extinction to make the observed colors match. At a
distance of $2.7\,$kpc, $L_{X}=4.3\times10^{31}\,{\rm ergs\,s^{-1}}$,
which, given the orbital period, favors a quiescent BH
accretor over a NS \citep{Rea11,Garcia01}. This object is a candidate eclipsing qLMXB, and spectroscopic 
follow-up is necessary to distinguish between CV and qLMXB interpretations.

\subsection{CX774 - CV or qLMXB plus an Interloper}

There are 2 variable stars within the X-ray error circle for
CX774 as well. One undergoes a smooth decline of $0.04$ magnitudes
over the 8 days of our observations, which is consistent with OGLE-IV 
observations showing a period of $43.478$ days \citep{Udalski12}. 
This object is very bright in the infrared in the 2MASS survey \citep{Skrutskie06}, 
with $J-K=2.38$ and $K=7.60$, suggesting this object is heavily absorbed. For 
Bulge distance and reddening for 2MASS data, $M_{K}=-7.7$ and $J-K=1.28$. 
This suggests that this object is a red giant most of the way to the Bulge, 
behind most of the dust in this line of sight. Absorbed 
$\frac{F_{X}}{F_{opt}} = 10^{-2}$, while assuming reddening equivalent to Bulge 
distance yields $\frac{F_{X}}{F_{opt}} = 2\times10^{-4}$, and 
$L_{X}=4\times10^{32} \, (\frac{d}{8\,{\rm kpc}})^{2}\,{\rm ergs\, cm^{-2}\, s^{-1}}$. If this is the true 
counterpart, it could be a giant star in contact with a compact
object, such that the variations observed are ellipsoidal modulations.

The other object shows ellipsoidal
modulations with a period of $0.362$ days, or $8.72$ hours. The
original lightcurve is shown in Figure \ref{lcchunk5} , and the folded
lightcurve is shown in Figure \ref{foldchunk1}. The flickering in the
lightcurve argues against this object being a W UMa. The
variable showing ellipsoidal variations has a magnitude of $r'=19.8$,
which means it is likely below the X-ray detection limit if it were a W UMa
star. If it is the X-ray source, absorbed $\frac{F_{X}}{F_{opt}} = 0.6$, which 
is consistent with CVs and 
qLMXBs. The orbital period detected is also consistent with either 
interpretation. This object is swamped by the light from the IR-bright star
discussed above in VVV data, and it is not in the 2MASS catalog. Assuming 
reddening equal to that at Bulge distance, $\frac{F_{X}}{F_{opt}} = 10^{-2}$, 
which is lower than typical values of NS qLMXBs, while a 
CV is unlikely to be as bright as
$L_{X}=4\times10^{32} \,{\rm ergs\, cm^{-2}\, s^{-1}}$. A BH qLMXB with this orbital period is unlikely to remain at this luminosity for very long as well \citep{Garcia01,Kong02,Lasota08,Rea11,Jonker12}, so if this is the true
counterpart, it is more likely a CV or qLMXB in the foreground rather than all the way to the Bulge.

The X-ray source could
realistically be either variable, but the
one with the short period is certainly a close binary, and the
flickering observed on top of the periodic changes is an indicator of
accretion. If the short period variable is
the X-ray source, then it could be a qLMXB or CV. We suspect that the
second object discussed is the true counterpart, but spectroscopic follow
up is necessary to 
fully determine which variable star is the true counterpart.

\subsection{CX820 - RS CVn}

CX820 shows smooth sinusoidal variations on a period of
$2.242$ days and an amplitude of only $0.02$ magnitudes, shown in
Figure \ref{lcchunk5}. The folded
lightcurve is displayed in Figure \ref{foldchunk1}. Absorbed
$\frac{F_{X}}{F_{opt}} = 0.1$ which is consistent with RS CVns, active
M stars, qLMXBs, and CVs. The small amplitude of variation, lack of
flickering, and multiday period together is suggestive of an RS CVn. The 
infrared colors of this possible counterpart are also consistent with an 
RS CVn \citep{Greiss13}. This object is primarily of note in this work 
as a demonstration of how well the photometry methods can work even for 
very small amplitude changes. This object is suggested to be a chromospheric active star or binary after spectroscopic observations in \citet{Torres14} pending further analysis, which agrees with our photometric classification.

\subsection{CX855 - CV or qLMXB}
\label{sec:855}

\begin{figure*}[p!]
\begin{minipage}{0.9\textwidth}
\centering
\parbox{\textwidth}{
\subfigure{\includegraphics[width=0.4\textwidth,angle=90]{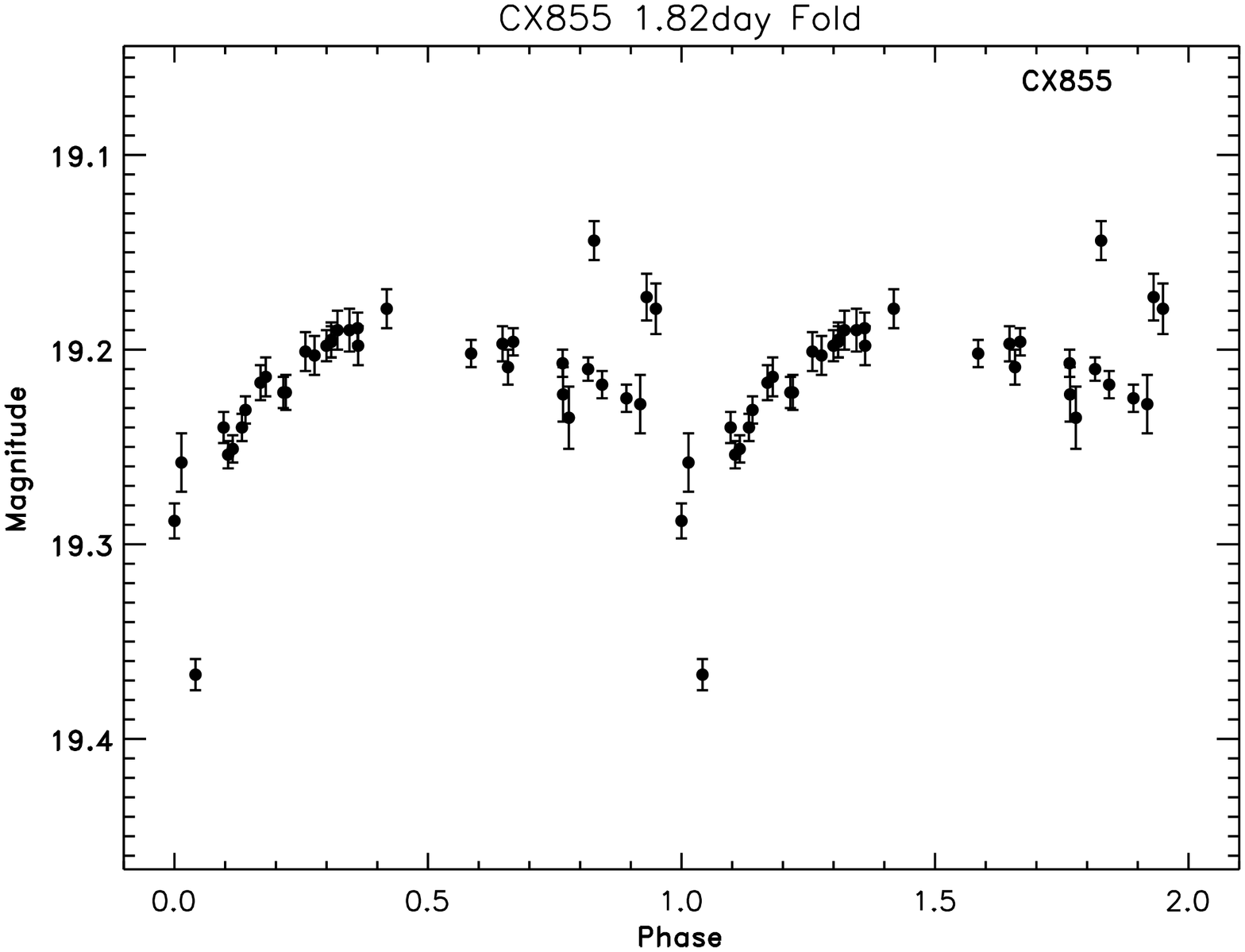}}\quad
\subfigure{\includegraphics[width=0.4\textwidth,angle=90]{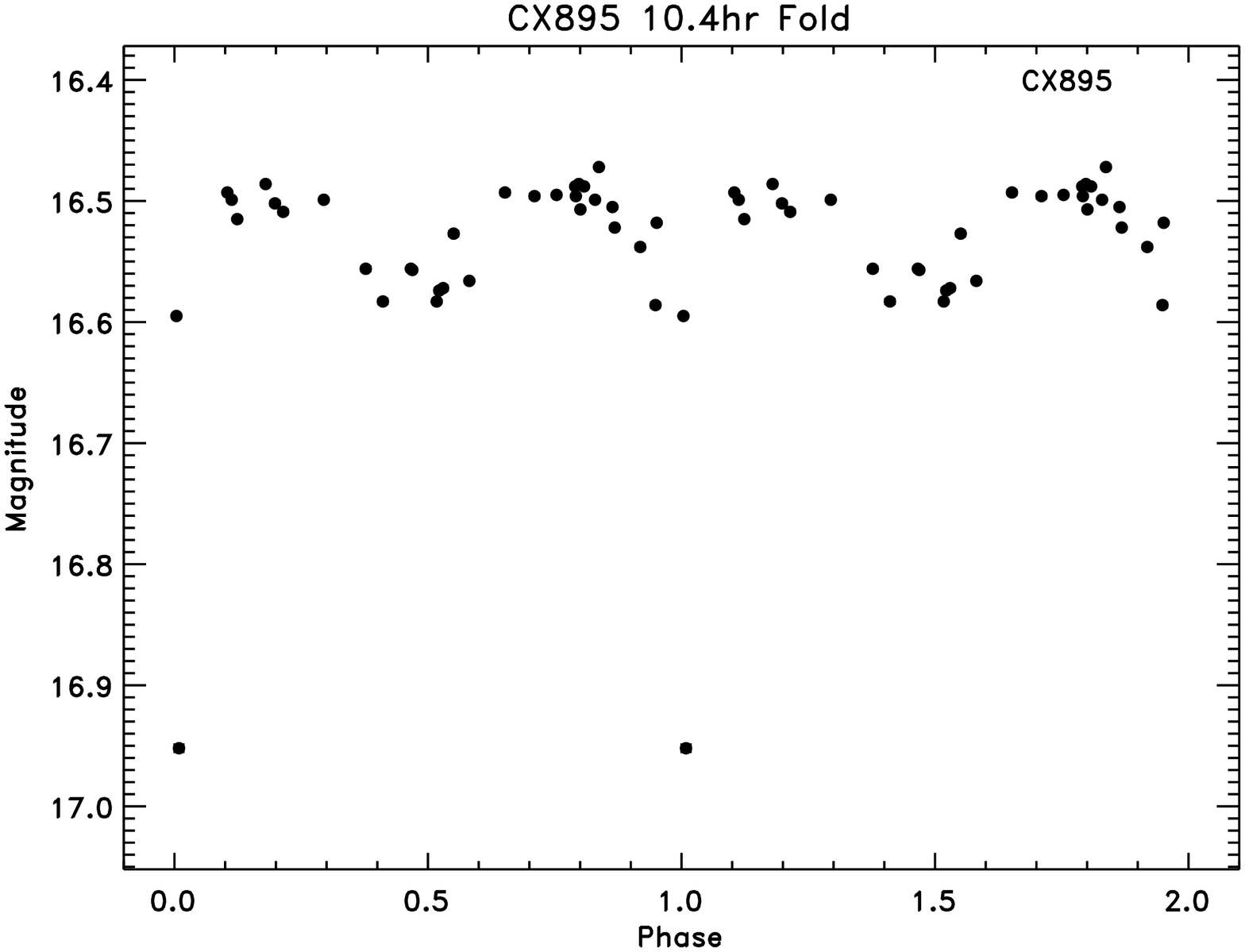}} \\
\subfigure{\includegraphics[width=0.4\textwidth,angle=90]{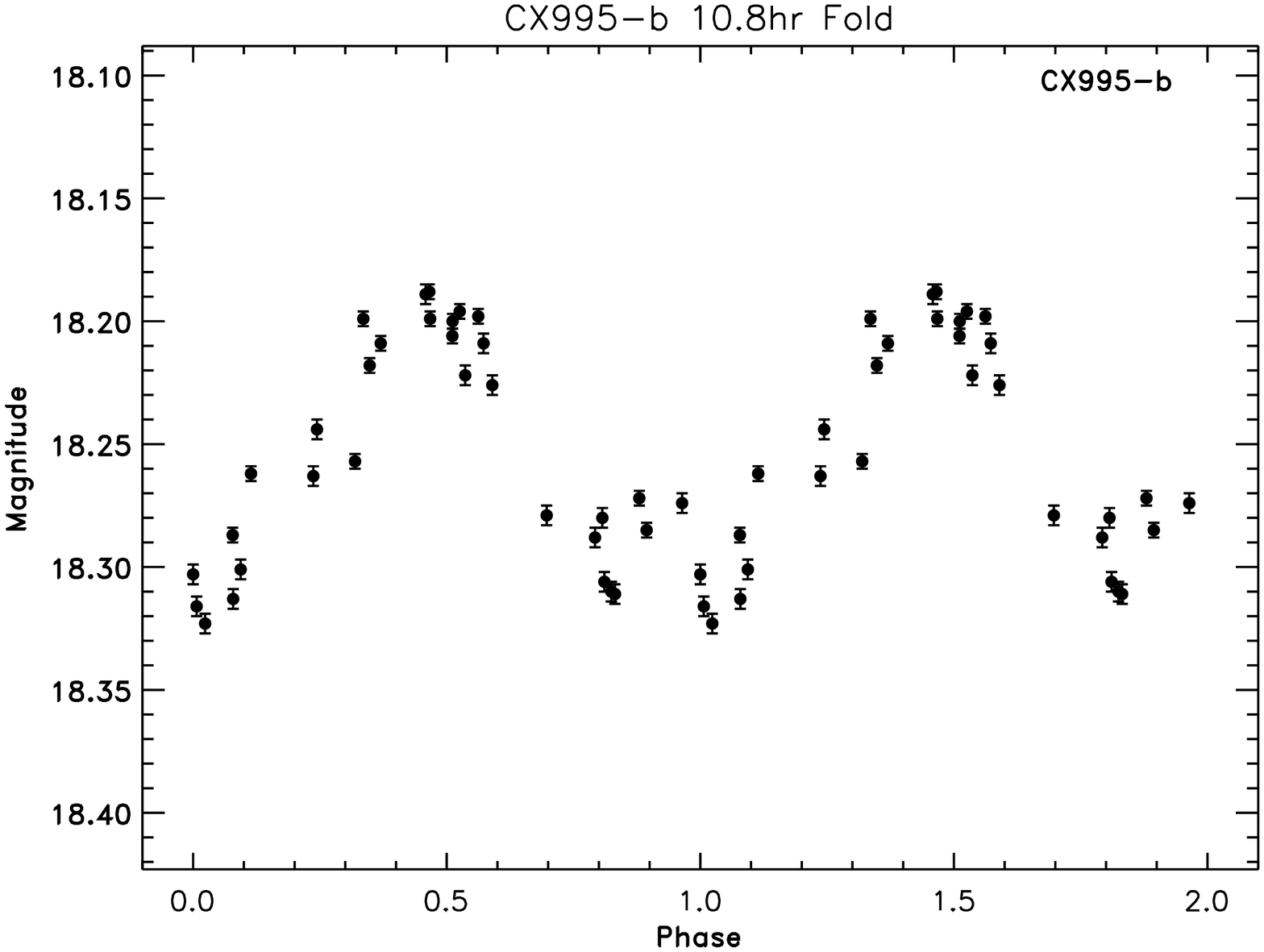}}\quad 
\subfigure{\includegraphics[width=0.4\textwidth,angle=90]{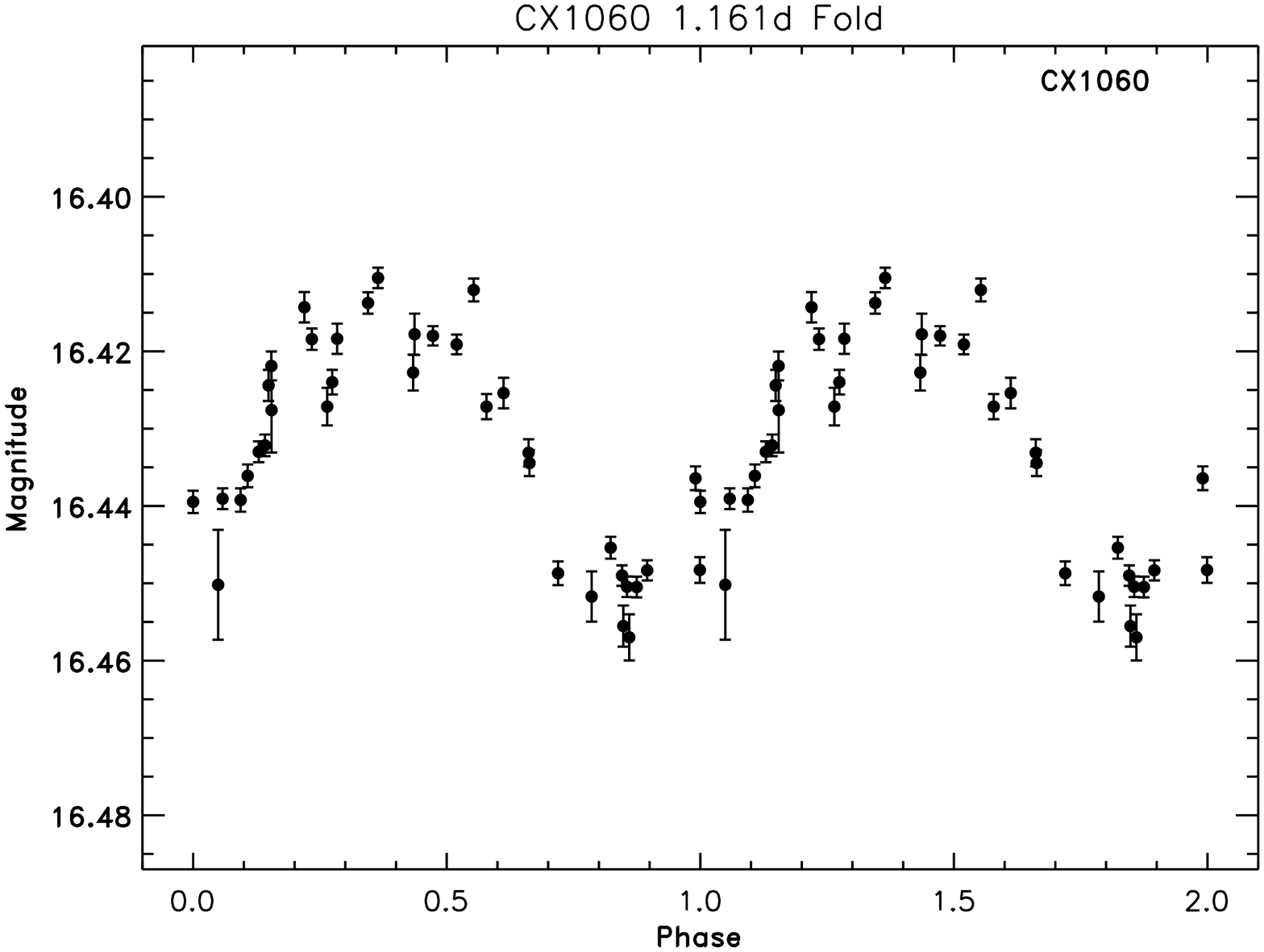}} \\
\subfigure{\includegraphics[width=0.4\textwidth,angle=90]{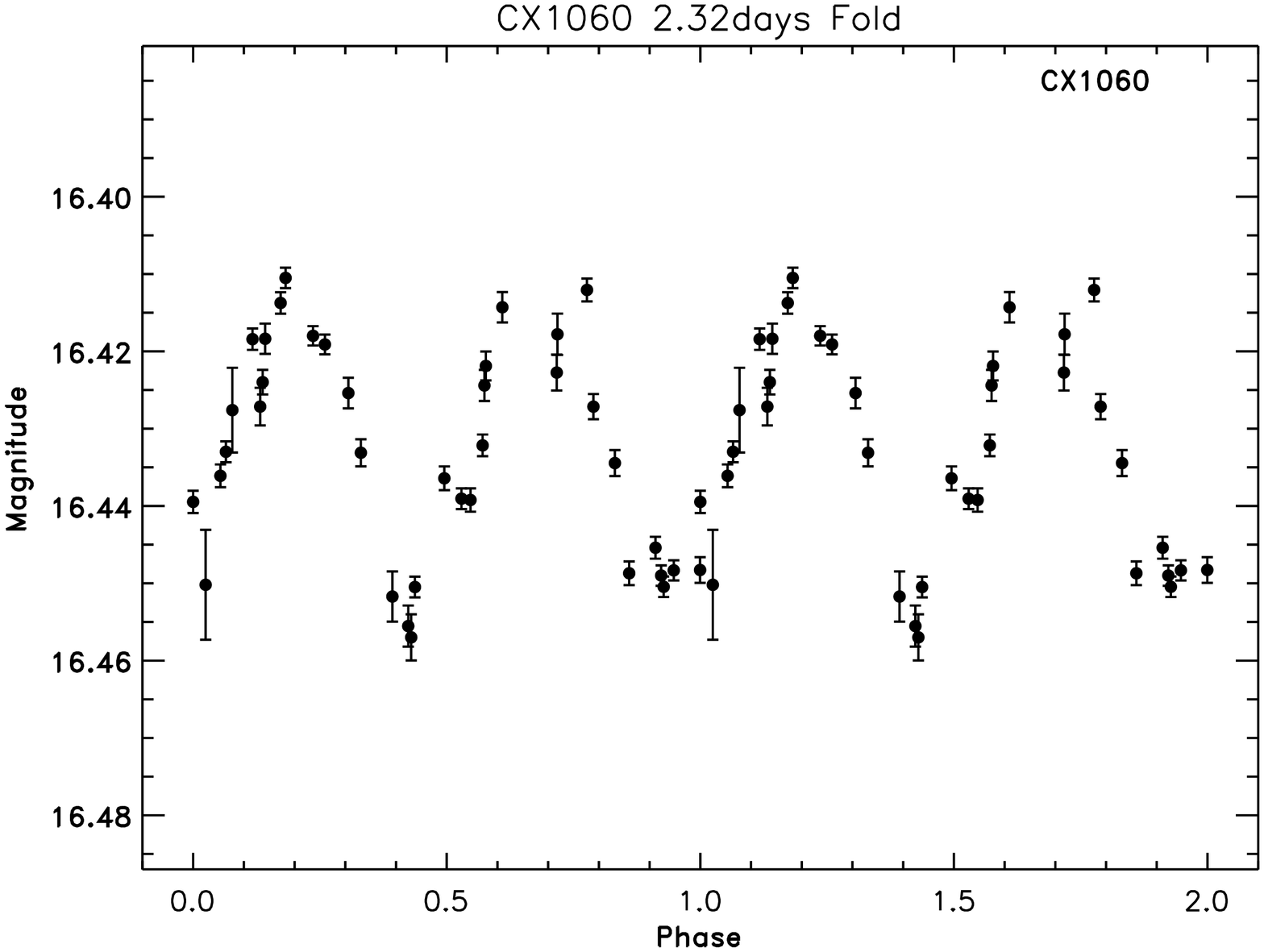}}\quad
\subfigure{\includegraphics[width=0.4\textwidth,angle=90]{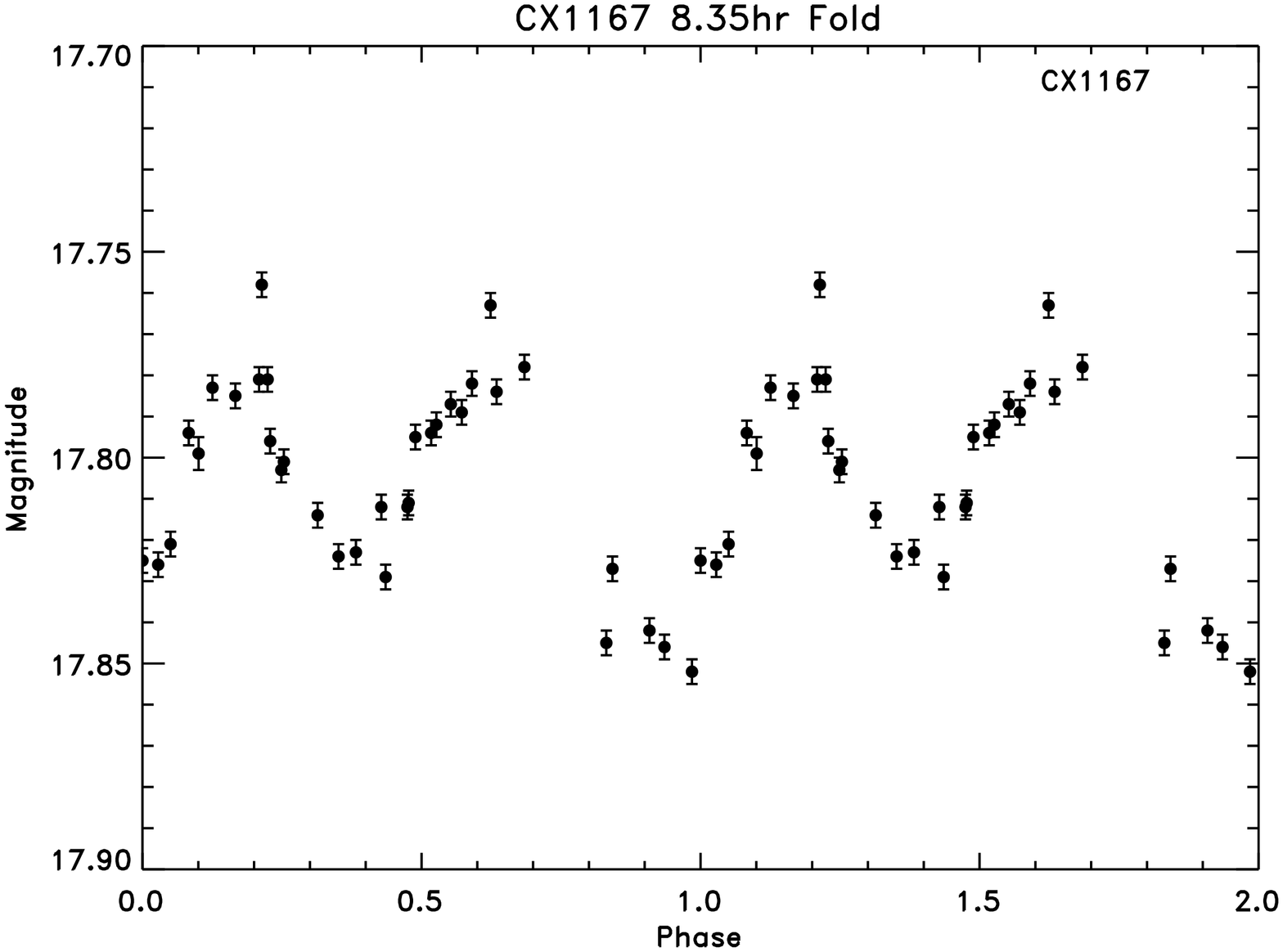}}
}
\caption{{\it Top Left:} Lightcurve of CX855 folded on a 1.8 day period.
{\it Top Right:} Lightcurve of CX895 folded on a 10.4 hour period.
{\it Middle Left:} Lightcurve of the second possible counterpart to CX995 folded on a 0.45 day period.
{\it Middle Right:} Lightcurve of one possible counterpart to CX1060
 folded on a 1.161 day period. 
{\it Bottom Left:} The same possible counterpart of CX1060 plotted with 
 twice the period. 
{\it Bottom Right:} Lightcurve of the likely counterpart to CX1167
  folded on a 8.35 hour period. }
\label{foldchunk2}
\end{minipage}
\end{figure*}

CX855, shown in Figure \ref{lcchunk5}, has an orbital period of $1.82\pm0.05$
days and a shallow, brief eclipse of at least $0.12$ magnitudes, though the 
depth is not well constrained. The
rise and fall are asymmetric, so the lightcurve is not very well fit by
a sine wave. The folded lightcurve is shown in Figure \ref{foldchunk2}. 
All of the observations on night 8 are slightly higher
than other observations at the same phase, suggesting some intrinsic
brightening of the source. Such level changes are commonly seen in accreting 
systems. Absorbed
$\frac{F_{X}}{F_{opt}} = 0.4$, which is consistent with qLMXBs and
CVs. From VVV data, $J-K_{S}=0.814$ and $K_{S}=15.36$. The multiday
orbital period implies an evolved donor for a Roche Lobe filling
companion to the compact obect. An absolute magnitude $M_{K}=2$,
assuming that reddening is linear with distance, implies a distance of
$\sim4$\,kpc, and $\frac{F_{X}}{F_{opt}} =0.04$, which is also
consistent with CVs and qLMXBs, with 
$L_{X}=2.6\times10^{32}\,(\frac{d}{8\,{\rm kpc}})^{2}\,{\rm  ergs\,s^{-1}}$. 
Phase resolved spectroscopy is needed to determine the mass of
the primary, which could be realistically either a WD, NS, or BH accretor, a task which is 
greatly simplified by the presence of eclipses which constrain
$\sin i$.

\subsection{CX860 - DN}

CX860 shows an outburst of at least $2.2$ magnitudes shown in Figure
\ref{lcchunk5}, from $r'=23$ to at
least $r'=20.8$. The counterpart is rising in brightness on night 7 of
observations and is falling again on night 8. The outburst peaks
during the day between these observations. An outburst lasting only
2-3 days is fast for a DN, though it is not without precedent. The
dramatic increase in luminosity is certainly consistent with a DN
outburst. Absorbed $\frac{F_{X}}{F_{opt}} = 10$, which is consistent
with CVs undergoing DNe outburst, especially once reddening is taken
into consideration. This object does not appear in VVV data \citep{Greiss13}, 
which is expected for an object as intrinsically blue as a CV that is this
faint in the optical. This appears to be a CV undergoing a DN outburst.

\subsection{CX895 - Eclipsing CV, qLMXB, or binary M dwarfs}

\begin{figure*}[p!]
\begin{minipage}{0.9\textwidth}
\centering
\parbox{\textwidth}{
\subfigure{\includegraphics[width=0.4\textwidth,angle=90]{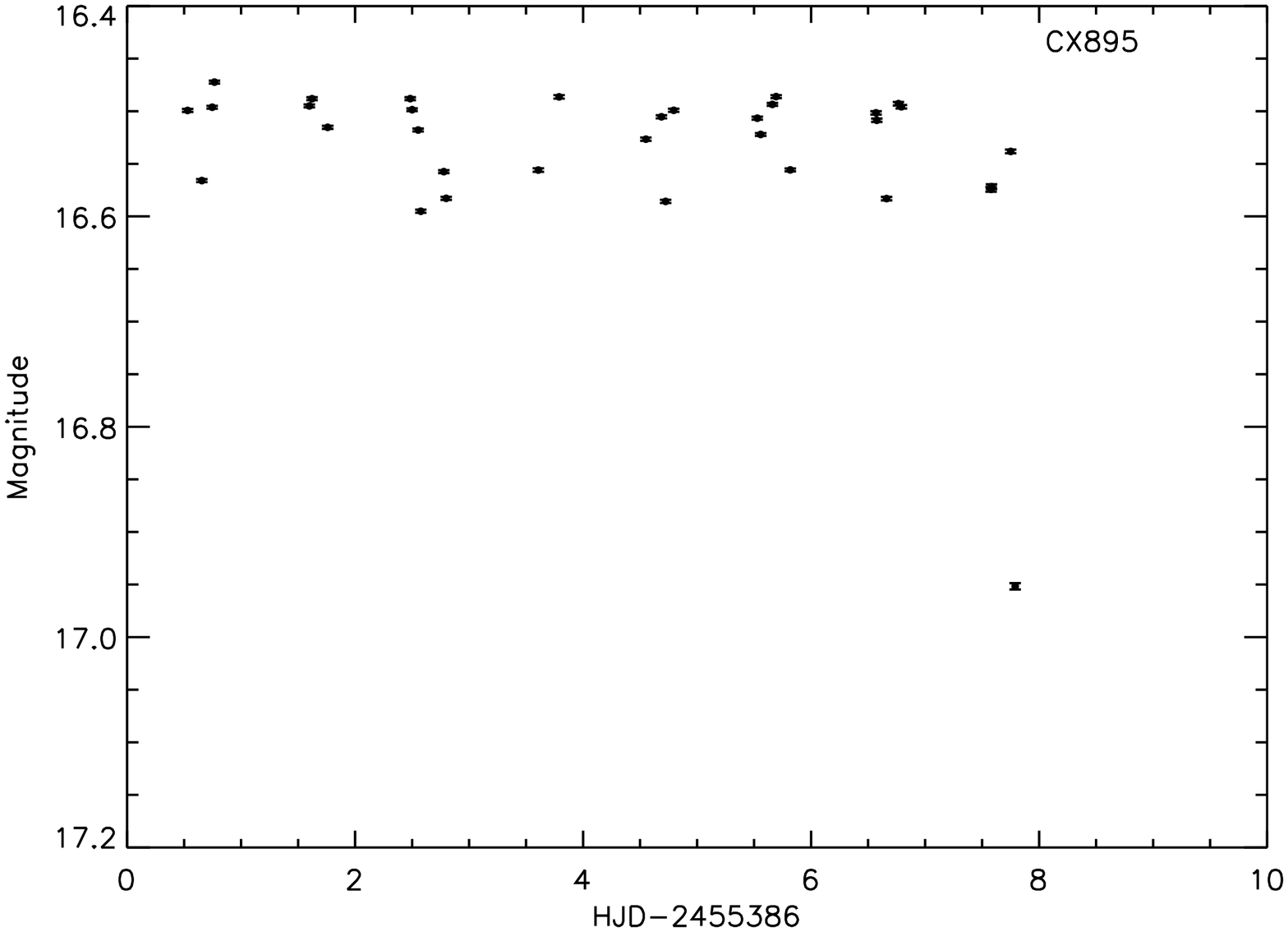}}\quad
\subfigure{\includegraphics[width=0.4\textwidth,angle=90]{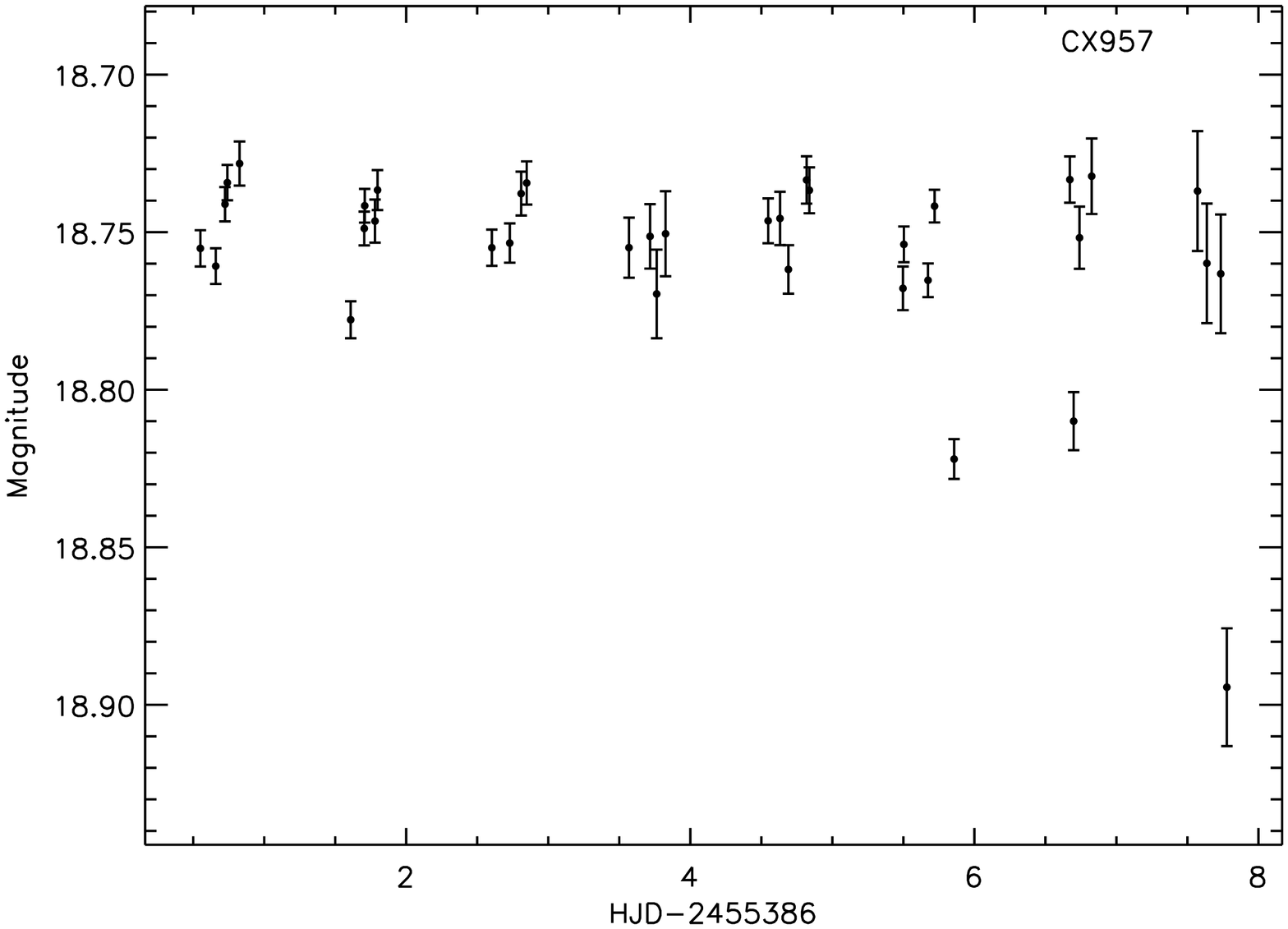}} \\
\subfigure{\includegraphics[width=0.4\textwidth,angle=90]{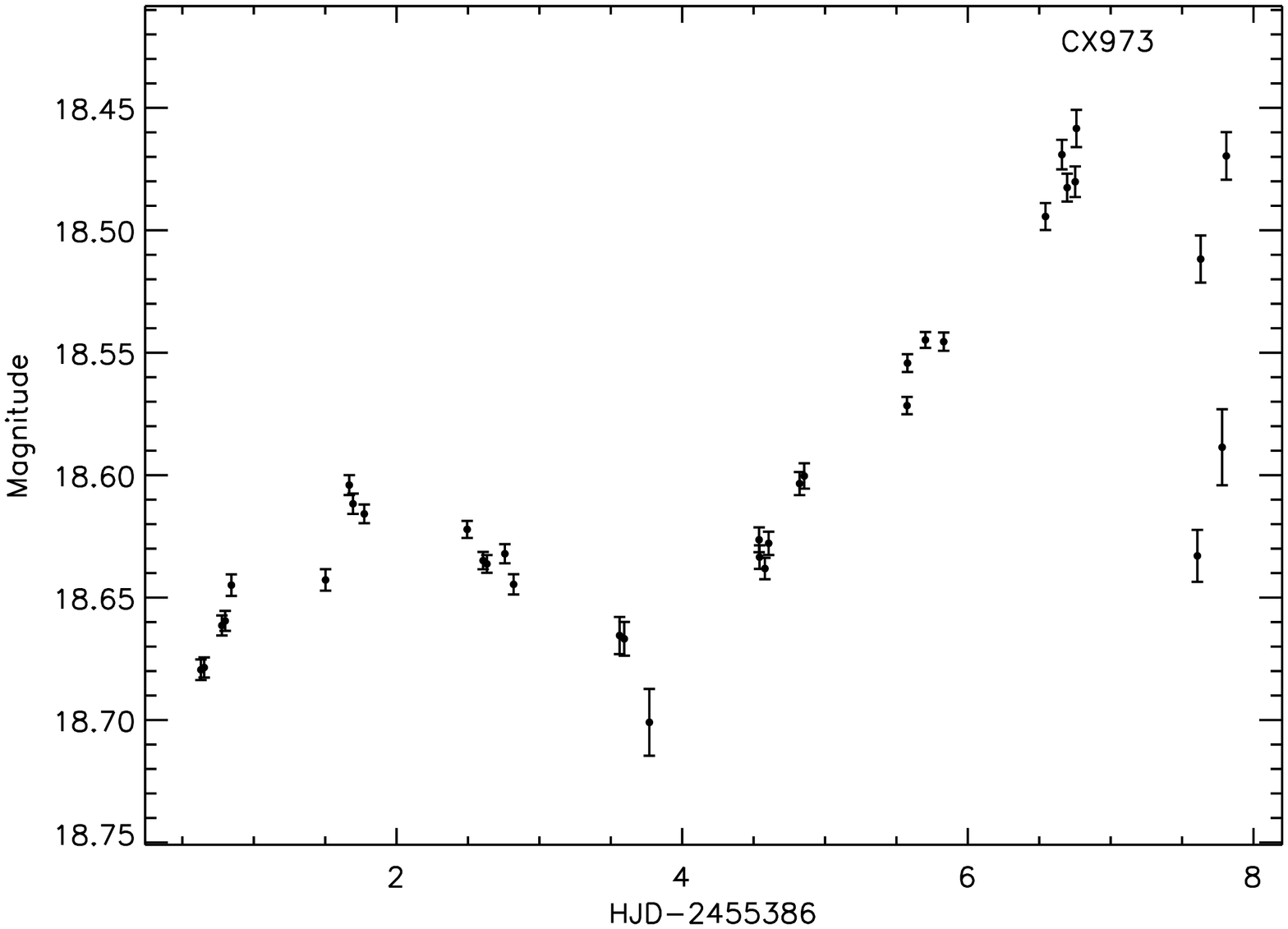}}\quad 
\subfigure{\includegraphics[width=0.4\textwidth,angle=90]{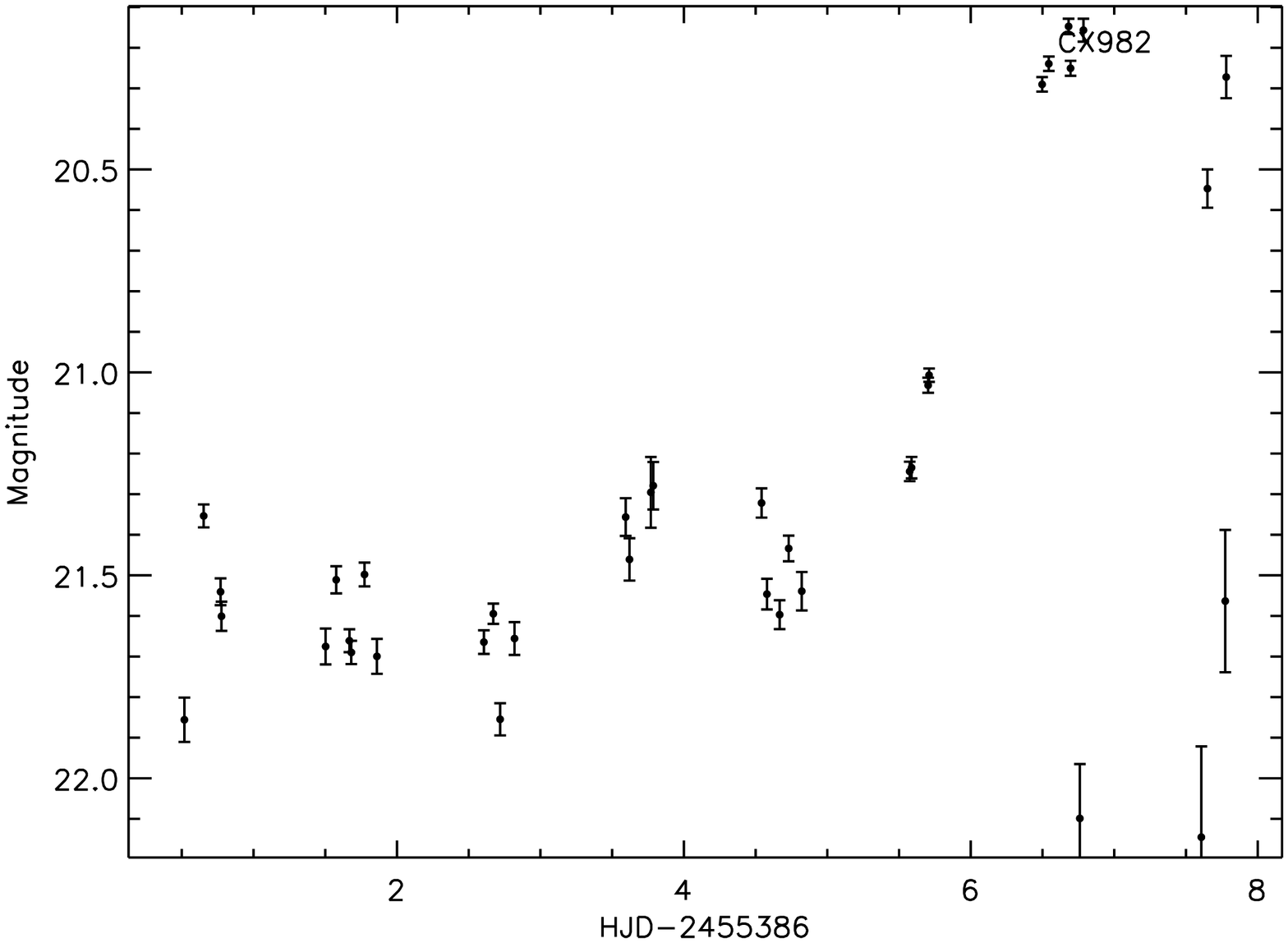}} \\
\subfigure{\includegraphics[width=0.4\textwidth,angle=90]{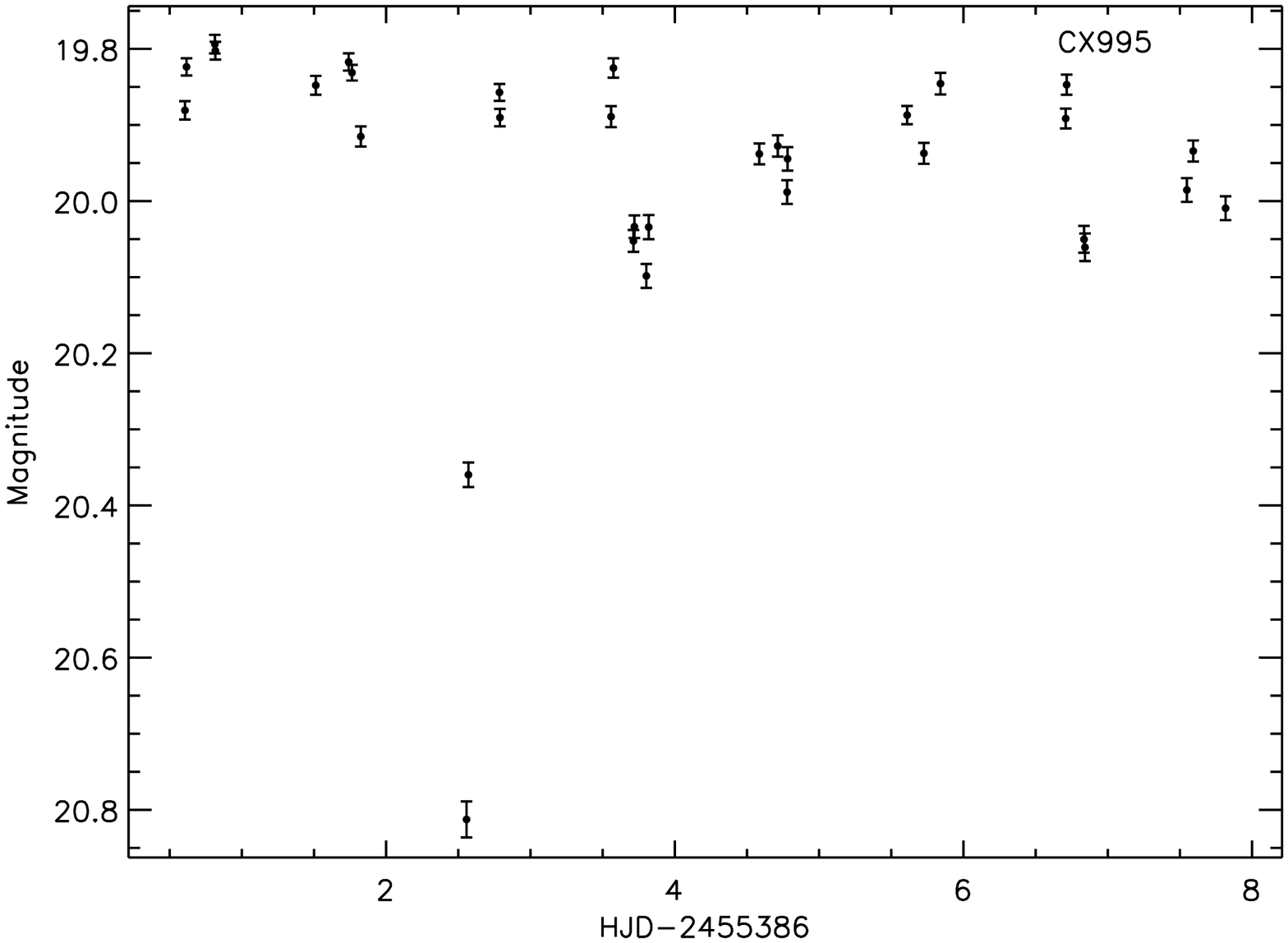}}\quad
\subfigure{\includegraphics[width=0.4\textwidth,angle=90]{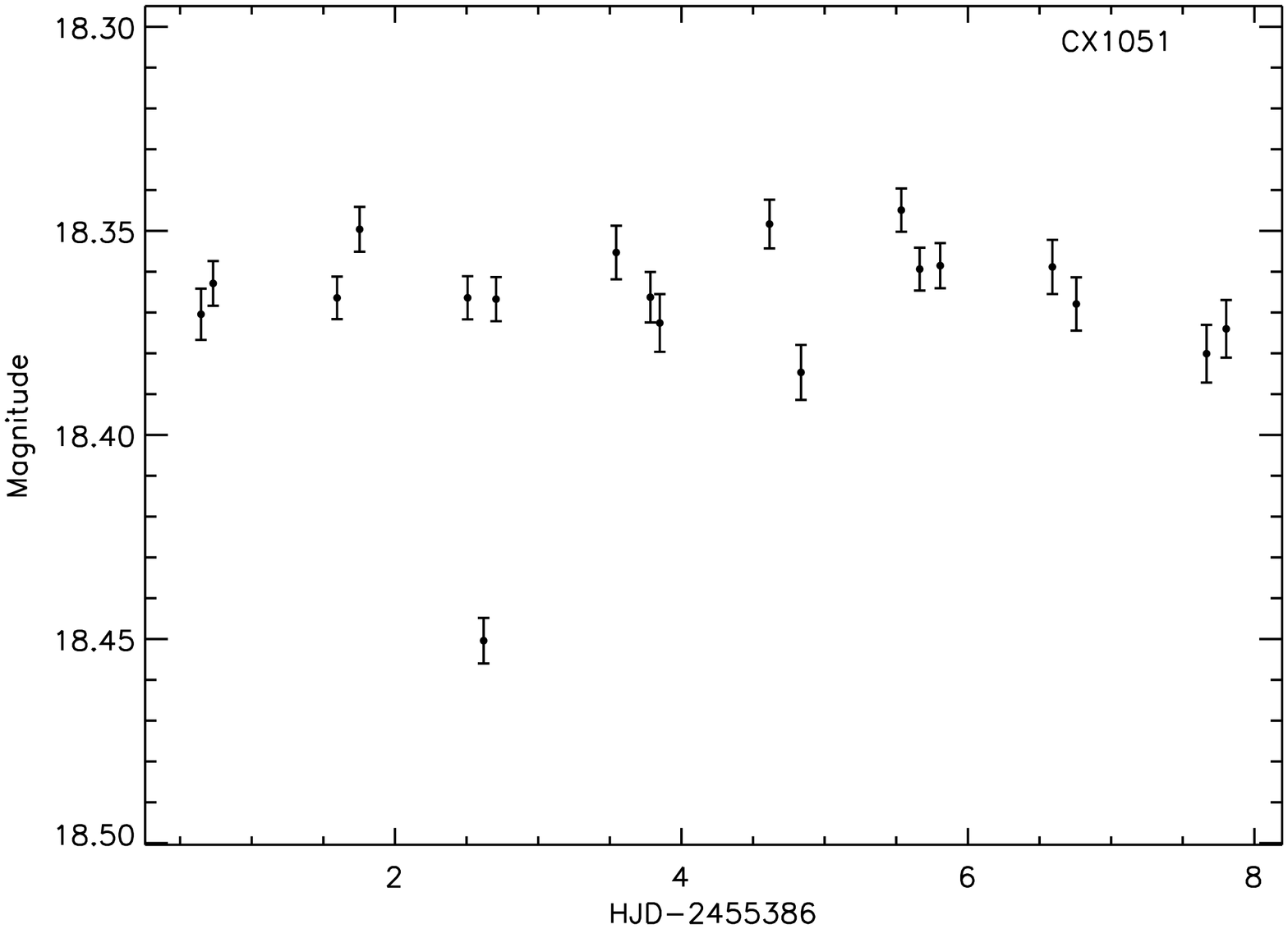}}
}
\caption{CX895, CX957, CX973, CX982, CX995, CX1051 Mosaic-II Lightcurves}
\label{lcchunk6}
\end{minipage}
\end{figure*}

CX895, shown in Figure \ref{lcchunk6}, shows a possible period of
$0.434\pm0.005$ days, or $10.4$ hours, with a FAP of $1.2\%$ with 
ellipsoidal modulations, shown in Figure \ref{foldchunk2}. This period 
was found with Phase Dispersion Minimization rather than the Lomb-Scargle
statistic, and agrees with the period found in \citet{Udalski12} of 
$0.42973$ days. \citet{Udalski12} also classify this object as eclipsing, 
which we confirm after observing an eclipse with a depth of at least
$0.35$ magnitudes. There is 
some flickering in the lightcurve on 
the order of 0.03 magnitudes, which argues strongly against this being a 
W UMa or RS CVn. The asymmetry between minima also argues strongly
against CX895 being a W UMa. This flickering is apparent in OGLE-IV data as
well. This source is optically bright, near the non-linear 
regime, and the relative photometric errors 
are generally much less than the observed dispersion. It is present in VVV
photometry, with $K_{S}=13.73$ and $J-K_{S}=0.89$ \citep{Greiss13}, which is
consistent with an M dwarf, RS CVn or a compact binary where the continuum light
is dominated by a cool companion. The ellipsoidal modulations indicate that
if this is a compact binary, then the donor star is a significant contributor to the continuum light. 
Absorbed $\frac{F_{X}}{F_{opt}} = \frac{1}{40}$, which is consistent with 
CVs, M dwarfs, or qLMXBs. This object is suggested to be a chromospheric active
star or binary in spectroscopic observations in \citet{Torres14}
pending further analysis or observation, which could be a result of
having a cool donor that dominates the optical spectrum. There are M
dwarf binaries known with shorter periods than CX895, and we cannot
rule out this interpretation without further observation.

\subsection{CX957 - CV or qLMXB}

CX957 shows no variability except for 3 points out of 36 dipping in
what could be part of eclipses,
shown in Figure \ref{lcchunk6}. The eclipse
is brief, lasting no more than 2 hours as the observations before and
after those in eclipse are back to the steady state. We can place an
upper limit on the orbital period of 1 day, assuming that phases are evenly 
sampled. The eclipses are
at least $0.13$ magnitudes deep. This object is quite red, as it is near
saturation in VVV $K_{S}$ band photometry \citep{Greiss13}. It is in the 2MASS
catalog with $J-K=1.83$, which becomes $J-K=0.81$ assuming reddening values
at Bulge distance. Absorbed $\frac{F_{X}}{F_{opt}} = 0.2$
which is consistent with qLMXBs and CVs, while at Bulge reddening this drops to
$\frac{F_{X}}{F_{opt}} =\frac{1}{100}$, which favors CVs and BH
qLMXBs, but those are likely too X-ray faint at allowed periods to be detected 
at Bulge distance. This is likely a compact binary in quiescence with a very 
cool donor. 

\subsection{CX973 - M dwarf, CV, or qLMXB}

CX973 appears in OGLE-IV data \citep{Udalski12} with a period of 10.352
days. Our Mosaic-II data are consistent with this, showing smooth
variations over several days shown in Figure
\ref{lcchunk6} with a period of roughly $\sim11$ days. The
first hump is $0.1$ magnitudes in amplitude, while the second is
$0.15$ magnitudes above that. We observe only one minimum. 
Absorbed $\frac{F_{X}}{F_{opt}} = 0.2$
which is consistent with qLMXBs and CVs or active M dwarfs, while at Bulge 
reddening,
$\frac{F_{X}}{F_{opt}} = 0.002$ and 
$L_{X}=2\times 10^{32}\,(\frac{d}{8\,{\rm kpc}})^{2}\,{\rm ergs\,s^{-1}}$. 
The possible long period and
low X-ray to optical flux ratio at the Bulge could be indicative of an evolved 
companion to an X-ray Binary or CV. At the Bulge, $M_{r'}=-1.7$ which is
consistent with a giant star. It is also bright in the infrared, with
$K_{S}=12.79$ in VVV data \citep{Greiss13}. The X-ray
brightness is too high for coronal activity at Bulge distance, while
$\frac{F_{X}}{F_{opt}}$ is too high for coronal activity at nearer
distances except for M dwarfs which emit most of their radiation in
the IR. This could be a CV or qLMXB with an evolved donor or a nearby
M dwarf with a 10.352 day rotation period.  Spectroscopy can quickly
differentiate between the case of a cool spotted star or accreting compact binary.

\subsection{CX982 - DN?}

The lightcurve morphology of CX982 is unique. There is an outburst of
$>1.3$ magnitudes
starting on night 6 of observations, but on night 7 it drops 2
magnitudes from the peak of the outburst within $1.5$ hours. It returns to
its peak brightness at the next observation 36 minutes later. The full
lightcurve is shown in Figure \ref{lcchunk6}. There is nothing wrong
with the images showing it back to or below its quiescent brightness, and
there are 3 observations of it at low level within the apparent
outburst. If it were not for these 3 points, it
would appear to be a fairly typical DN outburst. It is possible that
this is a high inclination system and the donor star blocks the
rapidly accreting disk in these points, which would explain the sudden
drops and increases in observed luminosity. It does not appear in VVV 
\citep{Greiss13}, which is consistent with an intrinsically blue object of this
magnitude such as a CV undergoing DNe outbursts.

\subsection{CX995 - CV or qLMXB plus Interloper}

\begin{figure}
\begin{center}
\includegraphics[width=0.35\textwidth,angle=0]{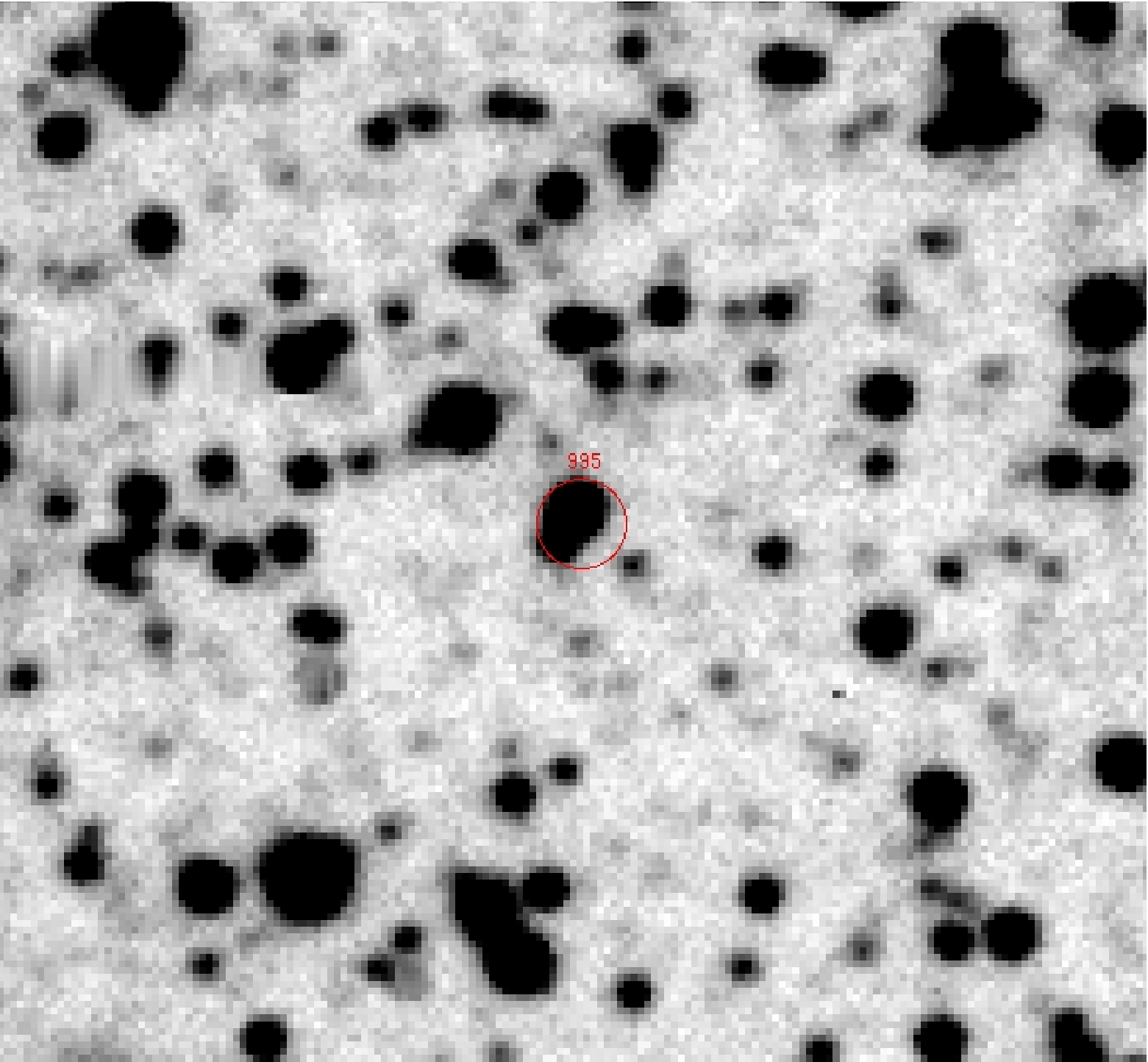}
\includegraphics[width=0.35\textwidth,angle=0]{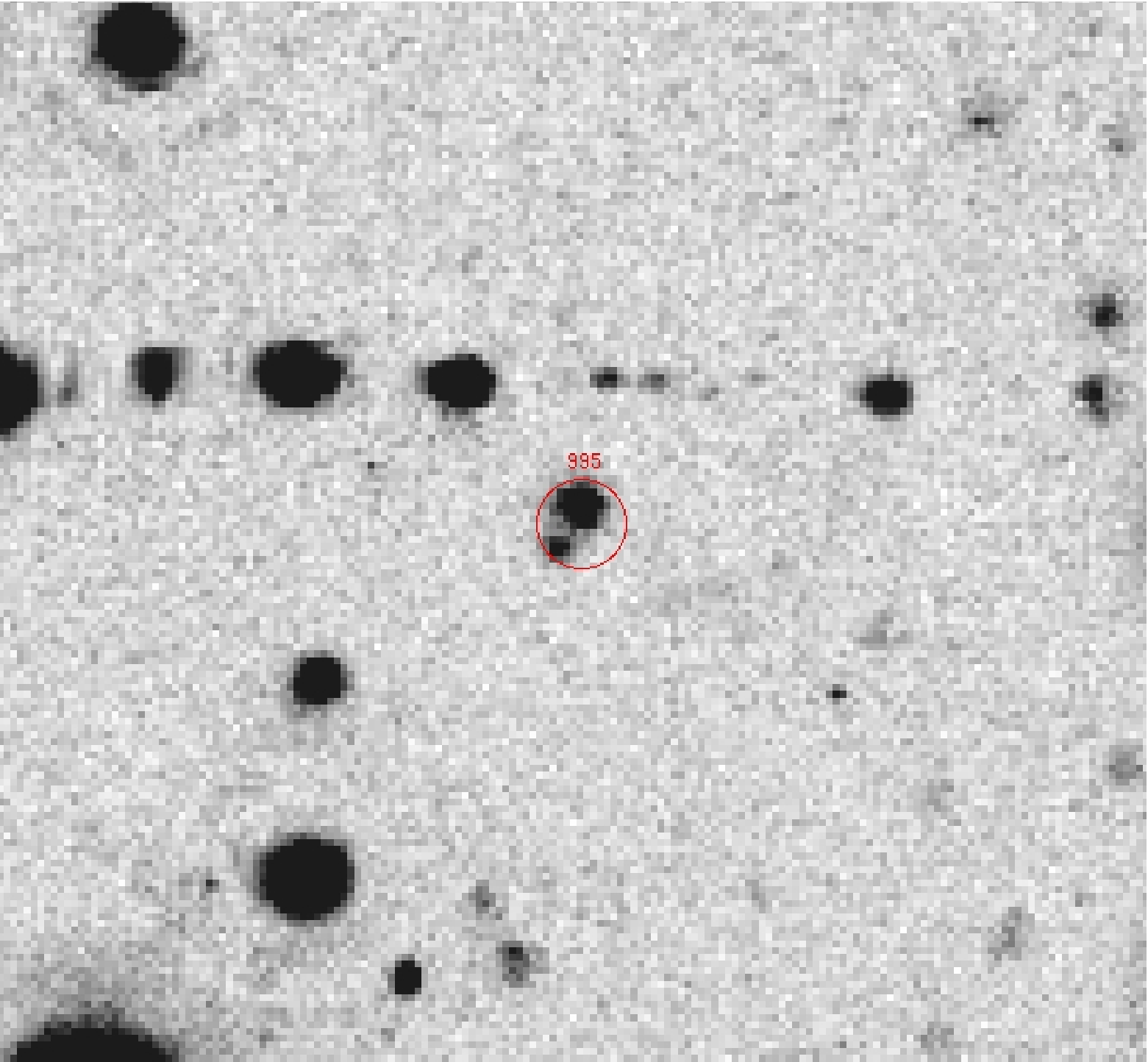}
\caption{Left: Finder chart for CX995 with X-ray position plotted in
  red. Right: Variance map of Mosaic-II images of this field.}
\label{fig:995var}
\end{center}
\end{figure}

CX995 also has two variables inside the X-ray confidence region. The reference image and variance image are
shown in Figure \ref{fig:995var}. The fainter star of the two
shows an eclipse $\sim1$ magnitude deep, with no
significant period, shown in Figure \ref{lcchunk6}. The eclipse lasts
over two observations separated
by 20 minutes. The eclipse comes at the start of the night's
observations, so there is no way to be sure exactly how deep the eclipse is or
how long it lasts. A lower limit for the eclipse duration can be
estimated if we assume that we see the eclipse at its full depth; it should 
last $1.3$ hours. It is also possible that the first data point is
taken as the source goes into eclipse, in which case the eclipse could
be substantially shorter.

The second variable in the X-ray confidence
region has a
period of either $0.4517$ days or 
$0.9034$ days, depending upon whether the variations are ellipsoidal or
sinusoidal. The $0.4517$ day fold is shown in Figure
\ref{foldchunk2}. Our observations cannot adequately distinguish between the
two in this case. Some M
dwarfs have rotation periods that fast. The second variable is
marginally more
likely to be the counterpart based on the proximity to the center of
the X-ray position, while the first shows flickering and a deep eclipse which 
makes it the more likely candidate.

These two objects are blended in VVV data, with a color of $J-K_{S}=1.38$ and 
$K_{S}=13.54$ \citep{Greiss13}, suggesting that at least one of these objects is
fairly red. The flickering and eclipse in the fainter variable star are more 
easily explained as a compact binary than a field star, while
the brighter second variable could be responsible for the red color in VVV as 
a spotted star in the field. 

The eclipsing possible counterpart would have an absorbed X-ray to optical flux 
ratio of $0.5$, which is consistent with non-magnetic CVs, qLMXBs, and magnetic
CVs. The second variable would have an absorbed flux ratio of $0.1$,
which is also consistent with CVs, qLMXBs, and active M
dwarfs.  Spectroscopy is needed to distinguish between
possibilities, as well as to confirm which of these two variables is
the true counterpart to the X-ray source.

\subsection{CX1051 - CV or qLMXB?}

CX1051 has a dip that could be the ingress or egress of a brief
eclipse at least $0.1$ magnitudes deep, 
lasting no more than 2 hours,  shown in Figure \ref{lcchunk6}. Only 1 
observation out of 19 is in eclipse. If we assume that we evenly sample all 
phases, and that the
eclipse lasts the full 2 hours possible and lasts only the $5\%$ of
the orbit indicated by seeing $\frac{1}{19}$ observations in eclipse,
this places an upper limit
of 40 hours on the orbital period. Absorbed
$\frac{F_{X}}{F_{opt}}=\frac{1}{7}$, which is consistent with
qLMXBs and CVs.

\subsection{CX1060 - CV or qLMXB, RS CVn?}

\begin{figure*}[p!]
\begin{minipage}{0.9\textwidth}
\centering
\parbox{\textwidth}{
\subfigure{\includegraphics[width=0.4\textwidth,angle=90]{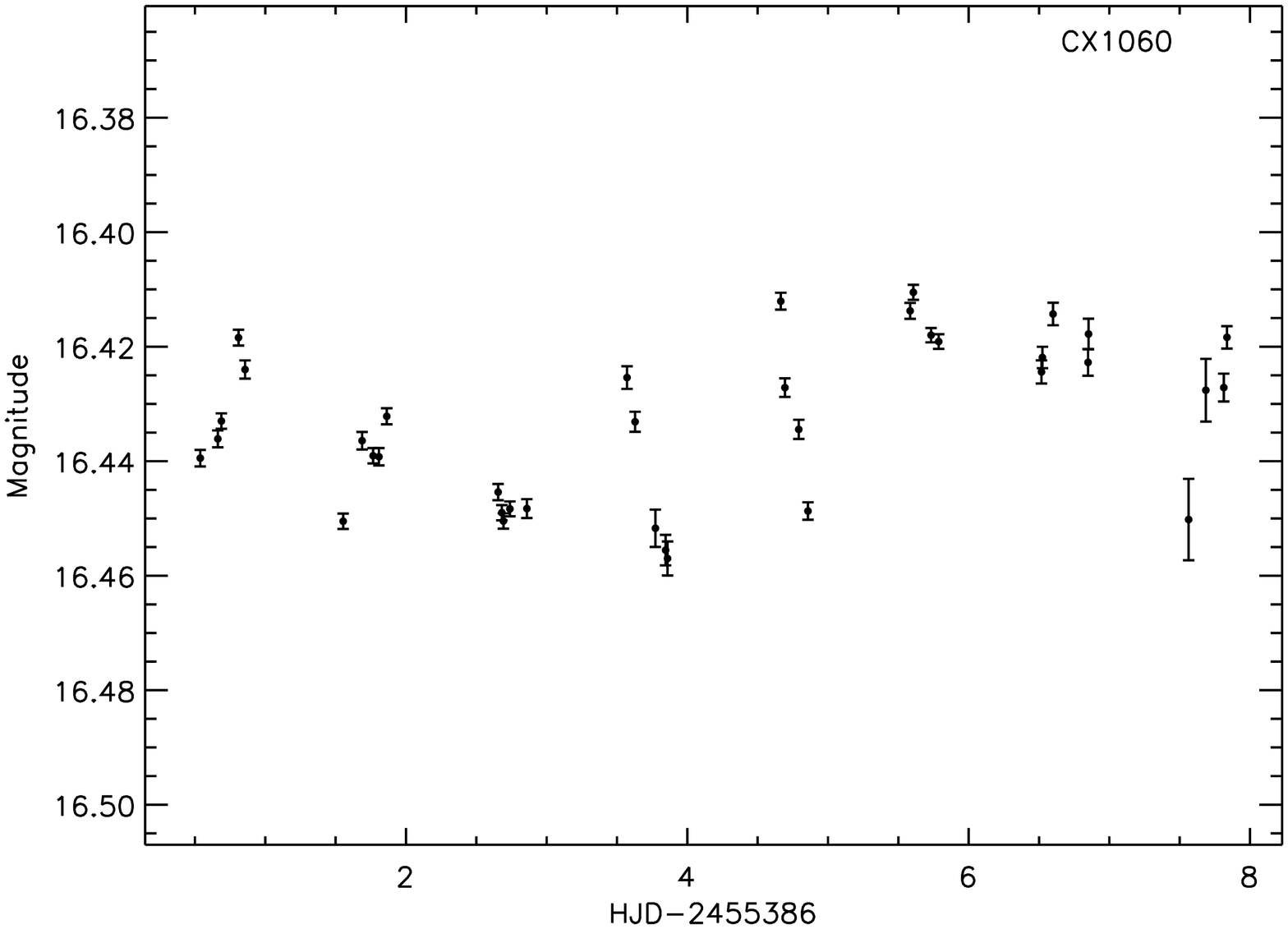}}\quad
\subfigure{\includegraphics[width=0.4\textwidth,angle=90]{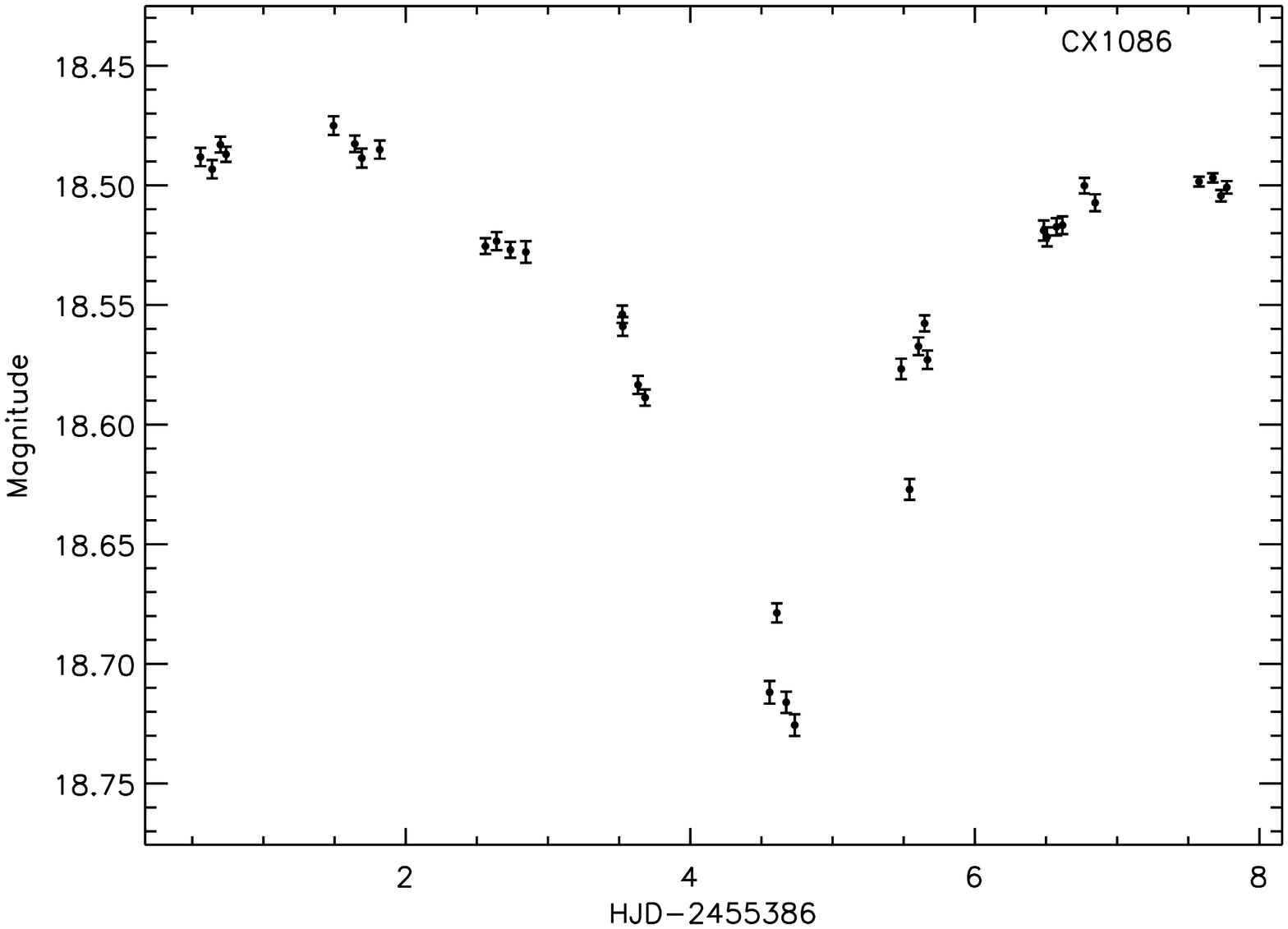}} \\
\subfigure{\includegraphics[width=0.4\textwidth,angle=90]{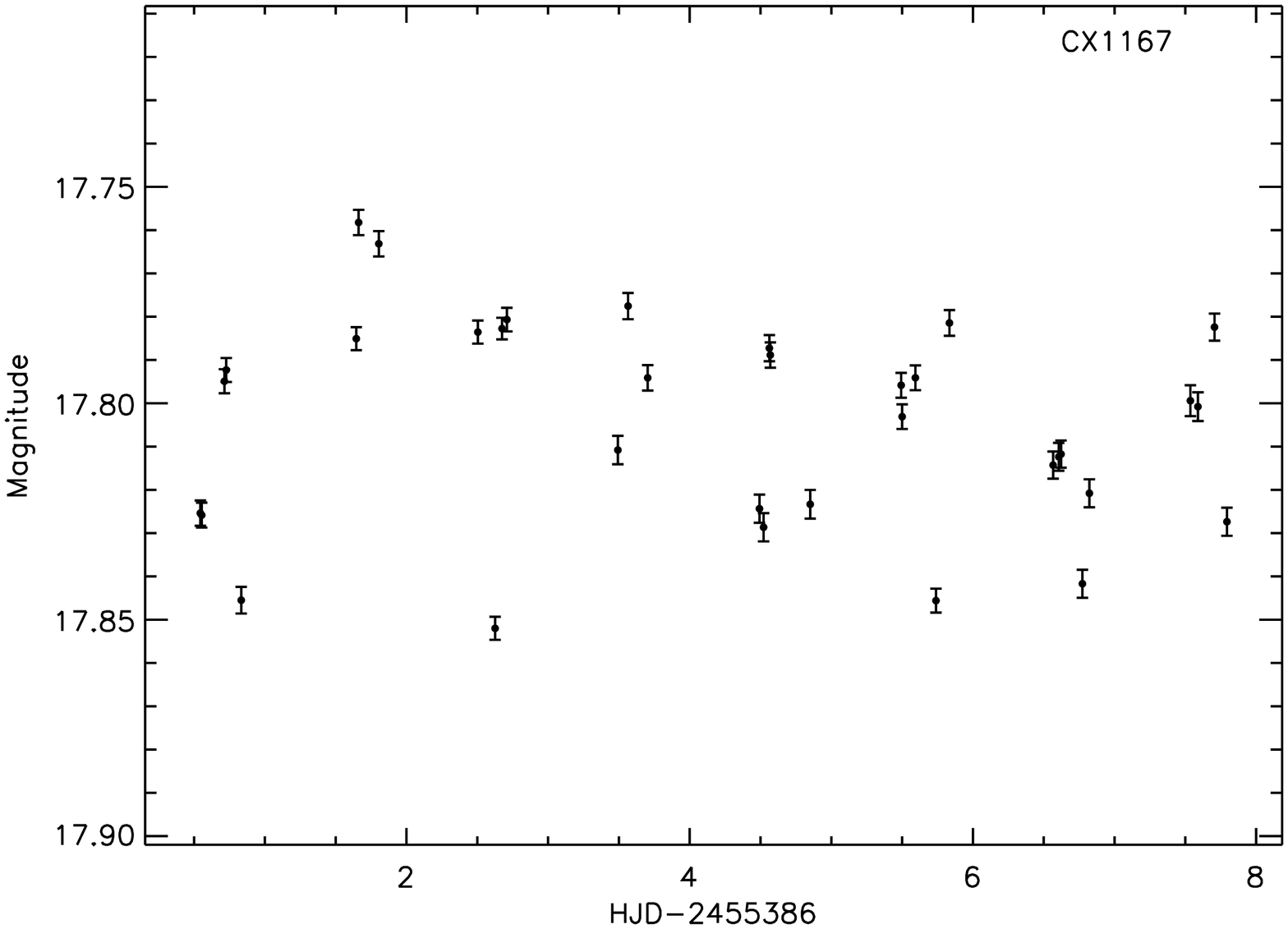}}\quad 
\subfigure{\includegraphics[width=0.4\textwidth,angle=90]{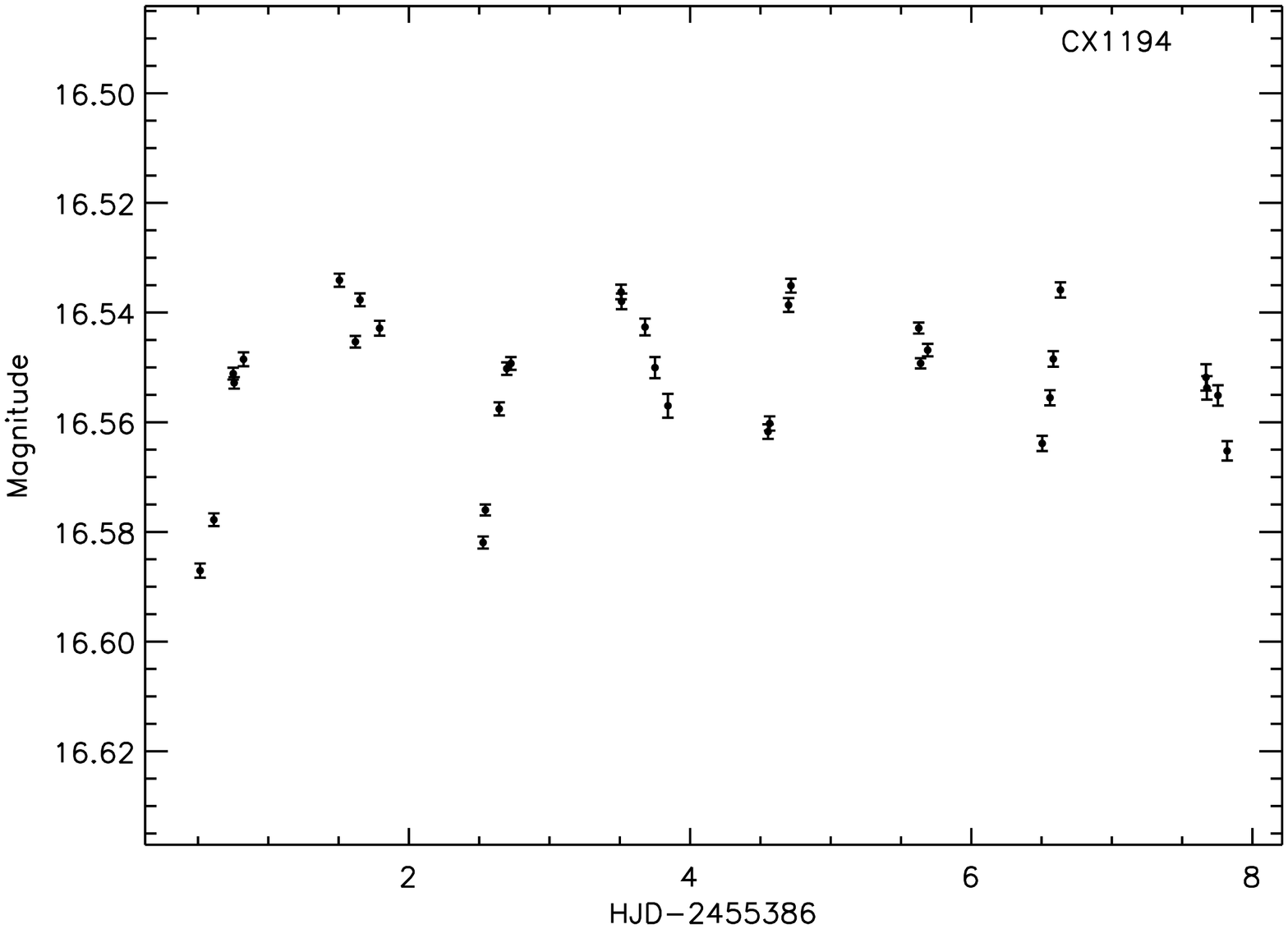}} \\
\subfigure{\includegraphics[width=0.4\textwidth,angle=90]{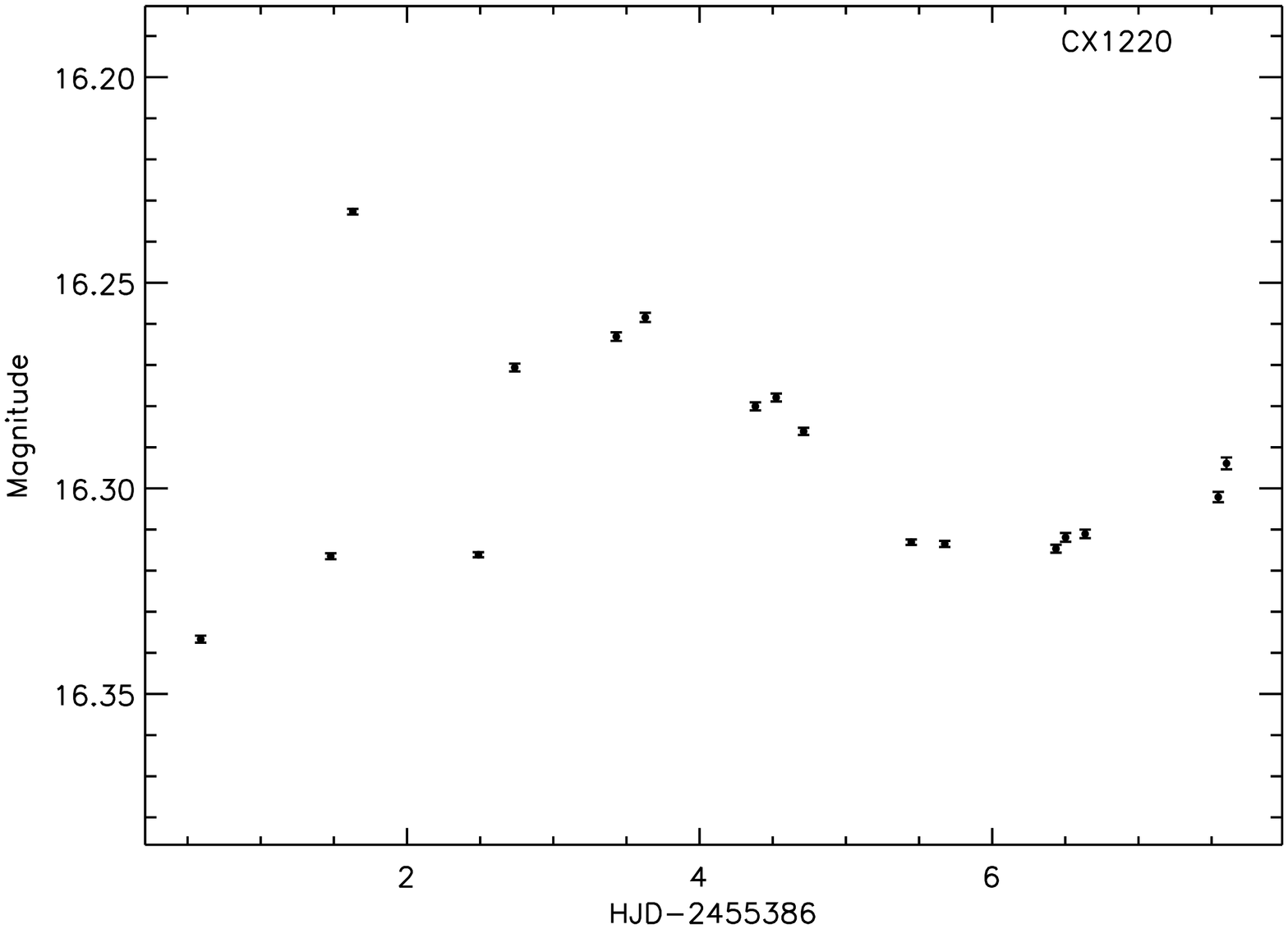}}\quad
\subfigure{\includegraphics[width=0.4\textwidth,angle=90]{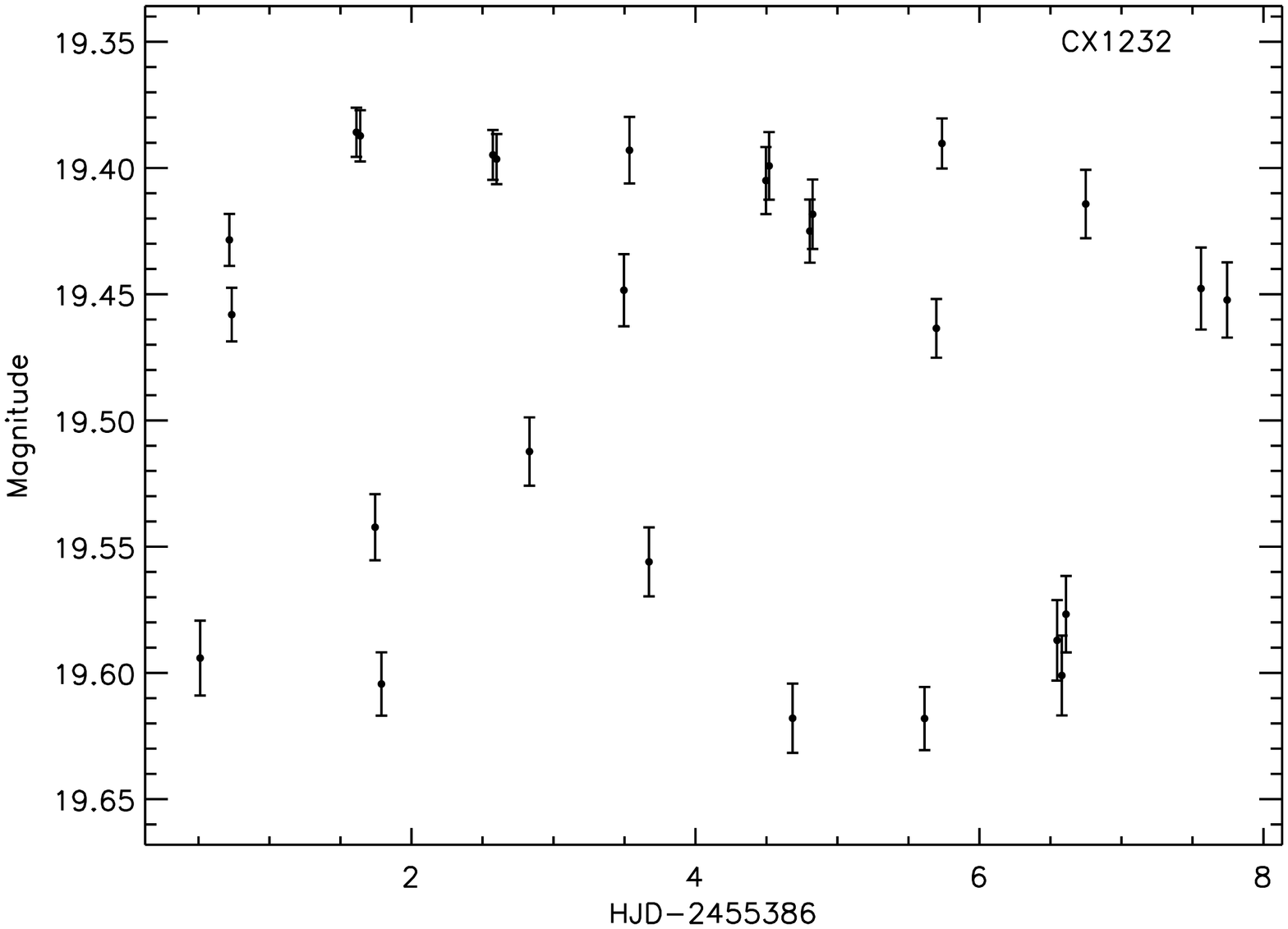}}
}
\caption{CX1060, CX1086, CX1167, CX1194, CX1220, CX1232 Mosaic-II Lightcurves}
\label{lcchunk7}
\end{minipage}
\end{figure*}

CX1060,  shown in Figure \ref{lcchunk7}, has periodic modulations with a fundamental period of 1.161
days, and an amplitude of only $0.05$ magnitudes. There is more
scatter around the peak on this period than in the rest of the
lightcurve. Doubling the period to $2.322$ days does not produce a 
significantly better fit. Each fold is shown
in Figure \ref{foldchunk2}. Absorbed
$\frac{F_{X}}{F_{opt}}=0.02$ which is consistent with an RS
CVn, CV, or qLMXB, and drops further to $\frac{F_{X}}{F_{opt}}=0.001$ assuming reddening at Bulge distance, indicating that this source is more likely in the foreground as the types of objects found at this period and that low X-ray to optical flux ratio are not luminous enough in the X-ray to be seen at the Bulge in our survey. The dispersion above what 
expected from statistical noise is inconsistent with an RS CVn which should 
not show flickering. This object is therefore a candidate CV or qLMXB. In the qLMXB interpretation, the low value of $\frac{F_{X}}{F_{opt}}$ even without extinction favors a BH primary. This 
object is blended with a nearby star separated by $1.5''$ in 
VVV photometry with $K_{S}=10.0$ \citep{Greiss13}. This IR-bright star is also
within the X-ray confidence region, but does not show significant variability 
in Sloan $r'$ band photometry.  The
variable star itself does not appear in VVV photometry. Spectroscopy is needed to differentiate between cases and to reject more firmly an RS CVn interpretation.

\subsection{CX1086 - CV, qLMXB, or M dwarf}

CX1086 shows what appears to be an eclipse with a depth
of $0.26$ magnitudes and FWHM duration of 2 days,  shown in Figure
\ref{lcchunk7}. In OGLE-IV data this object exhibits the 
same behavior on a period of $P=11.768$ days \citep{Udalski12}. In OGLE 
data, it is listed as an eclipsing source, with a minimum at phase 
$\phi=0.5$ that is not as deep as that at $\phi=0.0$ by $\Delta I=0.05$ 
magnitudes. The OGLE lightcurve is also consistent with ellipsoidal variations.
OGLE-IV observations span a period of 2 years, and the lack of changes in the 
lightcurve morphology cuts weakly against the idea that the modulations are produced
by starspots, which could produce changes in phase and amplitude, and even in
period, in that time as star spots appear, disappear, and move towards the 
equator from the poles. Absorbed $\frac{F_{X}}{F_{opt}}=\frac{1}{6}$, which 
is consistent with CVs,
qLMXBs, or M dwarfs but is too high for an RS CVn star, which is unlikely 
to be detected at sufficient distance to suffer the amount of extinction 
needed to lower $\frac{F_{X}}{F_{opt}}$ to a point consistent with RS CVns. An 
eclipse this broad or ellipsoidal
modulations lasting this long would imply an evolved donor. At the Bulge,
$E(B-V)=2.11$ in this line of sight, which would imply $M_{r'}=-1.77$,
consistent with giant stars. $\frac{F_{X}}{F_{opt}}=0.002$ at the
Bulge, which is also consistent with qLMXBs with giant donors. In VVV data, 
this object is also variable and quite red at $K_{S}=12.20$ and $J-K_{S}=1.52$. 
Correcting for reddening assuming Bulge distance, $J-K_{S}=0.39$ and $K_{S}=-3$,
which is too blue and too bright for an RGB star, but would be consistent 
for a distance somewhat before the bulge, experiencing less reddening than
the Red Clump stars used to estimate reddening in the VVV survey. This object
is consistent with a qLMXB with an evolved companion, though spectroscopy is
needed to completely rule out the scenario of a nearby coronally active M 
dwarf with a pattern of star spots which emulates either ellipsoidal
modulations or a broad eclipse.

\subsection{CX1167 - CV or qLMXB}

CX1167 is one of the few X-ray sources with a counterpart showing
ellipsoidal modulations at a period one expects for a qLMXB. The
lightcurve is shown in Figure \ref{lcchunk7} and is
shown in Figure \ref{foldchunk2} folded on $P_{orb}=8.35$\,hours. It has
absorbed $\frac{F_{X}}{F_{opt}}=0.08$, which drops to $0.002$ for
Bulge reddening. We would not see an object such as a W UMa or other
low luminosity X-ray emitter at the
bulge, while reddening equivalent to being at the Bulge distance is
needed to make the observed quantities consistent with a W UMa. The low X-ray 
to optical flux ratio at Bulge distance is also unlikely for CVs or qLMXBs, 
which favors the interpretation that CX1167 is closer than Bulge distance,
with the X-ray to optical flux ratio towards the higher end of the
possible range. In VVV photometry for this object \citep{Greiss13}, 
$K_{S}=14.4$ and $J-K_{S}=1.02$. This is consistent with
a qLMXB or CV with a cool main sequence donor and moderate reddening.

\subsection{CX1194}

\begin{figure*}
\begin{minipage}{0.9\textwidth}
\centering
\parbox{\textwidth}{
\subfigure{\includegraphics[width=0.4\textwidth,angle=90]{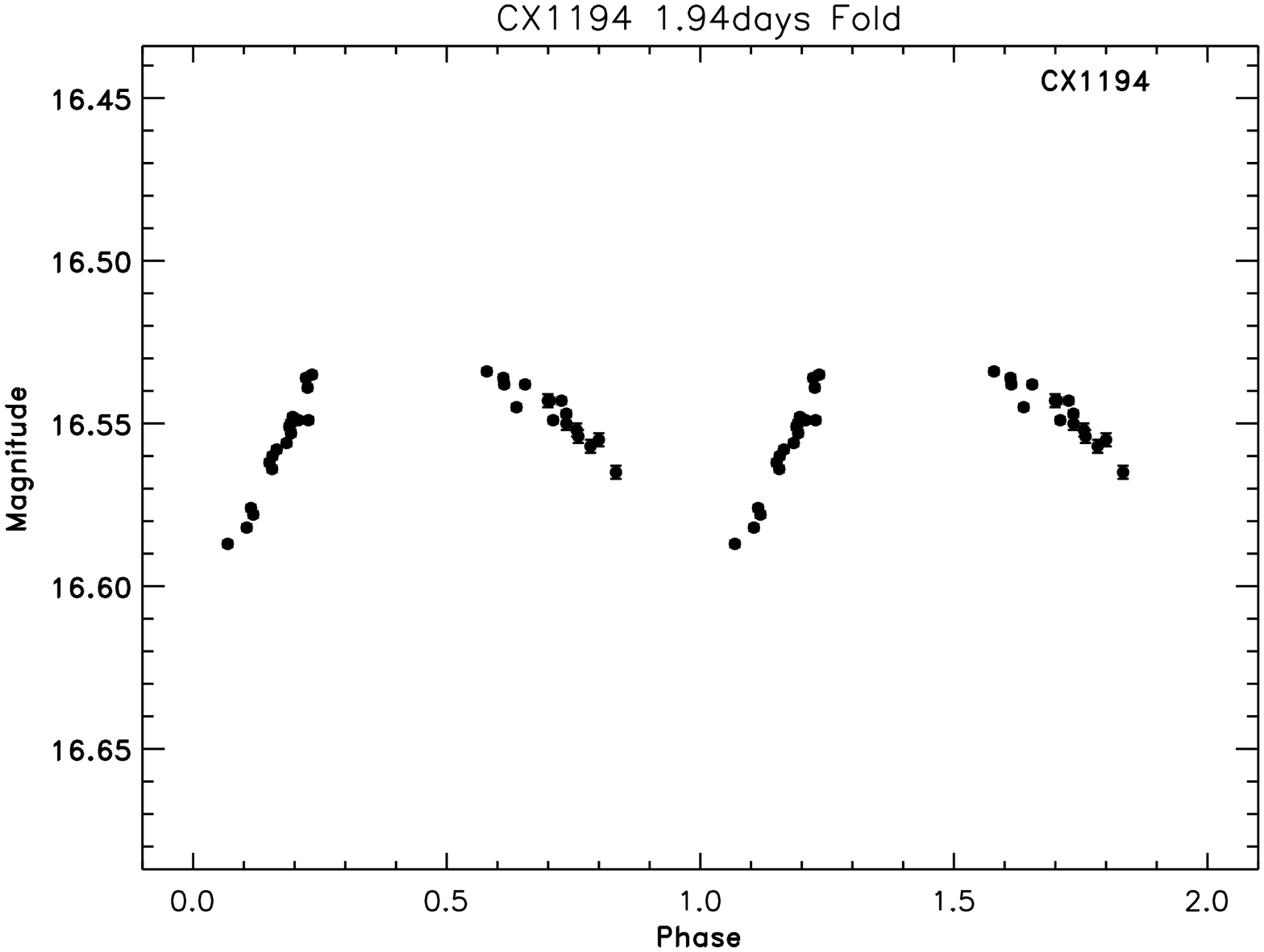}}\quad
\subfigure{\includegraphics[width=0.4\textwidth,angle=90]{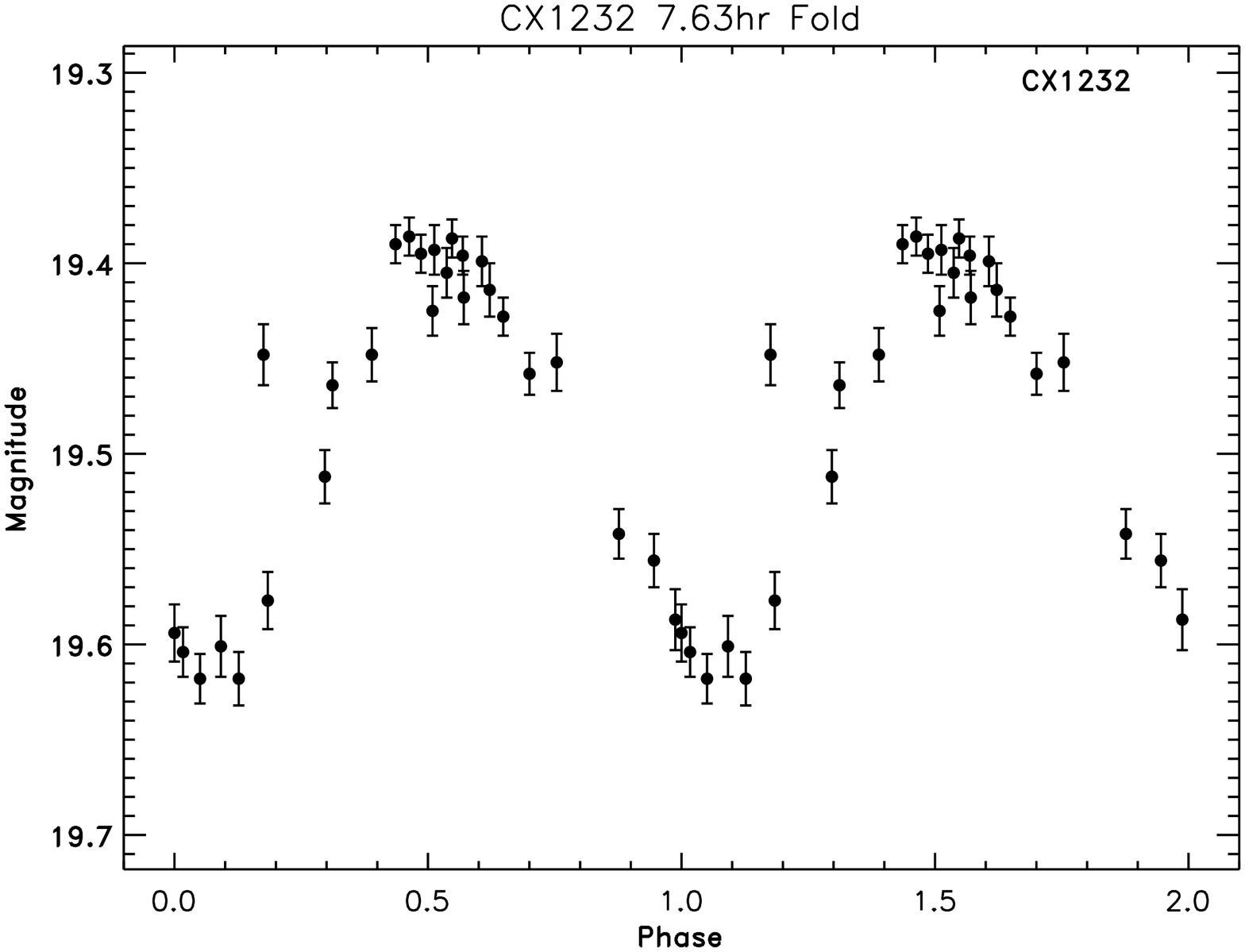}} \\
}
\caption{{\it Left:} Lightcurve of one possible counterpart to CX1194
  folded on a 1.94 day period.
{\it Right:} Lightcurve of the likely counterpart to CX1232
  folded on a 7.63 hour period. }
\label{foldchunk3}
\end{minipage}
\end{figure*}

CX1194, shown in Figure \ref{lcchunk7}, has a significant period of
$1.94$ days, which is of note 
because it is so close to an integer of 1 day, and is strongly
aliased by our nightly sampling rate. This aliasing is apparent in the
folded lightcurve in the form of large gaps at near $\phi=0.4$ and
$\phi=0.9$ seen in Figure \ref{foldchunk3}.
The gaps in phase coverage make it difficult to determine the exact
amplitude of the variations, but they are on the order of $0.06$
magnitudes. The lightcurve is asymmetric and 
single-humped, rising faster that it falls. In VVV data, $K_{S}=14.19$ and 
$J-K_{S}=0.73$ \citep{Greiss13}. Absorbed $\frac{F_{X}}{F_{opt}}=\frac{1}{40}$, 
which is consistent with RS CVns, CVs, or qLMXBs. If this object is a CV or
qLMXB, the period would suggest a somewhat evolved donor star. Spectroscopy is 
necessary to firmly distinguish between possibilities.

\subsection{CX1220 - Flaring M dwarf or RS CVn}

CX1220 has a counterpart that is very near the non-linear regime of
the CCD response at $r'=16.29$, and some observations in good seeing
conditions were removed for being in this regime. This source shows a
brief outburst of $0.1$ magnitudes and a rise and decline of $0.06$
magnitudes over several days, shown in Figure \ref{lcchunk7}. This could
be a flaring M dwarf or an RS
CVn. Absorbed $\frac{F_{X}}{F_{opt}}=\frac{1}{50}$ which is
consistent with RS CVns and active stars. This object appears in the 2MASS
catalog \citep{Skrutskie06} and is quite red with $J-K=1.347$ and $K=11.311$. 
The very red color and IR brightness of this object, as well as the apparent
flare, is consistent with a nearby flare star, but spectroscopy is necessary
to firmly classify this object.

\subsection{CX1228}

There is a variable star outside the $95\%$ X-ray confidence region
that is listed as a possible counterpart in \citet{Udalski12}. We
include it in Table \ref{table:varstats} to show that we confirm their
results, but based on our selection criteria, it should not be
considered as a likely counterpart to the
X-ray source without other evidence, such as from spectroscopy.

\subsection{CX1232 - CV or qLMXB}

CX1232 has a possible counterpart which is variable, but near the edge of the
X-ray $95\%$ confidence region. The lightcurve, shown in Figure \ref{lcchunk7},
has a significant period of $0.3179$ days, or $7.63$ hours, with an amplitude 
of $0.24$ magnitudes. The true period could be twice this if
the modulations are ellipsoidal, though the fit provided by doubling
the period is no better. The folded lightcurve is shown in Figure
\ref{foldchunk3}. Absorbed $\frac{F_{X}}{F_{opt}}=0.4$ which is consistent with 
CVs and qLMXBs. The period, amplitude of oscillations, and X-ray to optical 
flux ratio are certainly suggestive of a compact binary, but spectroscopy is 
needed to classify this object securely.

\section{Period Distribution of Optical Counterparts}

\begin{figure}
  \begin{center}
    \includegraphics[width=0.43\textwidth, angle=90]{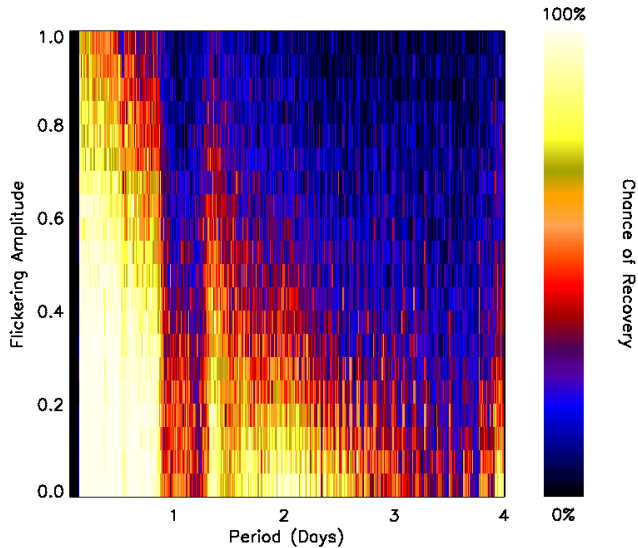}
  \caption{This plot
    shows the likelihood of recovering a given period
    at different levels of flickering imposed on the underlying
    periodicity for our Mosaic-II observations. As you can see, it is
    unlikely that we recover photometric periods below 2 hours, though
  we maintain better than even chances of recovering periods between 2
hours and 1 day through the RMS of random variations being 80\% of the
amplitude. Loss of sensitivity due to aliasing around 1 day is
apparent. For qLMXBs with periods above 1 day, the donor should
comprise a larger fraction of the continuum light because of its
larger radius, so random flickering originating in the disk should be smaller amplitude
compared to shorter period systems. For RS CVns and W UMas, there
should be no intrinsic flickering, with the only noise being
photometric error on the order of $<1\%$ for sources brighter than $r'=19$.}
  \label{fig:montecarloperiod}
  \end{center}
\end{figure}

Our observations have a high rate of period recovery for systems with
little to no flickering. As aperiodic fluctuations begin to overtake
the periodic, however, we lose the ability to reliably recover
periods with the limited sampling we have. To quantify this, we have run 
Monte Carlo
simulations to determine the rate of
period recovery for various periods with our average level of sampling
at different relative amounts of flickering, the results of which are
shown in Figure \ref{fig:montecarloperiod}. For qLMXB systems in
which $\gae \half$ of the light should come from the donor star rather
than the accretion disk, we expect to recover almost all periods
between 2 and 23 hours. We do suffer from aliasing around 1 day, which
is an expected and unavoidable result of being able to observe only at
night when only a single observatory is used. We cannot claim strong evidence of periods 
longer than 4 days
because our baseline of observations only extends 8 days and we
cannot, therefore, see a suspected period $>4$ days repeat. All
periods listed greater than 4 days should be treated as suspect
pending other observations. Some systems show a long rise or a single
crest or trough. After matching these to the recovered counterparts in
OGLE IV data \citep{Udalski12}, the majority of these systems are long
period variables with
$P_{orb} > 8\,{\rm days}$. Because larger stars will have longer periodicities, the shallower 
OGLE IV survey likely contains the majority of the long period objects in its field of view. Some 
of these objects are also irregular variables with
characteristic timescales of days or longer, such as CX332 \citep{Hynes14}. 

\begin{figure}
  \begin{center}
   \includegraphics[width=0.35\textwidth, angle=90]{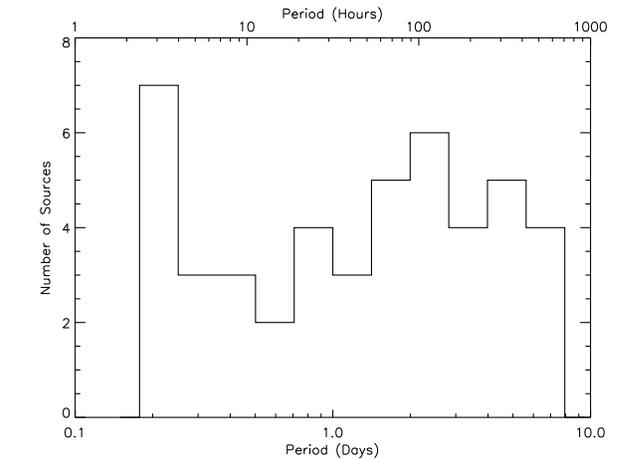}
  \caption{The period distribution of likely
    counterparts to X-ray sources in the GBS. There are only \vp\
    sources for which we suspect or confirm periods, compared to the \vble\
    sources seen to be variable. This includes \lpv\ sources which are suspected
    through overlap with OGLE IV fields \citep{Udalski12} to be long
    period variables with periods above 8 days that our
    observations are not sensitive to. Periods between 4 and 8 days
    are tentative and require confirmation from observations with a
    longer baseline. }
  \label{fig:perdistrib}
  \end{center}
\end{figure}

The period distribution, shown in Figure \ref{fig:perdistrib}, is a broad
indicator of the types of
sources we are finding. RS
CVns tend to have periods
of days, W UMas less than a
day, while the period of LMXBs depends on the mass and evolutionary
stage of the donor. The naive expectation from Kepler's Third Law is
that the mass of the primary will affect the period of the system, but
the Roche Lobe geometry changes as a
function of the mass ratio in such a way as to counter this
dependence. For main sequence donors, the period is on the
scale of hours with $M_{2}\approx0.11\,P_{hr}$ \citep{Frank02}, but periods of days
are possible for more evolved
donors. Of the
population of currently known LMXBs, the period
distribition peaks around 5-6 hours \citep{Lewin06}, though systems
with $P_{orb}>1$ day are certainly known. It is also important to note
that the current selection effects in finding LMXBs in outburst and
following them into quiescence select against short period systems
\citep{Knevitt14}. Non-magnetic CV systems, of which we
expect $\approx46$ with optical counterparts in the Mosaic-II area,
have a bimodal orbital period distribution
peaking at $1.5$ hours and $3.5$ hours for X-ray selected CVs, with a
substantial gap in between 2-3 hours that is known as the CV period
gap \citep{Gansicke05,Warner76}. We are unable to reliably recover with confidence 
periods below the CV period gap. We expect hundreds of RS
CVn systems \citep{Jonker11}.  The population assumptions used in
\citet{Jonker11}, converted into Sloan $r'$ for this work, predict
only 18 qLMXBs with $r'\le 21.5$ that we are likely to detect, so the
vast majority of systems with $P_{orb} > 1\,{\rm day}$ should be
RS CVns rather than LMXBs, though it is likely that a select few LMXBs
can be found at these periods. Indeed, qLMXBs at these periods should
be optically brighter, having larger Roche Lobe filling companions,
and should be detected as variables at a much higher efficiency than
shorter period qLMXBs. X-ray detected short period BH qLMXBs are
likely fainter in the X-ray
than the assumed X-ray luminosity in \citet{Jonker11} because energy is carried either through the event horizon by ADAFs \citep{Narayan97,Garcia01,Hameury03,Narayan08} or away by jets \citep{Fender03}. The optical light, however, eminates from the accretion disk and donor star and is comparable to NS qLMXBs at the same periods. Therefore, X-ray detected BH qLMXBs should be
nearer and therefore optically brighter and detected more efficiently as
variables than may be reflected in the rough population estimate
provided here. 

\section{Discussion}

\subsection{Likely Composition of X-ray Sources with Nonvariable Counterparts}

Assuming 10,000 LMXBs in the Galaxy with $L_{X}=10^{33}\,{\rm ergs \, cm^{-2} \, 
s^{-1}}$ and a mean $r'$ absolute magnitude of 5, \citet{Jonker11} predicted 
that the GBS should find 120 qLMXBs with optical counterparts with
$i'<23$, which translates to 71 qLMXBs with $r'<23$,
with 53 lying in the region imaged by Mosaic-II. These are mostly
expected to be quite faint, with only 18 brighter than magnitude
$21.5$. Using the known population 
as a guide, we expect these to have significant contributions in optical 
wavelengths from the light of their companion stars, and so to show 
ellipsoidal variations with periods of typically several hours. The 
amplitude of these variations depends upon $i$, being at maximum when
$i=90^{\circ}$ and zero when $i=0^{\circ}$, while sources are 
distributed on the sky uniformly in $\cos i$. For a random distribution of 
orbital inclinations, half of this population has $i>60^{\circ}$, so 
most should show measurable ellipsoidal variations. For intrinsic
variations of amplitude
$0.2$ magnitudes, for example, $82\%$ of sources would have $\Delta m>0.05$ 
magnitudes. For intrinsic variations $0.1$ magnitudes in amplitude, $59\%$ 
of sources would have $\Delta m>0.05$ magnitudes.
Our discovery efficiency is quite high for eclipsing sources and even
small amplitude flickering and periodic sources down to $r' \approx21.5$
\citep{Ratti13,Britt13,Torres14}. Some portion of counterparts with no detected
optical variability will be high inclination systems, but it is not possible 
that inclination effects account for a majority of the \nv\ X-ray sources
with candidate counterparts that do not vary. Still, it is difficult to place
constraints on the make-up of these non-variables because we have
spectroscopically confirmed that some systems that we would expect to show
either flickering or ellipsoidal modulations do not, in fact, vary at
all \citep[e.g.][CX561]{Britt13}. Given our
X-ray detection limit of 3 photons, our sensitivity to AGN is an unabsorbed flux in the $2-10\,{\rm keV}$ range of $\approx5\times10^{-14}\,{\rm ergs\,s^{-1}\,cm^{-2}}$ assuming an average $N_{H}$ of $2\times10^{22}\,{\rm cm^{-2}}$ and a photon index of $\Gamma=1.7$. From the observed distribution of AGN luminosities in \citet{Giacconi01}, we estimate there to be $\approx590$ AGN in the Mosaic-II region at this flux. Our observations are slightly softer, from $0.3-8\,{\rm keV}$, so that X-ray extinction is worse, though the number of AGN in our X-ray sample is still likely in the hundreds. As discussed in Section \ref{sec:results}, there are
256 X-ray sources which both contain no variable stars and have error
regions large enough to admit likely interlopers. It is likely that
many of these ``nonvariables'' are in fact background AGN with no
optical counterpart coincident with field
stars. 

\subsection{Likely Composition of Periodic Counterparts to X-ray Sources}

\begin{figure}
\begin{center}
 \includegraphics[width=0.35\textwidth, angle=90]{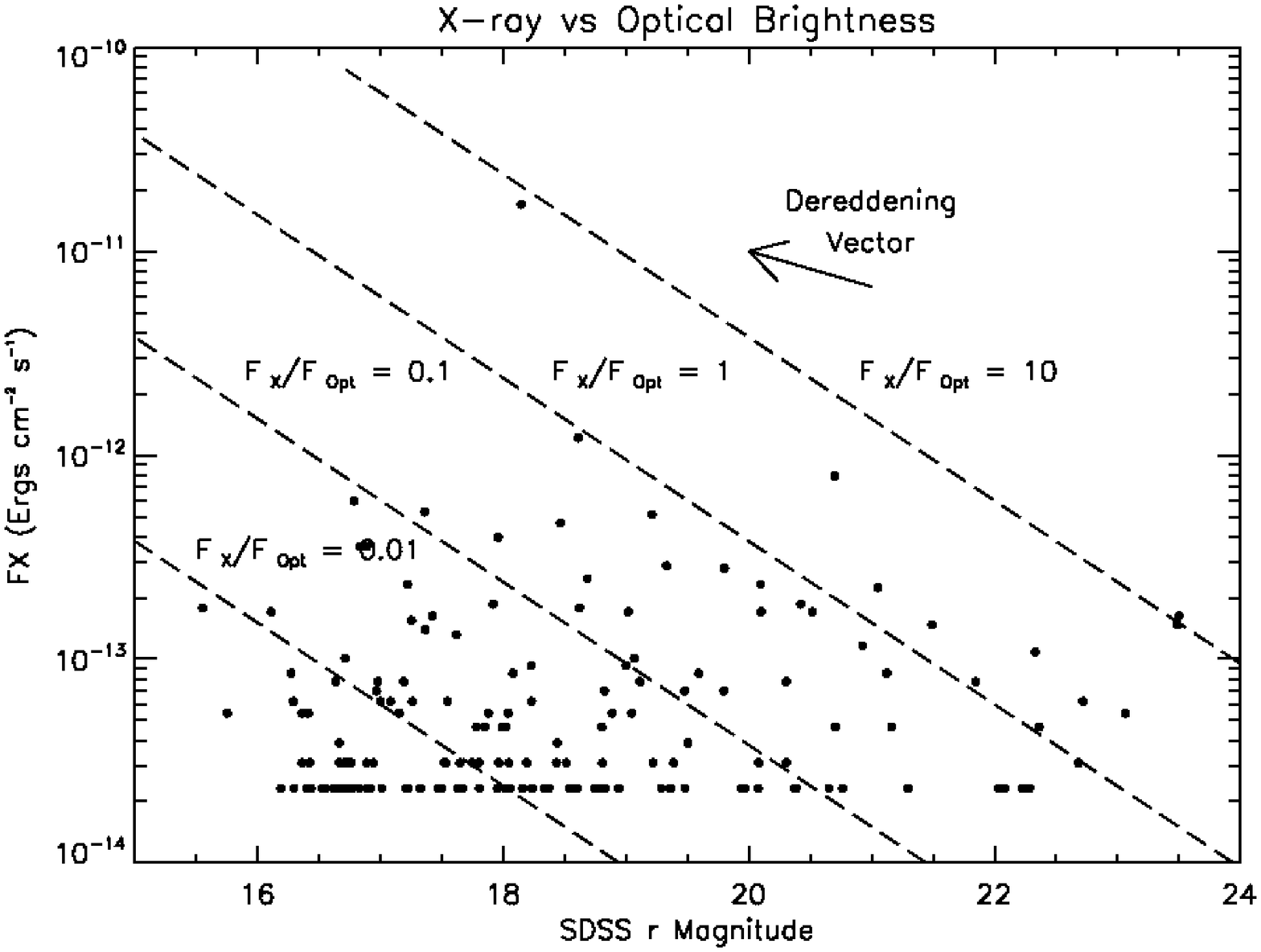}
\includegraphics[width=0.35\textwidth, angle=90]{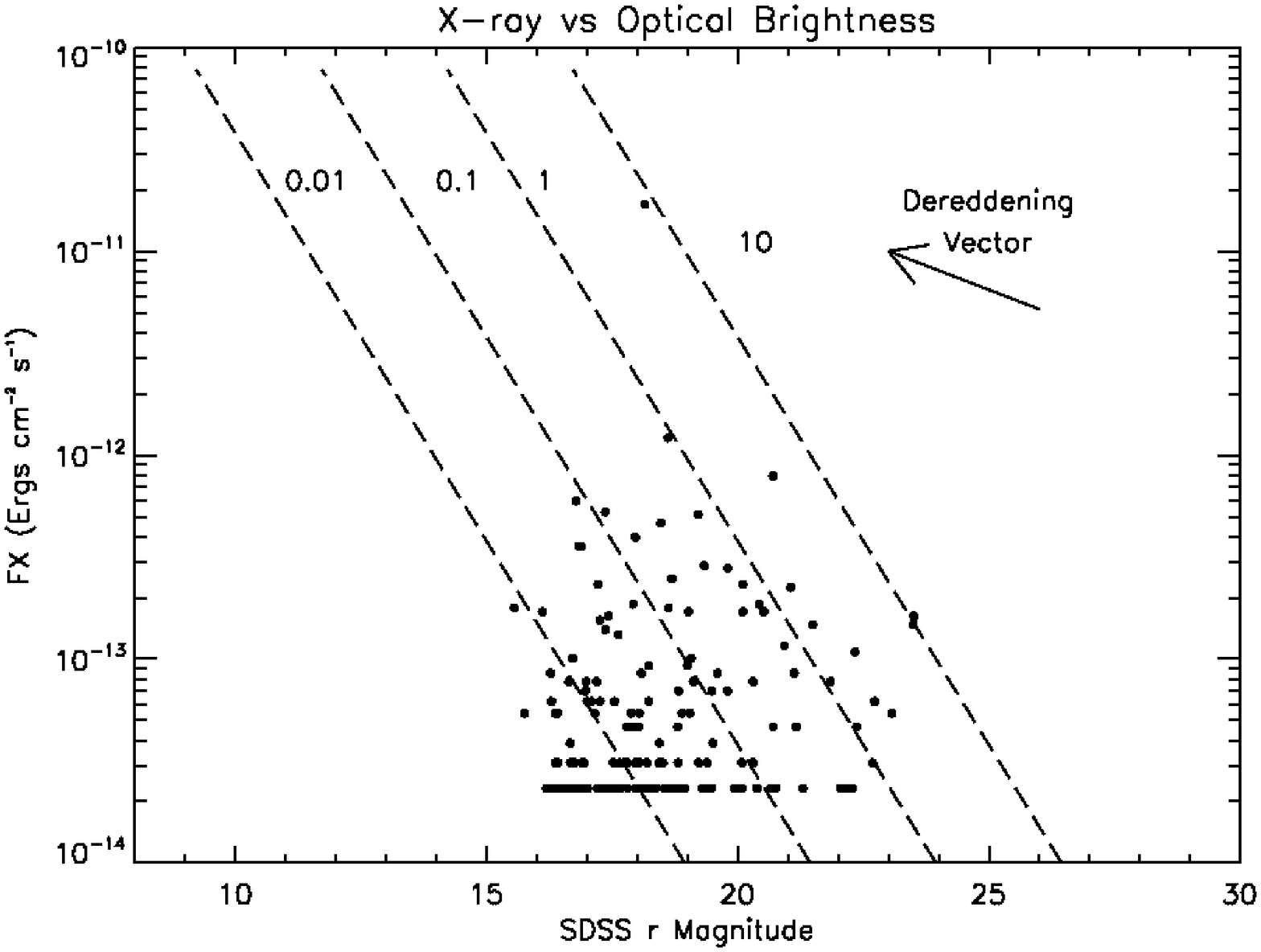}
  \caption{Each variable
    coincident with an X-ray source is plotted
    as magnitude vs flux. The observed values are plotted, with the direction of dereddening indicated by the arrow. The length of the arrow is arbitrary, as the amount of reddening at Bulge distance is heavily dependent on the line of sight. Each optical magnitude carries a systematic uncertainty of
    0.5 magnitudes as they were calibrated with the USNO catalog. Statistical
  errors for the X-ray flux are given as $\sigma_{N}=1+\sqrt{N+0.75}$
  \citep{Gehrels86}, though there
  are additional systematic uncertainties in the X-ray spectral shape.
  Uncertainties in extinction affect both X-ray and optical flux, with
  reddening moving sources in the direction indicated. The same
  information is plotted above
  and below, but the scale in the bottom figure has been matched to
  the population synthesis in \citet{Jonker11} in order to aid the reader
  in making comparisons.}
  \label{fig:fxfopt}
\end{center}
\end{figure}

We have not found $\sim11$ systems with $r'<21$ and
clear ellipsoidal variations with
periods below a day as predicted, or $\sim18$ systems with
$r'<21.5$. There are only 5 systems which show clear ellipsoidal
variations with or without eclipses, are unlikely to be W UMas, and that are not dynamically 
confirmed CVs (CX70, CX750, CX774, CX1060, CX1167), though there could be others
where the donor's gravity darkened inner face is heated through
reprocessing of X-rays enough to make the minima at $\phi=0$ and
$\phi=0.5$ appear to be of equal depths.  
There is a spike in the histogram at period of
3-4 hours, as is expected for CVs and the suspected population of short period 
qLMXBs, containing 7-10 sources. The sources with
periods above 1 day should be
largely RS CVns, though there are a few possible qLMXBs in this region as
well, e.g. CX855 (see Section \ref{sec:855}). Most RS CVns in the GBS are saturated in Mosaic-II
data, as extrapolated from the plot of X-ray versus optical luminosity
shown in Figure \ref{fig:fxfopt} in conjunction with the predictions
in \citet{Jonker11}. The fainter tail of the distribution is what we pick
up in our Mosaic-II data. Immediately of note in the differences between Figure
\ref{fig:fxfopt} and the predictions in Figure \ref{fig:rpopsynth} are the lack of X-ray 
bright LMXBs. There are only 2 active LMXBs in the GBS, CX1 and CX3. 
CX1 is one of the \offchip\ sources that falls off of the chip in every 
Mosaic-II observation. CX3 is optically faint and near brighter stars, and no 
significant variability could be recovered for it. The one object 
above $F_{X}=2\times10^{-12}\, {\rm ergs \, cm^{-2} \, s^{-1}}$ that appears 
as a variable is CX2, an AGN \citep{Maccarone12}. We also have contamination from W UMas,
which have periods $< 1$ day and these sources show variations in their
lightcurves that mimic ellipsoidal
variations. Short period ellipsoidal variations with superposed flickering 
or night to night variations in level are likely to be accreting compact binaries. 
W UMas likely comprise a large number of sources with
periods below a day, but in some cases they cannot be readily distinguished from
compact binaries without spectroscopy, especially since BH
qLMXBs can have quite low X-ray to optical flux ratios. With the large
errors in estimating this ratio, it is not necessarily a reliable
diagnostic on its own for differentiating between BH qLMXBs showing
ellipsoidal variations and W UMas. 

Though individual sources are difficult to classify exactly without
spectroscopy, we can make probabalistic statements about the
population as a whole.
The total number of GBS X-ray sources closely matches the number
predicted by the population model in \citet{Jonker11}. That
population model also predicted that $\sim400$ sources in the
region covered by Mosaic-II would not have an optical counterpart in
our variability survey. We found \noc\ sources without any visible star
inside the X-ray error circle in Sloan $r'$, while a further \nv\
objects had at least one star inside the error circle which was not
seen to vary. It is certain, given the stellar density in the region,
that many of these non-variable stars are
not the optical counterpart to the X-ray source, but rather are field
stars. The crowding in our fields in the Bulge region is, on average,
high enough that coincident alignments are expected, especially for
sources observed in the X-ray at a large off-axis angle. Some X-ray
sources observed far off-axis and with few counts have confidence
regions $10''$ across, with tens of candidate
counterparts. $256$ X-ray sources have an uncertainty in the position of $\ge
1.9''$, which admits probable chance alignment with at least one field
star. It is therefore possible that
the expected remaining
sources without optical counterparts, primarily AGN and including many
UCXBs, IPs, and qLMXBs could be ``hiding'' in the group
of non-variables. Still, some nonvariable objects have been identified through
spectroscopy to be interacting binaries, e.g. CX561 \citep{Britt13} and 
CX1004 \citep{Torres14}. It is also possible that these objects are in
fact variable, but less than what we can detect at a level considered
significant here.

\subsection{Active Stars}

RS CVns are much more common than LMXBs. They are comparatively well
studied, and there are far fewer uncertainties in estimates of their
population compared to qLMXBs. \citet{Jonker11} predicted $\approx447$ 
RS CVns in the GBS area covered by Mosaic-II, all with optical counterparts 
as they are relatively 
optically bright and X-ray faint. They also show no detectable
intrinsic aperiodic variations, apart from occasional flares, but instead a 
smooth lightcurve that
is a combination of variations on the same period produced by star 
spots which the Lomb-Scargle statistic is excellent at finding. 
We find \rscvns\ sources with periods above a day, and \lpv\ sources with what 
could be periodic behavior with periods above 8 days.
Based on Monte Carlo simulations with
noise on the scale of relative photometric errors, we estimate that we have
recovered periods for $\sim70\%$ of unsaturated RS CVns with periods
less than 8 days and suspected long period variables, which implies
that $~40$ RS CVns are unsaturated. Referring to the population estimate in
Figure \ref{fig:rpopsynth}, most RS CVns should indeed be saturated in our
Mosaic-II data. Because we only have \saturated\ sources aligned with a
saturated star which are likely RS CVns \citep[see]{Hynes12}, and only a 
further \chancesat\ sources either on bleed trails 
or close enough to a saturated star for photometry to be impossible, we 
have detected at maximum somewhat fewer RS CVns in the Mosaic-II region than 
the 447 expected, though the exact number fewer remains uncertain from these 
data.

We also find \shortper\ candidate counterparts with periods below a day. 
A similar estimate to that above with W UMas shows that,
 since there is no instrinsic flickering in W UMa systems on a scale
 that we can detect as there is in CVs and 
qLMXBs, we expect to recover nearly all of the periods for W UMas
which do not saturate. Crudely assuming that the spike in the period histogram
at 3-4 hours is due entirely to either CVs or qLMXBs and that W UMas
are responsible for all other detected periods below a day, we estimate
that we observe $\sim\WUMa$ unsaturated W UMa systems. 
Using the population estimate in Figure \ref{fig:rpopsynth}, we expect most W UMas 
in the GBS to saturate in Mosaic-II data.

Also among the nonvariables and saturated stars are coronally active single 
stars \citep{Hynes12}. These are typically faint X-ray sources, especially 
compared to their optical luminosity. While most main sequence stars with 
coronal emission should have $\frac{L_{X}}{L_{opt}} << 1$, M dwarfs peak in 
luminosity in infrared wavelengths such that $L_{opt} << L_{Bol}$. It is 
unsurprising, therefore, that this optically deep dataset contains 
15 objects that appear to be flare stars, showing small flares of $<1$ 
magnitude that decay over a few hours. Some of these also show small amplitude
sinusoidal periods which are consistent with sunspots on flare stars, e.g. 
CX161 and CX1220. 

\subsection{CVs and LMXBs}

We have 
76 sources which flicker without a large outburst or
periodicity. These are likely a mix of IPs, CVs, and qLMXBs. The data presented here are largely ambiguous in differentiating between BH and NS qLMXBs, and we treat their populations taken together as the number of BH qLMXBs should equal $\approx10\%$ of NS qLMXBs. Many IPs
in this optical dataset will not have visible
counterparts. The initial estimate of visible IPs in the survey area is
sensitive to the optical magnitude we reach. Referring to
Figure \ref{fig:rpopsynth}, most X-ray detected IPs are predicted to
be fainter than $r'=23$. We have difficulty
securely detecting variability below 0.2 magnitudes fainter
than $r'=22$, so a larger
proportion of IPs may be hiding among the sources with non-variable or
non-detected counterparts. Because of their high X-ray luminosity at
high accretion rate, the sources that show dramatic aperiodic
flickering of a magnitude or more are more likely to be IPs than
quiescent systems, but even quiescent CVs and qLMXBs can show large
amounts of flickering \citep{Jonker08}.

Eight counterparts to X-ray sources in our Mosaic-II data were observed to
undergo what appear to be dwarf nova outbursts. These are CX18, CX39, CX81, 
CX87, CX298, CX426, CX476, and CX860. One of these, CX18,
was confirmed through follow up photometric data and with
spectroscopic observations \citep{Britt13}. A ninth object, CX982, shows a 
candidate dwarf nova, but the unusual morphology of this object's lightcurve 
demands confirmation.

Quiescent CVs and qLMXBs can also show flickering, though
CVs and LMXBs in quiescence can also have very smooth lightcurves, for
example in the case of CX93 \citep{Ratti13}. Most of these systems
should have a period below a day, concentrated around a few to several
hours \citep{Gansicke05}, where indeed we see a spike in our period
histogram. Under the
assumption that qLMXBs show clear ellipsoidal modulations, we can
place a limit of 5 sources (CX70, CX750, CX774, CX1060, CX1167) that could be qLMXBs in our optical data
set, as one of the \ellip\ sources showing ellipsoidal modulations, CX93, 
is confirmed
as a CV. Some qLMXBs, though, are doubtless flickering objects without a recovered
period. Indeed, a further 19 GBS objects discussed in this work and in
\citet{Britt13,Torres14,Jianfeng14} could potentially be qLMXBs. It is
notoriously difficult to distinguish between quiescent CVs and qLMXBs, however,
and these 24 objects will contain a mix of both. For 17 of these 24,
spectroscopic data are still to be obtained or analyzed. (For many
objects which could be candidate qLMXBs from photometry alone,
spectroscopy suggests a CV \citep{Torres14}). 
Only 5 possible qLMXBs have evident ellipsoidal modulations, out
of the $\approx18$ qLMXBs which should be present going by Figure
\ref{fig:rpopsynth}, and
all 5 are potentially CVs as well. Although, if our baseline
in these data were longer, it is likely that some sources which
appear to flicker could be shown to have ellipsoidal variations
underlying the flickering. For example, CX19 is periodic in OGLE-IV data but
flickers enough that our observations did not independently recover a period.
Following up these selected possible qLMXBs
to confirm the expected presence of ellipsoidal modulations would
provide a critical check for what large future surveys should
detect. The simple assumption that qLMXBs should have clear
ellipsoidal modulations does not hold for a large fraction of
objects, at least those which are X-ray selected. 

\subsection{Comparison to Population Synthesis models}

The ratio of quiescent LMXBs to active LMXBs in Globular Clusters has
been observed to be $\approx 10$ \citep{Heinke03}, though these studies may
underestimate the number of qLMXBs by half \citep{Heinke05}. There is an observed deficit of BH LMXBs in Globular Clusters \citep{Kalogera04}. The
Galactic Bulge may not produce qLMXBs with the same duty cycle as in
Globular Clusters, however, because those in Globular Clusters
are formed predominantly by different mechanisms than in the
Bulge, in addition to having older donor stars in general to binaries in the
Milky Way. It is perhaps worth noting that we observe 2 active LMXBs in
the GBS (CX1, CX3), which would imply $\sim40$ qLMXBs in the
Mosaic-II data if Bulge
LMXBs have the same quiescent to active proportions as those in
Globular Clusters, of which $\sim10$ should be detectable in the optical at $r'<23$. 

The prediction actually used in \citet{Jonker11} relied upon X-ray binary
formation rates and lifetimes. The estimates of \citet{Jonker11} used a
formation rate of $10^{-5}$ NS binaries formed per year
\citep{Zwart97,Kiel06}. Using a typical lifetime for LMXB systems of 1 Gyr
yields $10^{4}$ LMXB systems in the Galaxy \citep{Jonker11}, which combined
with 140 active LMXBs at a given time, yields a ratio of quiescent to
active LMXBs of 70. \citet{Pfahl03} predict $\approx10^{3}-10^{5}$
LMXBs by assuming many LMXBs are descendants of Intermediate Mass
X-ray Binaries, with the major driver of the different population
sizes being a substantial dependence on the
structure of the Common Envelope. The population estimate in
\citet{Jonker11} fits in the center of this range. The period
distribution predicted by \citet{Pfahl03} for the median CE parameter
peaks around $\log P_{orb} {\rm (days)}=0.25$,
which, referring to Figure \ref{fig:montecarloperiod}, we are only
likely to recover with flickering of $20\%$ of the amplitude of orbital 
variability or less. 

The favored
population model of \citet{Kiel06}, however, predicts a somewhat lower
number of 1900 LMXBs, both active and quiescent, with a ratio between
the two closer to 13. The primary differences between the models in
\citet{Kiel06} and \citet{Pfahl03} are an updated model for the CE, an
inclusion of the helium star's mass-loss from wind after the CE phase,
metallicity of the binary, and the inclusion of tidal forces in the
code. The resulting population of
LMXBs would lead to $\sim17$
qLMXBs with optical counterparts in the GBS region surveyed with the
Mosaic-II instrument ($\sim6$ with $r'<21.5$) by using the same luminosity and 
extinction assumptions as in \citet{Jonker11}, which
matches well with the number of ellipsoidal variables identified, although they remain unconfirmed as qLMXBs in this work.

If every object that could be a qLMXB based on the photometric results presented here were indeed to turn out to be a qLMXB rather than a CV, it is still difficult to support the presence of $10^5$ qLMXBs in the Milky Way. It is possible for the qLMXB population in the Milky Way to be somewhat larger than we find it for the following reasons: 
\begin{itemize}
\item The initial estimate in \citet{Jonker11} may have overestimated the X-ray
  luminosity of qLMXBs. 
  If quiescent neutron star LMXBs that have not
  undergone a recent outburst are fainter than the known population of
  LMXBs, which have been followed into quiescence from outburst, 
  they might not have been detected in the X-ray. 
\item Some qLMXBs with optical counterparts may be too faint to notice
  flickering or variability. A source with ellipsoidal modulations
  like the CV CX93 \citep{Ratti13} would likely
  be overlooked as a variable at $r'=22$ because the photometric errors
  are on par with the amplitude of the variability. We have attempted to
  control for this by limiting our analysis to the predicted population with 
  counterparts brighter than $r'=21.5$.
\item Some qLMXBs could also be intrinsically nonvariable, showing no flickering
  or ellipsoidal modulations. A candidate BH qLMXB based on
  spectroscopy, CX561, shows no variability despite being at
  $r'=20$. If this is the case, our complementary spectroscopic
  survey \citep[e.g.][]{Torres14} will be the best way to find qLMXB 
  systems.
\item Some qLMXBs may flicker too much ($\gae 50\%$ of periodic variations) to 
  recover periods reliably. This
  cannot bring us up to a Galactic population as high as $10^5$ qLMXBs for 
  several reasons. First, as shown through Monte Carlo
  simulations in Figure \ref{fig:montecarloperiod}, we can reliably find periods
  below 1 day even with intrinsic flickering up to $50\%$. Second, IPs
  in the survey should be flickering, and the observed number of
  flickering sources roughly matches the number of expected IPs. While
  a few qLMXBs are likely hiding in this
  population, there are not enough flickering sources to account for
  {\it both} the expected population of IPs and a large unseen population 
  of qLMXBs. This seems at best a partial solution.
\item The initial population estimate in \citet{Jonker11} assumed that 
  the spatial distribution of qLMXBs followed the distribution of stars in
  the Milky Way. Kicks imparted to the system by both mass loss (Blaauw kicks)
  and by asymmetry in supernovae should increase the velocity dispersion of
  qLMXBs and therefore extend the scale height of the distribution above and
  below the Galaxy. If these kicks are large enough, the column density of
  qLMXBs will noticeably decline. The spatial distribution of qLMXBs 
  above and below the Galactic plane remains poorly understood because of
  selection effects in the systems we have discovered, but it is clear that
  known systems do not exactly trace stellar density but instead more than
  half are found more than $0.5\,$kpc above or below the Plane, with almost 
  $20\%$ of BH transients found between $1\,{\rm kpc} < z < 1.5\,{\rm kpc}$ 
  above or below the Galactic Plane \citep{Jonker04,Repetto12}. It is unclear
  whether a population discovered in quiescence would have the same observed
  distribution.
\end{itemize}
The lower to median theoretical estimates of $10^3-10^4$ fit reasonably well the current population as we have found it, while spectroscopic campaigns currently underway will better define the discovered population of qLMXBs in the Milky Way.

\begin{acknowledgements}
This work was supported by the National Science Foundation under Grant No. 
AST-0908789, by the Louisiana Board of Regents Fellowship, by the 
NAS/Louisiana Board
of Regents grant NNX07AT62A/LEQSF(2007-2010) Phase 3-02. P.G.J. and M.A.P.T. 
acknowledge support from the Netherlands Organisation for Scientific Research. 
This research has made use of NASA's Astrophysics Data System Bibliographic 
Services and of SAOImage DS9, developed by Smithsonian Astrophysical 
Observatory. T.J.M. thanks Reba Bandyopadhyay for useful discussions,
especially illustrating the possibility of SyXBs.
\end{acknowledgements}

\bibliographystyle{apj}
\bibliography{Britt}

\LongTables
\input{vartable.tex}

\end{document}

%% file: vartable.tex
\begin{longtable*}{l l l l l l c c c c c c c c}
\caption{Each GBS source with a likely variable optical counterpart. The columns are as follows: (1) Catalog ID (2) Right Ascension (3) Declination (4) $r'$ magnitude (5) Average relative photometric error (6) RMS variation of the lightcurve (7) Distance of variable star from X-ray position (8) $95\%$ confidence radius of X-ray position (9) Number of observations in the lightcurve (10) Log of the X-ray to optical flux ratio without correcting for extinction (11) $E(B-V)$ for the line of sight at Bulge distance using the reddening maps from \citet{Gonzalez12} (12) X-ray to optical flux ratio after correcting for reddening, assuming Bulge distance (13) X-ray hardness ratio as calculated in \citet{Jonker11}, only presented for sources with $>20$ photons (14) False Alarm Probability for a variable star at the given distance from the X-ray position, as calculated in Section \ref{sec:results}.}
\label{table:varstats} \\
CX ID & RA & DEC & $r'$ mag & $<{\rm error}>$ & RMS & Distance & $95\%$ CR & N & Absorbed  & $E(B-V)$ & $\log\frac{F_{X}}{F_{opt}}$ & HR & FAP \\
  &   &   &    &     &    & (arcsec) & (arcsec) &  & $\log\frac{F_{X}}{F_{opt}}$ &   &  &  &  \\
\smallskip \\
\hline
\endfirsthead
CX ID & RA & DEC & $r'$ mag & $<{\rm error}>$ & RMS & Distance & $95\%$ CR & N & Absorbed  & $E(B-V)$ & $\log\frac{F_{X}}{F_{opt}}$ & HR & FAP \\
  &   &   &    &     &    & (arcsec) & (arcsec) &  & $\log\frac{F_{X}}{F_{opt}}$  &   &  &  & \\
\smallskip \\
\hline
\endhead \\
\hline
\multicolumn{13}{|r|}{Continued on Next Page} \\
\hline
\endfoot \\
\hline
\endlastfoot
      2  & 264.36832  & -29.13384  & 18.14  & 0.004  & 0.027  & 0.192  & 0.839  &      36  &  1.9  & 1.72  &  0.3  &  0.00 &  2.12E-05 \\
\smallskip \\
      5  & 265.03806  & -28.79050  & 18.61  & 0.005  & 0.084  & 0.162  & 0.802  &      28  &  1.0  & 2.73  & -1.6  &  0.50 &  8.74E-06 \\
\smallskip \\
     11  & 265.46423  & -27.03995  & 20.70  & 0.025  & 0.164  & 1.603  & 1.605  &      33  &  1.6  & 2.10  & -0.4  &  0.64 &  9.55E-03 \\
\smallskip \\
     18  & 264.89896  & -27.49324  & 17.36  & 0.003  & 0.577  & 0.406  & 1.017  &      19  &  0.0  & 1.45  & -1.3  & -0.20 &  2.84E-04 \\
\smallskip \\
     19  & 267.47760  & -29.72652  & 19.21  & 0.007  & 0.243  & 0.763  & 1.133  &      20  &  0.8  & 1.99  & -1.0  &  0.89 &  1.67E-03 \\
\smallskip \\
     21  & 265.39072  & -28.67623  & 18.46  & 0.004  & 0.371  & 0.703  & 1.440  &      18  &  0.5  & 2.64  & -2.0  & -0.38 &  1.30E-03 \\
\smallskip \\
     23  & 265.63152  & -27.73004  & 17.96  & 0.003  & 0.025  & 0.983  & 1.115  &      36  &  0.2  & 2.06  & -1.7  &  0.39 &  3.28E-03 \\
\smallskip \\
     28  & 264.94583  & -27.30242  & 16.89  & 0.002  & 0.204  & 0.803  & 0.892  &      28  & -0.3  & 1.47  & -1.6  &  0.50 &  2.12E-03 \\
\smallskip \\
     29  & 268.42447  & -28.06488  & 16.83  & 0.001  & 0.036  & 0.520  & 0.893  &      19  & -0.3  & 2.75  & -2.9  & -0.14 &  5.92E-04 \\
\smallskip \\
     37  & 264.37158  & -29.46776  & 19.33  & 0.009  & 0.187  & 0.689  & 1.424  &      33  &  0.6  & 1.89  & -1.1  &  0.90 &  1.23E-03 \\
\smallskip \\
     39  & 265.41681  & -27.29389  & 19.80  & 0.018  & 0.413  & 0.472  & 1.491  &      29  &  0.8  & 1.86  & -0.9  & -0.35 &  5.02E-04 \\
\smallskip \\
     44  & 268.92844  & -28.30259  & 18.68  & 0.004  & 0.055  & 0.469  & 1.461  &      19  &  0.2  & 1.48  & -1.1  & -0.19 &  4.90E-04 \\
\smallskip \\
     45  & 263.91068  & -28.88103  & 20.45  & 0.032  & 0.392  & 0.574  & 0.844  &      26  &  0.9  & 1.75  & -0.6  &  0.66 &  8.25E-04 \\
\smallskip \\
     48  & 263.99130  & -30.17292  & 17.21  & 0.002  & 0.010  & 0.277  & 1.018  &      32  & -0.4  & 2.06  & -2.2  &  0.37 &  9.97E-05 \\
\smallskip \\
     49  & 262.94528  & -30.05237  & 20.09  & 0.024  & 0.105  & 0.690  & 0.837  &      18  &  0.7  & 1.15  & -0.2  &  0.26 &  1.44E-03 \\
\smallskip \\
     54  & 265.07709  & -27.11973  & 21.05  & 0.048  & 0.207  & 0.653  & 0.884  &      19  &  1.1  & 1.52  & -0.2  &  0.17 &  1.19E-03 \\
\smallskip \\
     67  & 266.39377  & -26.31584  & 17.92  & 0.002  & 0.015  & 0.666  & 2.472  &      28  & -0.2  & 1.82  & -1.8  & -1.12 &  1.28E-03 \\
\smallskip \\
     70  & 263.89621  & -29.99487  & 20.42  & 0.025  & 0.082  & 1.319  & 2.411  &      35  &  0.8  & 1.60  & -0.6  &  0.63 &  5.69E-03 \\
\smallskip \\
     73  & 266.19797  & -27.01706  & 18.62  & 0.006  & 0.076  & 0.069  & 1.024  &      27  &  0.1  & 1.98  & -1.7  &  0.22 &  4.35E-06 \\
\smallskip \\
     74  & 266.22629  & -25.97759  & 15.54  & 0.001  & 0.071  & 0.152  & 0.847  &      17  & -1.2  & 1.43  & -2.4  & -0.48 &  1.08E-05 \\
\smallskip \\
     80  & 267.03888  & -26.02062  & 16.11  & 0.001  & 0.025  & 0.165  & 1.050  &       9  & -0.9  & 3.19  & -3.9  & -0.40 &  3.02E-05 \\
\smallskip \\
     81  & 266.10968  & -27.32389  & 20.10  & 0.036  & 1.503  & 0.496  & 1.699  &      18  &  0.7  & 2.47  & -1.6  & -0.50 &  5.95E-04 \\
\smallskip \\
     83  & 266.34747  & -31.51387  & 20.52  & 0.022  & 0.061  & 0.748  & 0.811  &      26  &  0.9  & 3.00  & -2.0  &  0.86 &  1.84E-03 \\
\smallskip \\
     84  & 264.55350  & -29.10341  & 19.02  & 0.007  & 0.029  & 0.057  & 0.847  &      35  &  0.2  & 1.98  & -1.6  & -0.27 &  8.64E-07 \\
\smallskip \\
     87  & 264.20071  & -29.61090  & 23.50  & 0.248  & 1.625  & 0.318  & 1.292  &      22  &  2.0  & 1.81  &  0.3  &  0.31 &  1.84E-04 \\
\smallskip \\
     88  & 263.28349  & -29.63126  & 17.42  & 0.002  & 0.008  & 0.481  & 0.901  &      16  & -0.5  & 1.62  & -1.9  & -0.74 &  4.63E-04 \\
\smallskip \\
     93  & 266.18674  & -26.05830  & 17.25  & 0.003  & 0.031  & 0.587  & 1.005  &      35  & -0.5  & 1.61  & -2.0  &  0.00 &  8.26E-04 \\
\smallskip \\
     96  & 266.73950  & -30.80195  & 23.49  & 0.197  & 0.612  & 0.324  & 1.398  &      37  &  2.0  & 3.34  & -1.2  &  0.00 &  2.05E-04 \\
\smallskip \\
    102  & 268.33871  & -28.53332  & 21.49  & 0.041  & 0.111  & 0.718  & 0.919  &      37  &  1.1  & 1.32  & -0.0  &  0.00 &  1.54E-03 \\
\smallskip \\
    105  & 266.63806  & -31.37048  & 17.36  & 0.002  & 0.040  & 1.807  & 2.815  &      19  & -0.5  & 2.96  & -3.3  &  0.00 &  1.14E-02 \\
\smallskip \\
    118  & 264.70923  & -28.80241  & 17.62  & 0.003  & 0.169  & 0.391  & 1.643  &      34  & -0.4  & 1.98  & -2.2  &  0.00 &  3.47E-04 \\
\smallskip \\
    128  & 265.11829  & -27.19329  & 20.92  & 0.044  & 0.259  & 0.301  & 0.833  &      18  &  0.8  & 1.55  & -0.5  &  0.00 &  9.52E-05 \\
\smallskip \\
    142  & 266.01562  & -31.38474  & 22.33  & 0.088  & 0.177  & 0.442  & 2.222  &      16  &  1.5  & 4.22  & -2.7  &  0.00 &  5.23E-04 \\
\smallskip \\
    161  & 264.96777  & -28.57350  & 19.07  & 0.008  & 0.083  & 0.179  & 0.930  &      32  &  0.0  & 2.34  & -2.1  &  0.00 &  2.61E-05 \\
\smallskip \\
    168  & 266.72791  & -30.93660  & 16.71  & 0.001  & 0.031  & 0.398  & 1.053  &      37  & -0.9  & 3.03  & -3.7  &  0.00 &  2.73E-04 \\
\smallskip \\
    184  & 266.82608  & -30.96338  & 19.00  & 0.005  & 0.039  & 0.512  & 1.306  &      37  & -0.0  & 2.55  & -2.4  &  0.00 &  5.78E-04 \\
\smallskip \\
    190  & 263.93500  & -28.90008  & 18.23  & 0.005  & 0.042  & 0.861  & 1.094  &      36  & -0.4  & 1.72  & -1.9  &  0.00 &  2.34E-03 \\
\smallskip \\
    195  & 263.78143  & -29.31806  & 16.27  & 0.001  & 0.063  & 0.274  & 0.982  &      22  & -1.2  & 1.86  & -2.8  &  0.00 &  9.17E-05 \\
\smallskip \\
    196  & 267.51392  & -29.19727  & 18.08  & 0.005  & 0.036  & 1.258  & 2.089  &      17  & -0.4  & 3.41  & -3.7  &  0.00 &  5.15E-03 \\
\smallskip \\
    207  & 266.60535  & -26.52615  & 21.12  & 0.064  & 0.880  & 1.222  & 4.295  &      18  &  0.8  & 2.07  & -1.1  &  0.00 &  5.02E-03 \\
\smallskip \\
    208  & 266.00229  & -27.50574  & 19.59  & 0.011  & 0.022  & 3.201  & 4.635  &      36  &  0.2  & 2.11  & -1.8  &  0.00 &  3.76E-02 \\
\smallskip \\
    209  & 265.21691  & -27.81823  & 21.12  & 0.043  & 0.070  & 0.848  & 0.904  &      36  &  0.7  & 1.59  & -0.6  &  0.00 &  2.44E-03 \\
\smallskip \\
    220  & 269.48364  & -27.38840  & 20.30  & 0.013  & 0.038  & 0.103  & 1.156  &      14  &  0.4  & 1.35  & -0.8  &  0.00 &  1.36E-05 \\
\smallskip \\
    222  & 266.02271  & -27.00365  & 16.63  & 0.001  & 0.008  & 0.254  & 0.832  &      26  & -1.1  & 1.91  & -2.8  &  0.00 &  5.28E-05 \\
\smallskip \\
    239  & 264.23932  & -28.94192  & 19.11  & 0.018  & 0.121  & 1.586  & 2.156  &      67  & -0.1  & 1.68  & -1.6  &  0.00 &  8.77E-03 \\
\smallskip \\
    243  & 268.71939  & -28.22630  & 16.98  & 0.001  & 0.003  & 0.903  & 2.072  &      36  & -0.9  & 1.57  & -2.3  &  0.00 &  2.40E-03 \\
\smallskip \\
    246  & 266.71359  & -30.75761  & 17.19  & 0.002  & 0.006  & 0.267  & 1.540  &      36  & -0.8  & 3.68  & -4.3  &  0.00 &  1.46E-04 \\
\smallskip \\
    247  & 265.61743  & -28.24335  & 21.85  & 0.070  & 0.138  & 0.481  & 1.353  &      18  &  1.0  & 2.26  & -1.0  &  0.00 &  5.02E-04 \\
\smallskip \\
    251  & 266.50726  & -31.87567  & 16.97  & 0.001  & 0.016  & 1.392  & 4.748  &      17  & -0.9  & 3.31  & -4.1  &  0.00 &  6.64E-03 \\
\smallskip \\
    279  & 264.86536  & -27.35311  & 18.82  & 0.010  & 0.223  & 0.324  & 1.675  &      18  & -0.3  & 1.40  & -1.5  &  0.00 &  2.35E-04 \\
\smallskip \\
    282  & 264.14355  & -28.89267  & 19.79  & 0.016  & 0.102  & 0.371  & 0.865  &      24  &  0.1  & 1.68  & -1.4  &  0.00 &  1.99E-04 \\
\smallskip \\
    291  & 267.63403  & -29.60579  & 19.47  & 0.014  & 0.085  & 1.017  & 1.755  &      35  &  0.0  & 1.59  & -1.4  &  0.00 &  3.18E-03 \\
\smallskip \\
    298  & 265.90741  & -26.27883  & 22.72  & 0.205  & 0.748  & 0.715  & 5.555  &      26  &  1.2  & 1.44  & -0.0  &  0.00 &  1.76E-03 \\
\smallskip \\
    300  & 263.46497  & -30.10851  & 16.29  & 0.001  & 0.007  & 0.756  & 1.074  &      20  & -1.3  & 2.08  & -3.2  &  0.00 &  1.66E-03 \\
\smallskip \\
    316  & 266.46586  & -31.58334  & 18.39  & 0.003  & 0.065  & 3.007  & 8.132  &      36  & -0.4  & 2.96  & -3.2  &  0.00 &  3.29E-02 \\
\smallskip \\
    321  & 266.02939  & -31.52280  & 18.56  & 0.003  & 0.011  & 0.077  & 1.166  &       8  & -0.3  & 3.98  & -4.2  &  0.00 &  7.65E-06 \\
\smallskip \\
    330  & 264.18280  & -28.35604  & 17.26  & 0.002  & 0.067  & 0.481  & 2.506  &      18  & -0.9  & 2.07  & -2.8  &  0.00 &  6.53E-04 \\
\smallskip \\
    331  & 264.09781  & -29.37620  & 18.23  & 0.005  & 0.037  & 0.242  & 1.230  &      33  & -0.5  & 1.71  & -2.1  &  0.00 &  9.38E-05 \\
\smallskip \\
    332  & 264.08429  & -29.56079  & 17.54  & 0.002  & 0.011  & 0.563  & 4.308  &      24  & -0.8  & 1.86  & -2.5  &  0.00 &  1.04E-03 \\
\smallskip \\
    335  & 263.97943  & -28.79802  & 17.08  & 0.002  & 0.023  & 2.498  & 5.794  &      12  & -1.0  & 1.59  & -2.4  &  0.00 &  2.24E-02 \\
\smallskip \\
    336  & 263.88098  & -30.40647  & 17.00  & 0.002  & 0.062  & 0.431  & 5.411  &      19  & -1.0  & 1.77  & -2.6  &  0.00 &  6.36E-04 \\
\smallskip \\
    346  & 264.42276  & -28.91604  & 15.75  & 0.001  & 0.023  & 0.515  & 0.907  &      25  & -1.6  & 1.71  & -3.1  &  0.00 &  5.72E-04 \\
\smallskip \\
    349  & 269.16055  & -27.09665  & 23.06  & 0.117  & 0.388  & 0.230  & 1.857  &      17  &  1.4  & 2.10  & -0.6  &  0.00 &  1.24E-04 \\
\smallskip \\
    354  & 268.46982  & -28.42028  & 16.41  & 0.001  & 0.089  & 0.894  & 2.718  &      23  & -1.4  & 1.39  & -2.6  &  0.00 &  2.44E-03 \\
\smallskip \\
    361  & 267.78189  & -29.67702  & 17.15  & 0.002  & 0.023  & 0.232  & 6.367  &      34  & -1.0  & 1.49  & -2.3  &  0.00 &  1.89E-04 \\
\smallskip \\
    368  & 266.68481  & -26.01658  & 18.04  & 0.004  & 0.088  & 0.508  & 0.888  &      34  & -0.7  & 1.74  & -2.2  &  0.00 &  5.51E-04 \\
\smallskip \\
    376  & 265.95166  & -26.53639  & 16.36  & 0.001  & 0.013  & 1.524  & 2.125  &      35  & -1.4  & 1.54  & -2.7  &  0.00 &  8.02E-03 \\
\smallskip \\
    377  & 265.81891  & -27.76027  & 18.89  & 0.009  & 0.060  & 0.461  & 1.918  &      36  & -0.3  & 2.01  & -2.1  &  0.00 &  5.36E-04 \\
\smallskip \\
    399  & 268.72684  & -27.58814  & 17.88  & 0.002  & 0.015  & 0.284  & 2.175  &      38  & -0.7  & 2.21  & -2.7  &  0.00 &  2.09E-04 \\
\smallskip \\
    411  & 264.29290  & -29.43636  & 18.02  & 0.003  & 0.026  & 5.884  & 6.035  &      34  & -0.7  & 1.90  & -2.4  &  0.00 &  1.27E-01 \\
\smallskip \\
    420  & 268.54672  & -27.72951  & 18.80  & 0.005  & 0.041  & 0.340  & 3.807  &      19  & -0.4  & 2.43  & -2.7  &  0.00 &  3.68E-04 \\
\smallskip \\
    426  & 268.15021  & -29.32773  & 17.99  & 0.003  & 0.320  & 0.044  & 0.923  &      19  & -0.8  & 1.37  & -2.0  &  0.00 &  1.08E-06 \\
\smallskip \\
    437  & 267.51099  & -30.32872  & 22.36  & 0.123  & 0.798  & 0.486  & 1.105  &      17  &  1.0  & 1.79  & -0.6  &  0.00 &  4.82E-04 \\
\smallskip \\
    439  & 267.16891  & -30.93149  & 20.70  & 0.013  & 0.142  & 3.032  & 3.328  &      18  &  0.3  & 2.15  & -1.6  &  0.00 &  3.45E-02 \\
\smallskip \\
    446  & 266.61322  & -25.83115  & 21.16  & 0.062  & 0.166  & 0.293  & 1.139  &      37  &  0.5  & 1.56  & -0.9  &  0.00 &  1.33E-04 \\
\smallskip \\
    450  & 266.47653  & -31.35476  & 17.78  & 0.003  & 0.060  & 0.192  & 1.428  &      26  & -0.8  & 3.19  & -3.8  &  0.00 &  6.81E-05 \\
\smallskip \\
    476  & 264.77896  & -28.56475  & 23.13  & 0.275  & 1.039  & 0.717  & 2.776  &      33  &  1.3  & 2.12  & -0.6  &  0.00 &  1.54E-03 \\
\smallskip \\
    484  & 264.33148  & -29.55660  & 17.85  & 0.003  & 0.018  & 0.481  & 2.786  &      33  & -0.8  & 1.88  & -2.5  &  0.00 &  6.77E-04 \\
\smallskip \\
    490  & 263.95304  & -30.06397  & 18.80  & 0.012  & 0.367  & 2.512  & 3.340  &      33  & -0.4  & 2.13  & -2.4  &  0.00 &  2.31E-02 \\
\smallskip \\
    536  & 267.81430  & -29.41001  & 16.66  & 0.001  & 0.016  & 1.962  & 2.117  &      14  & -1.4  & 1.80  & -3.0  &  0.00 &  1.43E-02 \\
\smallskip \\
    585  & 265.95374  & -31.41655  & 19.50  & 0.010  & 0.511  & 3.052  & 6.715  &      16  & -0.1  & 3.80  & -3.8  &  0.00 &  3.38E-02 \\
\smallskip \\
    645  & 266.63937  & -26.38723  & 18.44  & 0.006  & 0.632  & 0.526  & 1.136  &      21  & -0.6  & 2.03  & -2.5  &  0.00 &  6.01E-04 \\
\smallskip \\
    691  & 268.35190  & -28.45276  & 16.76  & 0.001  & 0.006  & 2.976  & 3.505  &      10  & -1.5  & 1.28  & -2.5  &  0.00 &  3.30E-02 \\
\smallskip \\
    705  & 267.81726  & -29.39559  & 17.04  & 0.002  & 0.062  & 0.800  & 1.831  &      38  & -1.3  & 1.80  & -2.9  &  0.00 &  1.80E-03 \\
\smallskip \\
    718  & 267.45114  & -30.20607  & 19.78  & 0.008  & 0.151  & 4.125  & 12.372  &       7  & -0.2  & 1.85  & -1.9  &  0.00 &  6.25E-02 \\
\smallskip \\
    730  & 266.97705  & -30.14194  & 17.96  & 0.001  & 0.021  & 0.038  & 3.693  &      34  & -0.8  & 4.29  & -5.0  &  0.00 &  4.53E-06 \\
\smallskip \\
    740  & 266.62271  & -30.77737  & 18.19  & 0.003  & 0.013  & 0.692  & 3.185  &      25  & -0.7  & 3.93  & -4.6  &  0.00 &  1.48E-03 \\
\smallskip \\
    741  & 266.60675  & -30.79257  & 16.73  & 0.002  & 0.017  & 1.452  & 5.037  &      21  & -1.3  & 4.17  & -5.4  &  0.00 &  7.29E-03 \\
\smallskip \\
    744  & 266.57233  & -31.56136  & 17.52  & 0.002  & 0.042  & 0.188  & 2.513  &      19  & -1.1  & 2.63  & -3.5  &  0.00 &  9.72E-05 \\
\smallskip \\
    748  & 266.53049  & -30.89473  & 18.43  & 0.003  & 0.012  & 0.145  & 0.954  &      26  & -0.7  & 3.41  & -4.0  &  0.00 &  1.70E-05 \\
\smallskip \\
    750  & 266.50717  & -31.20721  & 18.14  & 0.003  & 0.114  & 6.430  & 11.876  &      29  & -0.8  & 3.35  & -4.0  &  0.00 &  1.47E-01 \\
\smallskip \\
    774  & 266.00571  & -27.27881  & 19.78  & 0.017  & 0.102  & 1.909  & 9.911  &      38  & -0.2  & 2.09  & -2.1  &  0.00 &  1.34E-02 \\
\smallskip \\
    781  & 265.79639  & -27.27264  & 17.89  & 0.004  & 0.087  & 0.613  & 5.157  &      38  & -1.0  & 1.87  & -2.6  &  0.00 &  1.28E-03 \\
\smallskip \\
    792  & 265.53076  & -27.57881  & 17.64  & 0.003  & 0.012  & 1.095  & 1.863  &      34  & -1.1  & 1.68  & -2.6  &  0.00 &  3.77E-03 \\
\smallskip \\
    794  & 265.42651  & -27.97504  & 20.08  & 0.018  & 0.496  & 0.579  & 2.965  &      28  & -0.1  & 1.97  & -1.9  &  0.00 &  1.01E-03 \\
\smallskip \\
    800  & 265.21494  & -27.76656  & 18.81  & 0.009  & 0.060  & 0.286  & 0.952  &      36  & -0.6  & 1.64  & -2.1  &  0.00 &  9.84E-05 \\
\smallskip \\
    816  & 264.80682  & -28.05227  & 17.80  & 0.004  & 0.022  & 2.149  & 2.563  &      30  & -1.0  & 1.52  & -2.3  &  0.00 &  1.69E-02 \\
\smallskip \\
    818  & 264.75804  & -28.04725  & 16.42  & 0.002  & 0.010  & 3.390  & 3.395  &      37  & -1.6  & 1.55  & -2.9  &  0.00 &  4.36E-02 \\
\smallskip \\
    819  & 264.73523  & -27.80927  & 18.04  & 0.005  & 0.025  & 0.277  & 1.762  &      28  & -0.9  & 1.45  & -2.2  &  0.00 &  1.75E-04 \\
\smallskip \\
    820  & 264.73431  & -28.18854  & 17.53  & 0.003  & 0.015  & 0.395  & 2.206  &      19  & -1.1  & 1.52  & -2.4  &  0.00 &  4.13E-04 \\
\smallskip \\
    837  & 264.19843  & -28.37271  & 18.51  & 0.006  & 0.020  & 2.468  & 9.475  &      18  & -0.7  & 2.09  & -2.6  &  0.00 &  2.23E-02 \\
\smallskip \\
    838  & 264.17538  & -29.13022  & 16.89  & 0.002  & 0.008  & 1.736  & 1.774  &      25  & -1.4  & 1.61  & -2.8  &  0.00 &  1.12E-02 \\
\smallskip \\
    847  & 263.90744  & -29.93362  & 16.94  & 0.002  & 0.011  & 0.827  & 1.081  &      35  & -1.3  & 1.85  & -3.0  &  0.00 &  2.11E-03 \\
\smallskip \\
    852  & 263.75925  & -30.06507  & 16.66  & 0.013  & 0.030  & 0.434  & 1.374  &      33  & -1.5  & 1.55  & -2.8  &  0.00 &  3.96E-04 \\
\smallskip \\
    853  & 263.69562  & -29.94207  & 16.71  & 0.002  & 0.051  & 0.401  & 1.247  &      32  & -1.5  & 1.48  & -2.8  &  0.00 &  3.09E-04 \\
\smallskip \\
    855  & 263.56445  & -30.16605  & 19.22  & 0.009  & 0.038  & 1.095  & 2.257  &      36  & -0.4  & 1.93  & -2.2  &  0.00 &  3.74E-03 \\
\smallskip \\
    858  & 263.36209  & -29.70317  & 16.36  & 0.001  & 0.027  & 0.393  & 2.070  &      26  & -1.6  & 1.48  & -2.9  &  0.00 &  3.97E-04 \\
\smallskip \\
    860  & 263.32257  & -30.26443  & 22.68  & 0.247  & 1.016  & 0.255  & 1.487  &      18  &  0.9  & 1.30  & -0.2  &  0.00 &  1.29E-04 \\
\smallskip \\
    870  & 265.96838  & -26.03933  & 20.30  & 0.030  & 0.128  & 0.445  & 1.890  &      36  & -0.0  & 1.58  & -1.4  &  0.00 &  4.94E-04 \\
\smallskip \\
    881  & 267.17218  & -30.19883  & 17.47  & 0.002  & 0.013  & 1.964  & 4.884  &      36  & -1.2  & 2.49  & -3.5  &  0.00 &  1.35E-02 \\
\smallskip \\
    884  & 266.71182  & -30.89109  & 22.03  & 0.083  & 0.136  & 0.194  & 1.904  &      37  &  0.6  & 3.31  & -2.5  &  0.00 &  8.89E-05 \\
\smallskip \\
    887  & 266.12207  & -26.05991  & 20.65  & 0.030  & 0.123  & 1.285  & 3.822  &      23  & -0.0  & 1.53  & -1.3  &  0.00 &  5.50E-03 \\
\smallskip \\
    895  & 263.52255  & -29.71214  & 16.53  & 0.002  & 0.028  & 0.810  & 1.119  &      34  & -1.7  & 1.47  & -3.0  &  0.00 &  1.97E-03 \\
\smallskip \\
    897  & 269.41718  & -27.38035  & 22.08  & 0.068  & 0.127  & 0.385  & 2.190  &      17  &  0.5  & 1.46  & -0.7  &  0.00 &  3.91E-04 \\
\smallskip \\
    905  & 269.10236  & -28.01431  & 20.37  & 0.022  & 0.162  & 1.691  & 3.774  &      12  & -0.2  & 1.08  & -1.1  &  0.00 &  9.78E-03 \\
\smallskip \\
    909  & 269.04129  & -27.58816  & 17.22  & 0.002  & 0.016  & 0.036  & 2.051  &      28  & -1.3  & 2.33  & -3.5  &  0.00 &  3.17E-06 \\
\smallskip \\
    910  & 269.03125  & -27.95736  & 16.53  & 0.001  & 0.018  & 0.165  & 1.285  &      19  & -1.7  & 1.36  & -2.8  &  0.00 &  4.36E-05 \\
\smallskip \\
    926  & 268.70605  & -28.12682  & 18.76  & 0.004  & 0.021  & 0.369  & 1.189  &      18  & -0.7  & 1.86  & -2.4  &  0.00 &  2.44E-04 \\
\smallskip \\
    929  & 268.62851  & -28.55983  & 17.50  & 0.002  & 0.024  & 0.349  & 1.301  &      17  & -1.3  & 1.12  & -2.3  &  0.00 &  2.29E-04 \\
\smallskip \\
    930  & 268.62094  & -28.53156  & 20.38  & 0.019  & 0.308  & 2.318  & 1.275  &      17  & -0.2  & 1.14  & -1.1  &  0.00 &  2.13E-02 \\
\smallskip \\
    944  & 268.22636  & -28.27495  & 17.66  & 0.003  & 0.041  & 2.469  & 2.487  &      18  & -1.2  & 1.61  & -2.6  &  0.00 &  2.31E-02 \\
\smallskip \\
    953  & 268.01581  & -28.56406  & 16.83  & 0.001  & 0.017  & 0.440  & 6.009  &      17  & -1.5  & 2.27  & -3.6  &  0.00 &  6.74E-04 \\
\smallskip \\
    955  & 267.96418  & -28.71287  & 22.28  & 0.066  & 0.121  & 0.203  & 1.320  &      18  &  0.7  & 2.22  & -1.4  &  0.00 &  6.99E-05 \\
\smallskip \\
    957  & 267.90463  & -29.37149  & 18.76  & 0.009  & 0.031  & 0.807  & 1.565  &      36  & -0.8  & 1.59  & -2.1  &  0.00 &  1.83E-03 \\
\smallskip \\
    966  & 267.42773  & -30.04566  & 18.78  & 0.006  & 0.044  & 1.836  & 3.854  &      24  & -0.7  & 2.22  & -2.7  &  0.00 &  1.17E-02 \\
\smallskip \\
    971  & 267.35327  & -30.55797  & 20.07  & 0.012  & 0.072  & 0.422  & 2.051  &      12  & -0.2  & 1.69  & -1.7  &  0.00 &  4.59E-04 \\
\smallskip \\
    973  & 267.34003  & -30.12748  & 18.60  & 0.006  & 0.069  & 0.438  & 1.191  &      36  & -0.8  & 2.11  & -2.7  &  0.00 &  3.76E-04 \\
\smallskip \\
    978  & 267.24841  & -29.66144  & 18.33  & 0.003  & 0.007  & 0.723  & 1.161  &      18  & -0.9  & 2.98  & -3.7  &  0.00 &  1.43E-03 \\
\smallskip \\
    982  & 267.19183  & -30.66064  & 21.30  & 0.047  & 0.556  & 0.415  & 1.243  &      37  &  0.3  & 1.99  & -1.5  &  0.00 &  3.36E-04 \\
\smallskip \\
    985  & 267.03864  & -30.86705  & 18.23  & 0.002  & 0.031  & 0.608  & 1.177  &       9  & -0.9  & 2.18  & -2.9  &  0.00 &  8.92E-04 \\
\smallskip \\
    987  & 266.97559  & -31.07145  & 16.42  & 0.012  & 0.035  & 0.486  & 1.939  &      18  & -1.6  & 2.71  & -4.2  &  0.00 &  6.03E-04 \\
\smallskip \\
    988  & 266.91620  & -31.09168  & 17.68  & 0.002  & 0.091  & 0.272  & 0.992  &      19  & -1.1  & 2.76  & -3.7  &  0.00 &  9.13E-05 \\
\smallskip \\
    990  & 266.88498  & -30.33887  & 17.95  & 0.002  & 0.018  & 1.006  & 2.044  &      18  & -1.0  & 4.05  & -4.9  &  0.00 &  3.07E-03 \\
\smallskip \\
    995  & 266.75504  & -26.01271  & 19.96  & 0.014  & 0.193  & 0.625  & 1.637  &      32  & -0.3  & 1.90  & -2.0  &  0.00 &  9.95E-04 \\
\smallskip \\
    997  & 266.70963  & -31.25989  & 17.32  & 0.002  & 0.020  & 0.214  & 2.044  &      38  & -1.3  & 2.59  & -3.7  &  0.00 &  1.13E-04 \\
\smallskip \\
    999  & 266.70531  & -25.97215  & 17.32  & 0.003  & 0.031  & 0.895  & 1.104  &      32  & -1.3  & 1.84  & -3.0  &  0.00 &  2.58E-03 \\
\smallskip \\
   1004  & 266.59787  & -31.09710  & 20.76  & 0.019  & 0.040  & 0.550  & 1.246  &      35  &  0.2  & 4.01  & -3.7  &  0.00 &  6.86E-04 \\
\smallskip \\
   1008  & 266.57822  & -26.54850  & 17.01  & 0.002  & 0.025  & 0.809  & 1.765  &      18  & -1.4  & 2.05  & -3.3  &  0.00 &  1.84E-03 \\
\smallskip \\
   1011  & 266.51968  & -31.28961  & 16.19  & 0.001  & 0.030  & 0.037  & 1.035  &      17  & -1.7  & 3.15  & -4.7  &  0.00 &  1.27E-06 \\
\smallskip \\
   1014  & 266.49261  & -31.26623  & 16.72  & 0.001  & 0.015  & 0.581  & 1.109  &      36  & -1.5  & 3.33  & -4.7  &  0.00 &  7.90E-04 \\
\smallskip \\
   1019  & 266.45236  & -26.84729  & 16.66  & 0.002  & 0.056  & 0.524  & 1.816  &       7  & -1.6  & 1.84  & -3.2  &  0.00 &  6.91E-04 \\
\smallskip \\
   1029  & 266.34171  & -26.01601  & 17.95  & 0.004  & 0.039  & 1.295  & 4.142  &      19  & -1.1  & 1.28  & -2.2  &  0.00 &  5.64E-03 \\
\smallskip \\
   1032  & 266.26807  & -26.28884  & 18.01  & 0.005  & 0.034  & 5.394  & 8.543  &      31  & -1.1  & 1.61  & -2.5  &  0.00 &  1.05E-01 \\
\smallskip \\
   1045  & 266.12482  & -26.33038  & 16.77  & 0.002  & 0.019  & 0.230  & 2.590  &      34  & -1.6  & 1.63  & -3.0  &  0.00 &  1.48E-04 \\
\smallskip \\
   1051  & 265.99564  & -27.52869  & 18.37  & 0.006  & 0.022  & 3.882  & 8.265  &      19  & -0.9  & 2.09  & -2.8  &  0.00 &  5.50E-02 \\
\smallskip \\
   1056  & 265.95566  & -26.78365  & 18.74  & 0.006  & 0.022  & 0.272  & 1.627  &      37  & -0.8  & 1.60  & -2.2  &  0.00 &  1.59E-04 \\
\smallskip \\
   1060  & 265.90619  & -26.03607  & 16.43  & 0.002  & 0.014  & 0.737  & 1.699  &      37  & -1.7  & 1.55  & -3.0  &  0.00 &  1.47E-03 \\
\smallskip \\
   1061  & 265.88643  & -26.75086  & 18.16  & 0.003  & 0.033  & 0.451  & 1.291  &      34  & -1.0  & 1.66  & -2.5  &  0.00 &  4.20E-04 \\
\smallskip \\
   1068  & 265.82571  & -25.97055  & 16.92  & 0.002  & 0.022  & 0.785  & 3.920  &      22  & -1.5  & 1.43  & -2.8  &  0.00 &  2.00E-03 \\
\smallskip \\
   1071  & 265.73792  & -27.77500  & 17.20  & 0.002  & 0.036  & 0.587  & 1.047  &      36  & -1.4  & 1.89  & -3.1  &  0.00 &  8.19E-04 \\
\smallskip \\
   1086  & 265.44363  & -27.83179  & 18.55  & 0.004  & 0.069  & 0.268  & 1.516  &      70  & -0.8  & 2.11  & -2.8  &  0.00 &  1.46E-04 \\
\smallskip \\
   1097  & 265.21237  & -27.72555  & 18.05  & 0.005  & 0.017  & 0.262  & 1.711  &      30  & -1.0  & 1.71  & -2.6  &  0.00 &  1.53E-04 \\
\smallskip \\
   1102  & 265.17502  & -28.12345  & 22.22  & 0.104  & 0.233  & 0.920  & 3.453  &      41  &  0.7  & 2.16  & -1.3  &  0.00 &  2.71E-03 \\
\smallskip \\
   1118  & 264.99039  & -28.93875  & 17.66  & 0.003  & 0.019  & 0.807  & 1.925  &      15  & -1.1  & 2.82  & -3.8  &  0.00 &  1.85E-03 \\
\smallskip \\
   1123  & 264.91211  & -27.73232  & 16.74  & 0.002  & 0.015  & 0.263  & 1.526  &      36  & -1.6  & 1.53  & -2.9  &  0.00 &  1.41E-04 \\
\smallskip \\
   1149  & 264.52695  & -29.57095  & 16.40  & 0.001  & 0.026  & 0.452  & 1.616  &      15  & -1.7  & 2.07  & -3.6  &  0.00 &  4.73E-04 \\
\smallskip \\
   1152  & 264.41760  & -28.52146  & 17.63  & 0.003  & 0.023  & 0.286  & 0.994  &      36  & -1.2  & 1.74  & -2.8  &  0.00 &  1.05E-04 \\
\smallskip \\
   1153  & 264.40826  & -28.44203  & 16.61  & 0.001  & 0.014  & 1.164  & 2.666  &      37  & -1.6  & 1.91  & -3.3  &  0.00 &  4.31E-03 \\
\smallskip \\
   1158  & 264.30957  & -29.64605  & 18.82  & 0.006  & 0.025  & 2.665  & 3.453  &      32  & -0.7  & 1.95  & -2.5  &  0.00 &  2.61E-02 \\
\smallskip \\
   1167  & 264.12970  & -28.88234  & 17.80  & 0.003  & 0.024  & 0.242  & 1.127  &      34  & -1.1  & 1.71  & -2.7  &  0.00 &  8.32E-05 \\
\smallskip \\
   1179  & 264.02588  & -29.92788  & 18.53  & 0.005  & 0.034  & 2.127  & 2.895  &      34  & -0.8  & 2.19  & -2.8  &  0.00 &  1.63E-02 \\
\smallskip \\
   1186  & 263.95166  & -30.32832  & 19.36  & 0.008  & 0.157  & 0.098  & 1.214  &      18  & -0.5  & 1.88  & -2.2  &  0.00 &  1.35E-05 \\
\smallskip \\
   1194  & 263.86462  & -30.23365  & 16.55  & 0.001  & 0.013  & 2.705  & 4.090  &      34  & -1.6  & 1.72  & -3.2  &  0.00 &  2.66E-02 \\
\smallskip \\
   1205  & 263.69189  & -30.23296  & 18.94  & 0.005  & 0.030  & 0.887  & 3.161  &      35  & -0.7  & 1.43  & -2.0  &  0.00 &  2.47E-03 \\
\smallskip \\
   1220  & 263.28711  & -30.16005  & 16.29  & 0.001  & 0.027  & 0.335  & 2.883  &      17  & -1.8  & 1.35  & -2.9  &  0.00 &  3.29E-04 \\
\smallskip \\
   1228  & 268.12955  & -28.96774  & 16.71  & 0.001  & 0.020  & 1.482  & 1.139  &      22  & -1.6  & 1.47  & -2.9  &  0.00 &  8.56E-03 \\
\smallskip \\
   1232  & 265.17218  & -27.63124  & 19.48  & 0.013  & 0.085  & 4.362  & 4.584  &      27  & -0.5  & 1.71  & -2.0  &  0.00 &  7.13E-02 \\
\smallskip \\
\end{longtable*}

%% file: ms3.bbl
\begin{thebibliography}{}
\expandafter\ifx\csname natexlab\endcsname\relax\def\natexlab#1{#1}\fi

\bibitem[{{Alard}(2000)}]{Alard00}
{Alard}, C. 2000, \aaps, 144, 363

\bibitem[{{Alard} \& {Lupton}(1998)}]{Alard98}
{Alard}, C., \& {Lupton}, R.~H. 1998, \apj, 503, 325

\bibitem[{{Bohlin} {et~al.}(1978){Bohlin}, {Savage}, \& {Drake}}]{Bohlin78}
{Bohlin}, R.~C., {Savage}, B.~D., \& {Drake}, J.~F. 1978, \apj, 224, 132

\bibitem[{{Britt} {et~al.}(2013){Britt}, {Torres}, {Hynes}, {Jonker},
  {Maccarone}, {Greiss}, {Steeghs}, {Groot}, {Knigge}, {Dieball}, {Nelemans},
  {Mikles}, \& {Gossen}}]{Britt13}
{Britt}, C.~T., {Torres}, M.~A.~P., {Hynes}, R.~I., {et~al.} 2013, \apj, 769,
  120

\bibitem[{{Cardelli} {et~al.}(1989){Cardelli}, {Clayton}, \&
  {Mathis}}]{Clayton89}
{Cardelli}, J.~A., {Clayton}, G.~C., \& {Mathis}, J.~S. 1989, \apj, 345, 245

\bibitem[{{Collins} \& {Wheatley}(2010)}]{Collins10}
{Collins}, D.~J., \& {Wheatley}, P.~J. 2010, \mnras, 402, 1816

\bibitem[{{Corral-Santana} {et~al.}(2013){Corral-Santana}, {Casares},
  {Mu{\~n}oz-Darias}, {Rodr{\'{\i}}guez-Gil}, {Shahbaz}, {Torres}, {Zurita}, \&
  {Tyndall}}]{Corral13}
{Corral-Santana}, J.~M., {Casares}, J., {Mu{\~n}oz-Darias}, T., {et~al.} 2013,
  Science, 339, 1048

\bibitem[{{DeWitt} {et~al.}(2010){DeWitt}, {Bandyopadhyay}, {Eikenberry},
  {Blum}, {Olsen}, {Sellgren}, \& {Sarajedini}}]{DeWitt10}
{DeWitt}, C., {Bandyopadhyay}, R.~M., {Eikenberry}, S.~S., {et~al.} 2010, \apj,
  721, 1663

\bibitem[{{Ducati} {et~al.}(2001){Ducati}, {Bevilacqua}, {Rembold}, \&
  {Ribeiro}}]{Ducati01}
{Ducati}, J.~R., {Bevilacqua}, C.~M., {Rembold}, S.~B., \& {Ribeiro}, D. 2001,
  \apj, 558, 309

\bibitem[{{Fender} {et~al.}(2003){Fender}, {Gallo}, \& {Jonker}}]{Fender03}
{Fender}, R.~P., {Gallo}, E., \& {Jonker}, P.~G. 2003, \mnras, 343, L99

\bibitem[{{Frank} {et~al.}(2002){Frank}, {King}, \& {Raine}}]{Frank02}
{Frank}, J., {King}, A., \& {Raine}, D.~J. 2002, {Accretion Power in
  Astrophysics: Third Edition}

\bibitem[{{G{\"a}nsicke}(2005)}]{Gansicke05}
{G{\"a}nsicke}, B.~T. 2005, in Astronomical Society of the Pacific Conference
  Series, Vol. 330, The Astrophysics of Cataclysmic Variables and Related
  Objects, ed. J.-M. {Hameury} \& J.-P. {Lasota}, 3

\bibitem[{{Garcia} {et~al.}(2001){Garcia}, {McClintock}, {Narayan}, {Callanan},
  {Barret}, \& {Murray}}]{Garcia01}
{Garcia}, M.~R., {McClintock}, J.~E., {Narayan}, R., {et~al.} 2001, \apjl, 553,
  L47

\bibitem[{{Gehrels}(1986)}]{Gehrels86}
{Gehrels}, N. 1986, \apj, 303, 336

\bibitem[{{Giacconi} {et~al.}(2001){Giacconi}, {Rosati}, {Tozzi}, {Nonino},
  {Hasinger}, {Norman}, {Bergeron}, {Borgani}, {Gilli}, {Gilmozzi}, \&
  {Zheng}}]{Giacconi01}
{Giacconi}, R., {Rosati}, P., {Tozzi}, P., {et~al.} 2001, \apj, 551, 624

\bibitem[{{Gonzalez} {et~al.}(2012){Gonzalez}, {Rejkuba}, {Zoccali}, {Valenti},
  {Minniti}, {Schultheis}, {Tobar}, \& {Chen}}]{Gonzalez12}
{Gonzalez}, O.~A., {Rejkuba}, M., {Zoccali}, M., {et~al.} 2012, \aap, 543, A13

\bibitem[{{Greiss} {et~al.}(2014){Greiss}, {Steeghs}, {Jonker}, {Torres},
  {Maccarone}, {Hynes}, \& {Britt}}]{Greiss13}
{Greiss}, S., {Steeghs}, D.~T.~H., {Jonker}, P.~G., {et~al.} 2014, \mnras, {In
  Press}

\bibitem[{{Grindlay} {et~al.}(2005){Grindlay}, {Hong}, {Zhao}, {Laycock}, {van
  den Berg}, {Koenig}, {Schlegel}, {Cohn}, {Lugger}, \& {Rogel}}]{Grindlay05}
{Grindlay}, J.~E., {Hong}, J., {Zhao}, P., {et~al.} 2005, \apj, 635, 920

\bibitem[{{Hameury} {et~al.}(2003){Hameury}, {Barret}, {Lasota}, {McClintock},
  {Menou}, {Motch}, {Olive}, \& {Webb}}]{Hameury03}
{Hameury}, J.-M., {Barret}, D., {Lasota}, J.-P., {et~al.} 2003, \aap, 399, 631

\bibitem[{{Hands} {et~al.}(2004){Hands}, {Warwick}, {Watson}, \&
  {Helfand}}]{Hands04}
{Hands}, A.~D.~P., {Warwick}, R.~S., {Watson}, M.~G., \& {Helfand}, D.~J. 2004,
  \mnras, 351, 31

\bibitem[{{Heinke} {et~al.}(2005){Heinke}, {Grindlay}, \& {Edmonds}}]{Heinke05}
{Heinke}, C.~O., {Grindlay}, J.~E., \& {Edmonds}, P.~D. 2005, \apj, 622, 556

\bibitem[{{Heinke} {et~al.}(2003){Heinke}, {Grindlay}, {Lugger}, {Cohn},
  {Edmonds}, {Lloyd}, \& {Cool}}]{Heinke03}
{Heinke}, C.~O., {Grindlay}, J.~E., {Lugger}, P.~M., {et~al.} 2003, \apj, 598,
  501

\bibitem[{{Hong} {et~al.}(2005){Hong}, {van den Berg}, {Schlegel}, {Grindlay},
  {Koenig}, {Laycock}, \& {Zhao}}]{Hong05}
{Hong}, J., {van den Berg}, M., {Schlegel}, E.~M., {et~al.} 2005, \apj, 635,
  907

\bibitem[{{Hynes} {et~al.}(2009){Hynes}, {Brien}, {Mullally}, \&
  {Ashcraft}}]{Hynes09}
{Hynes}, R.~I., {Brien}, K.~O., {Mullally}, F., \& {Ashcraft}, T. 2009, \mnras,
  399, 281

\bibitem[{{Hynes} {et~al.}(2012){Hynes}, {Wright}, {Maccarone}, {Jonker},
  {Greiss}, {Steeghs}, {Torres}, {Britt}, \& {Nelemans}}]{Hynes12}
{Hynes}, R.~I., {Wright}, N.~J., {Maccarone}, T.~J., {et~al.} 2012, \apj, 761,
  162

\bibitem[{{Hynes} {et~al.}(2014){Hynes}, {Torres}, {Heinke}, {Maccarone},
  {Mikles}, {Britt}, {Knigge}, {Greiss}, {Jonker}, {Steeghs}, {Nelemans},
  {Bandyopadhyay}, \& {Johnson}}]{Hynes14}
{Hynes}, R.~I., {Torres}, M.~A.~P., {Heinke}, C.~O., {et~al.} 2014, \apj, 780,
  11

\bibitem[{{Johnson} {et~al.}(2014){Johnson}, {Hynes}, {Britt}, {Jonker},
  {Bassa}, {Nelemans}, {Steeghs}, {Torres}, {Maccarone}, {Greiss}, \&
  {Dieball}}]{Johnson14}
{Johnson}, C.~C., {Hynes}, R.~I., {Britt}, C.~T., {et~al.} 2014, \apjs, In
  Preparation

\bibitem[{{Jonker} {et~al.}(2012){Jonker}, {Miller-Jones}, {Homan}, {Tomsick},
  {Fender}, {Kaaret}, {Markoff}, \& {Gallo}}]{Jonker12}
{Jonker}, P.~G., {Miller-Jones}, J.~C.~A., {Homan}, J., {et~al.} 2012, \mnras,
  423, 3308

\bibitem[{{Jonker} \& {Nelemans}(2004)}]{Jonker04}
{Jonker}, P.~G., \& {Nelemans}, G. 2004, \mnras, 354, 355

\bibitem[{{Jonker} {et~al.}(2008){Jonker}, {Torres}, \& {Steeghs}}]{Jonker08}
{Jonker}, P.~G., {Torres}, M.~A.~P., \& {Steeghs}, D. 2008, \apj, 680, 615

\bibitem[{{Jonker} {et~al.}(2011){Jonker}, {Bassa}, {Nelemans}, {Steeghs},
  {Torres}, {Maccarone}, {Hynes}, {Greiss}, {Clem}, {Dieball}, {Mikles},
  {Britt}, {Gossen}, {Collazzi}, {Wijnands}, {In't Zand}, {M{\'e}ndez}, {Rea},
  {Kuulkers}, {Ratti}, {van Haaften}, {Heinke}, {{\"O}zel}, {Groot}, \&
  {Verbunt}}]{Jonker11}
{Jonker}, P.~G., {Bassa}, C.~G., {Nelemans}, G., {et~al.} 2011, \apjs, 194, 18

\bibitem[{{Jonker} {et~al.}(2014){Jonker}, {Torres}, {Hynes}, {Maccarone},
  {Steeghs}, {Greiss}, {Britt}, {Wu}, {Johnson}, {Nelemans}, \&
  {Heinke}}]{Jonker14}
{Jonker}, P.~G., {Torres}, M.~A.~P., {Hynes}, R.~I., {et~al.} 2014, \apjs, 210,
  18

\bibitem[{{Kalogera} {et~al.}(2004){Kalogera}, {King}, \& {Rasio}}]{Kalogera04}
{Kalogera}, V., {King}, A.~R., \& {Rasio}, F.~A. 2004, \apjl, 601, L171

\bibitem[{{Kiel} \& {Hurley}(2006)}]{Kiel06}
{Kiel}, P.~D., \& {Hurley}, J.~R. 2006, \mnras, 369, 1152

\bibitem[{{Knevitt} {et~al.}(2014){Knevitt}, {Wynn}, {Vaughan}, \&
  {Watson}}]{Knevitt14}
{Knevitt}, G., {Wynn}, G.~A., {Vaughan}, S., \& {Watson}, M.~G. 2014, \mnras,
  437, 3087

\bibitem[{{Kong} {et~al.}(2002){Kong}, {McClintock}, {Garcia}, {Murray}, \&
  {Barret}}]{Kong02}
{Kong}, A.~K.~H., {McClintock}, J.~E., {Garcia}, M.~R., {Murray}, S.~S., \&
  {Barret}, D. 2002, \apj, 570, 277

\bibitem[{{Kuulkers} {et~al.}(2013){Kuulkers}, {Kouveliotou}, {Belloni},
  {Cadolle Bel}, {Chenevez}, {D{\'{\i}}az Trigo}, {Homan}, {Ibarra}, {Kennea},
  {Mu{\~n}oz-Darias}, {Ness}, {Parmar}, {Pollock}, {van den Heuvel}, \& {van
  der Horst}}]{Kuulkers13}
{Kuulkers}, E., {Kouveliotou}, C., {Belloni}, T., {et~al.} 2013, \aap, 552, A32

\bibitem[{{Lasota}(2008)}]{Lasota08}
{Lasota}, J.-P. 2008, \nar, 51, 752

\bibitem[{{Lewin} \& {van der Klis}(2006)}]{Lewin06}
{Lewin}, W.~H.~G., \& {van der Klis}, M. 2006, {Compact Stellar X-ray Sources}

\bibitem[{{Lomb}(1976)}]{Lomb76}
{Lomb}, N.~R. 1976, \apss, 39, 447

\bibitem[{{Lu} {et~al.}(2009){Lu}, {Kong}, {Bassa}, {Verbunt}, {Lewin},
  {Anderson}, \& {Pooley}}]{Lu09}
{Lu}, T.-N., {Kong}, A.~K.~H., {Bassa}, C., {et~al.} 2009, \apj, 705, 175

\bibitem[{{Lucas} {et~al.}(2008){Lucas}, {Hoare}, {Longmore}, {Schr{\"o}der},
  {Davis}, {Adamson}, {Bandyopadhyay}, {de Grijs}, {Smith}, {Gosling},
  {Mitchison}, {G{\'a}sp{\'a}r}, {Coe}, {Tamura}, {Parker}, {Irwin}, {Hambly},
  {Bryant}, {Collins}, {Cross}, {Evans}, {Gonzalez-Solares}, {Hodgkin},
  {Lewis}, {Read}, {Riello}, {Sutorius}, {Lawrence}, {Drew}, {Dye}, \&
  {Thompson}}]{Lucas08}
{Lucas}, P.~W., {Hoare}, M.~G., {Longmore}, A., {et~al.} 2008, \mnras, 391, 136

\bibitem[{{Maccarone} {et~al.}(2012){Maccarone}, {Torres}, {Britt}, {Greiss},
  {Hynes}, {Jonker}, {Steeghs}, {Wijnands}, \& {Nelemans}}]{Maccarone12}
{Maccarone}, T.~J., {Torres}, M.~A.~P., {Britt}, C.~T., {et~al.} 2012, \mnras,
  426, 3057

\bibitem[{{Mason} {et~al.}(1980){Mason}, {Seitzer}, {Tuohy}, {Hunt},
  {Middleditch}, {Nelson}, \& {White}}]{Mason80}
{Mason}, K.~O., {Seitzer}, P., {Tuohy}, I.~R., {et~al.} 1980, \apjl, 242, L109

\bibitem[{{Mauerhan} {et~al.}(2009){Mauerhan}, {Muno}, {Morris}, {Bauer},
  {Nishiyama}, \& {Nagata}}]{Mauerhan09}
{Mauerhan}, J.~C., {Muno}, M.~P., {Morris}, M.~R., {et~al.} 2009, \apj, 703, 30

\bibitem[{{Monet} {et~al.}(2003){Monet}, {Levine}, {Canzian}, {Ables}, {Bird},
  {Dahn}, {Guetter}, {Harris}, {Henden}, {Leggett}, {Levison}, {Luginbuhl},
  {Martini}, {Monet}, {Munn}, {Pier}, {Rhodes}, {Riepe}, {Sell}, {Stone},
  {Vrba}, {Walker}, {Westerhout}, {Brucato}, {Reid}, {Schoening}, {Hartley},
  {Read}, \& {Tritton}}]{Monet03}
{Monet}, D.~G., {Levine}, S.~E., {Canzian}, B., {et~al.} 2003, \aj, 125, 984

\bibitem[{{Motch} {et~al.}(2010){Motch}, {Warwick}, {Cropper}, {Carrera},
  {Guillout}, {Pineau}, {Pakull}, {Rosen}, {Schwope}, {Tedds}, {Webb},
  {Negueruela}, \& {Watson}}]{Motch10}
{Motch}, C., {Warwick}, R., {Cropper}, M.~S., {et~al.} 2010, \aap, 523, A92

\bibitem[{{Muno} {et~al.}(2003){Muno}, {Baganoff}, {Bautz}, {Brandt},
  {Garmire}, \& {Ricker}}]{Muno03}
{Muno}, M.~P., {Baganoff}, F.~K., {Bautz}, M.~W., {et~al.} 2003, \apj, 599, 465

\bibitem[{{Narayan} {et~al.}(1997){Narayan}, {Garcia}, \&
  {McClintock}}]{Narayan97}
{Narayan}, R., {Garcia}, M.~R., \& {McClintock}, J.~E. 1997, \apjl, 478, L79

\bibitem[{{Narayan} \& {McClintock}(2008)}]{Narayan08}
{Narayan}, R., \& {McClintock}, J.~E. 2008, \nar, 51, 733

\bibitem[{{Patterson} \& {Raymond}(1985)}]{Patterson85}
{Patterson}, J., \& {Raymond}, J.~C. 1985, \apj, 292, 535

\bibitem[{{Peacock} {et~al.}(2009){Peacock}, {Maccarone}, {Waters}, {Kundu},
  {Zepf}, {Knigge}, \& {Zurek}}]{Peacock09}
{Peacock}, M.~B., {Maccarone}, T.~J., {Waters}, C.~Z., {et~al.} 2009, \mnras,
  392, L55

\bibitem[{{Pfahl} {et~al.}(2003){Pfahl}, {Rappaport}, \&
  {Podsiadlowski}}]{Pfahl03}
{Pfahl}, E., {Rappaport}, S., \& {Podsiadlowski}, P. 2003, \apj, 597, 1036

\bibitem[{{Pooley} \& {Hut}(2006)}]{Pooley06}
{Pooley}, D., \& {Hut}, P. 2006, \apjl, 646, L143

\bibitem[{{Pooley} {et~al.}(2003){Pooley}, {Lewin}, {Anderson}, {Baumgardt},
  {Filippenko}, {Gaensler}, {Homer}, {Hut}, {Kaspi}, {Makino}, {Margon},
  {McMillan}, {Portegies Zwart}, {van der Klis}, \& {Verbunt}}]{Pooley03}
{Pooley}, D., {Lewin}, W.~H.~G., {Anderson}, S.~F., {et~al.} 2003, \apjl, 591,
  L131

\bibitem[{{Portegies Zwart} {et~al.}(1997){Portegies Zwart}, {Verbunt}, \&
  {Ergma}}]{Zwart97}
{Portegies Zwart}, S.~F., {Verbunt}, F., \& {Ergma}, E. 1997, \aap, 321, 207

\bibitem[{{Ratti} {et~al.}(2013){Ratti}, {van Grunsven}, {Jonker}, {Britt},
  {Hynes}, {Steeghs}, {Greiss}, {Torres}, {Maccarone}, {Groot}, {Knigge},
  {Gossen}, {Mikles}, {Villar}, \& {Collazzi}}]{Ratti13}
{Ratti}, E.~M., {van Grunsven}, T.~F.~J., {Jonker}, P.~G., {et~al.} 2013,
  \mnras, 428, 3543

\bibitem[{{Rea} {et~al.}(2011){Rea}, {Jonker}, {Nelemans}, {Pons}, {Kasliwal},
  {Kulkarni}, \& {Wijnands}}]{Rea11}
{Rea}, N., {Jonker}, P.~G., {Nelemans}, G., {et~al.} 2011, \apjl, 729, L21

\bibitem[{{Repetto} {et~al.}(2012){Repetto}, {Davies}, \&
  {Sigurdsson}}]{Repetto12}
{Repetto}, S., {Davies}, M.~B., \& {Sigurdsson}, S. 2012, \mnras, 425, 2799

\bibitem[{{Scargle}(1982)}]{Scargle82}
{Scargle}, J.~D. 1982, \apj, 263, 835

\bibitem[{{Schandl} {et~al.}(1997){Schandl}, {Meyer-Hofmeister}, \&
  {Meyer}}]{Schandl97}
{Schandl}, S., {Meyer-Hofmeister}, E., \& {Meyer}, F. 1997, \aap, 318, 73

\bibitem[{{Schlegel} {et~al.}(1998){Schlegel}, {Finkbeiner}, \&
  {Davis}}]{Schlegel98}
{Schlegel}, D.~J., {Finkbeiner}, D.~P., \& {Davis}, M. 1998, \apj, 500, 525

\bibitem[{{Servillat} {et~al.}(2012){Servillat}, {Grindlay}, {van den Berg},
  {Hong}, {Zhao}, \& {Allen}}]{Servillat12}
{Servillat}, M., {Grindlay}, J., {van den Berg}, M., {et~al.} 2012, \apj, 748,
  32

\bibitem[{{Shaw}(2009)}]{Shaw09}
{Shaw}, R.~A. 2009, {NOAO Data Handbook}

\bibitem[{{Skrutskie} {et~al.}(2006){Skrutskie}, {Cutri}, {Stiening},
  {Weinberg}, {Schneider}, {Carpenter}, {Beichman}, {Capps}, {Chester},
  {Elias}, {Huchra}, {Liebert}, {Lonsdale}, {Monet}, {Price}, {Seitzer},
  {Jarrett}, {Kirkpatrick}, {Gizis}, {Howard}, {Evans}, {Fowler}, {Fullmer},
  {Hurt}, {Light}, {Kopan}, {Marsh}, {McCallon}, {Tam}, {Van Dyk}, \&
  {Wheelock}}]{Skrutskie06}
{Skrutskie}, M.~F., {Cutri}, R.~M., {Stiening}, R., {et~al.} 2006, \aj, 131,
  1163

\bibitem[{{Soszy{\'n}ski} {et~al.}(2011{\natexlab{a}}){Soszy{\'n}ski},
  {Dziembowski}, {Udalski}, {Poleski}, {Szyma{\'n}ski}, {Kubiak},
  {Pietrzy{\'n}ski}, {Wyrzykowski}, {Ulaczyk}, {Koz{\l}owski}, \&
  {Pietrukowicz}}]{Soszynski11a}
{Soszy{\'n}ski}, I., {Dziembowski}, W.~A., {Udalski}, A., {et~al.}
  2011{\natexlab{a}}, \actaa, 61, 1

\bibitem[{{Soszy{\'n}ski} {et~al.}(2011{\natexlab{b}}){Soszy{\'n}ski},
  {Udalski}, {Pietrukowicz}, {Szyma{\'n}ski}, {Kubiak}, {Pietrzy{\'n}ski},
  {Wyrzykowski}, {Ulaczyk}, {Poleski}, \& {Koz{\l}owski}}]{Soszynski11b}
{Soszy{\'n}ski}, I., {Udalski}, A., {Pietrukowicz}, P., {et~al.}
  2011{\natexlab{b}}, \actaa, 61, 285

\bibitem[{{Stellingwerf}(1978)}]{Stellingwerf78}
{Stellingwerf}, R.~F. 1978, \apj, 224, 953

\bibitem[{{Stetson}(1987)}]{Stetson87}
{Stetson}, P.~B. 1987, \pasp, 99, 191

\bibitem[{{Swank} {et~al.}(1978){Swank}, {Boldt}, {Holt}, {Rothschild}, \&
  {Serlemitsos}}]{Swank78}
{Swank}, J.~H., {Boldt}, E.~A., {Holt}, S.~S., {Rothschild}, R.~E., \&
  {Serlemitsos}, P.~J. 1978, \apjl, 226, L133

\bibitem[{{Szyma{\'n}ski} {et~al.}(2011){Szyma{\'n}ski}, {Udalski},
  {Soszy{\'n}ski}, {Kubiak}, {Pietrzy{\'n}ski}, {Poleski}, {Wyrzykowski}, \&
  {Ulaczyk}}]{Szymanski11}
{Szyma{\'n}ski}, M.~K., {Udalski}, A., {Soszy{\'n}ski}, I., {et~al.} 2011,
  \actaa, 61, 83

\bibitem[{{Torres} {et~al.}(2014){Torres}, {Jonker}, {Bassa}, {Nelemans},
  {Steeghs}, {Maccarone}, {Hynes}, {Greiss}, {Dieball}, {Britt}, \&
  {Ratti}}]{Torres14}
{Torres}, M.~A.~P., {Jonker}, P.~G., {Bassa}, C.~G., {et~al.} 2014, \apj,
  Submitted to MNRAS

\bibitem[{{Udalski} {et~al.}(2012){Udalski}, {Kowalczyk}, {Soszy{\'n}ski},
  {Poleski}, {Szyma{\'n}ski}, {Kubiak}, {Pietrzy{\'n}ski}, {Koz{\l}owski},
  {Pietrukowicz}, {Ulaczyk}, {Skowron}, \& {Wyrzykowski}}]{Udalski12}
{Udalski}, A., {Kowalczyk}, K., {Soszy{\'n}ski}, I., {et~al.} 2012, AcA, 62,
  133

\bibitem[{{van den Berg} {et~al.}(2009){van den Berg}, {Hong}, \&
  {Grindlay}}]{vandenBerg09}
{van den Berg}, M., {Hong}, J.~S., \& {Grindlay}, J.~E. 2009, \apj, 700, 1702

\bibitem[{{van den Berg} {et~al.}(2012){van den Berg}, {Penner}, {Hong},
  {Grindlay}, {Zhao}, {Laycock}, \& {Servillat}}]{vandenBerg12}
{van den Berg}, M., {Penner}, K., {Hong}, J., {et~al.} 2012, \apj, 748, 31

\bibitem[{{Vaughan}(2010)}]{Vaughan10}
{Vaughan}, S. 2010, \mnras, 402, 307

\bibitem[{{Warner}(1976)}]{Warner76}
{Warner}, B. 1976, in IAU Symposium, Vol.~73, Structure and Evolution of Close
  Binary Systems, ed. P.~{Eggleton}, S.~{Mitton}, \& J.~{Whelan}, 85

\bibitem[{{Warner}(2003)}]{Warner03}
{Warner}, B. 2003, {Cataclysmic Variable Stars}

\bibitem[{{Wijnands} {et~al.}(2005){Wijnands}, {Heinke}, {Pooley}, {Edmonds},
  {Lewin}, {Grindlay}, {Jonker}, \& {Miller}}]{Wijnands05}
{Wijnands}, R., {Heinke}, C.~O., {Pooley}, D., {et~al.} 2005, \apj, 618, 883

\bibitem[{{Wu} {et~al.}(2014){Wu}, {Jonker}, {Torres}, {Britt}, {Hynes},
  {Greiss}, {Steeghs}, \& {Maccarone}}]{Jianfeng14}
{Wu}, J., {Jonker}, P.~G., {Torres}, M.~A.~P., {et~al.} 2014, \mnras, In
  Preparation

\bibitem[{{Wu} {et~al.}(2010){Wu}, {Yu}, {Li}, {Maccarone}, \& {Li}}]{Wu10}
{Wu}, Y.~X., {Yu}, W., {Li}, T.~P., {Maccarone}, T.~J., \& {Li}, X.~D. 2010,
  \apj, 718, 620

\bibitem[{{Zacharias} {et~al.}(2009){Zacharias}, {Finch}, {Girard}, {Hambly},
  {Wycoff}, {Zacharias}, {Castillo}, {Corbin}, {Divittorio}, {Dutta}, {Gaume},
  {Gauss}, {Germain}, {Hall}, {Hartkopf}, {Hsu}, {Holdenried}, {Makarov},
  {Martinez}, {Mason}, {Monet}, {Rafferty}, {Rhodes}, {Siemers}, {Smith},
  {Tilleman}, {Urban}, {Wieder}, {Winter}, \& {Young}}]{Zacharias09}
{Zacharias}, N., {Finch}, C., {Girard}, T., {et~al.} 2009, VizieR Online Data
  Catalog, 1315, 0

\bibitem[{{Zurita} {et~al.}(2008){Zurita}, {Durant}, {Torres}, {Shahbaz},
  {Casares}, \& {Steeghs}}]{Zurita08}
{Zurita}, C., {Durant}, M., {Torres}, M.~A.~P., {et~al.} 2008, \apj, 681, 1458

\end{thebibliography}
